\documentclass[12pt]{book}

\usepackage{latexsym}
\usepackage{amsmath}
\usepackage{amsfonts}
\usepackage{amssymb}
\usepackage[all]{xy}
\usepackage{amscd}
\usepackage{fancybox}
\usepackage{epsfig}

\usepackage{multirow}
\usepackage{fancyhdr}
\usepackage{lscape}

\fancypagestyle{plain}{\fancyhead{} 
}
\newcommand{\cK}{{\cal K}}
\newcommand{\cL}{{\cal L}}
\newcommand{\cM}{{\cal M}}
\newcommand{\cN}{{\cal N}}

\newcommand{\cP}{{\cal P}}
\newcommand{\cQ}{{\cal Q}}

\newcommand{\beq}{\begin{equation}}
\newcommand{\eeq}{\end{equation}}
\newcommand{\bi}{\begin{itemize}}
\newcommand{\ei}{\end{itemize}}
\newcommand{\bea}{\begin{eqnarray}}
\newcommand{\eea}{\end{eqnarray}}
\newcommand{\ba}{\begin{array}}
\newcommand{\ea}{\end{array}}
\newcommand{\bt}{\begin{tabular}}
\newcommand{\et}{\end{tabular}}
\newcommand{\bc}{\begin{center}}
\newcommand{\ec}{\end{center}}

\def\theequation{\arabic{section}.\arabic{equation}}

\newcommand{\vev}[1]{\langle#1\rangle}

\newcommand{\trans}[1]{{\vphantom{#1}}^t{#1}}
\newcommand{\ft}[2]{{\textstyle {\frac{#1}{#2}} }}
\newcommand{\ga}[3]{\Gamma^{#1}{}_{#3}{}^{#2}}
\newcommand{\gaquer}[3]{\overline{\Gamma}\vphantom{\Gamma}^{#1}{}_{#3}{}^{#2}}
\newcommand{\li}[1]{\hat{#1}}
\newcommand{\f}[1]{\boldsymbol{#1}}
\newcommand{\Oplus}{\hspace{0.3em}\oplus\hspace{0.3em}}

\newcommand{\Ga}{\alpha}

\newcommand{\GD}{\Delta}

\newcommand{\GTh}{\Theta}

\newcommand{\vl}{{\vphantom{[}}}

\begin{document}

\numberwithin{equation}{chapter}
\numberwithin{section}{chapter}

\thispagestyle{empty}

\newpage
\pagenumbering{roman}
\thispagestyle{plain}

\begin{titlepage}
\begin{center}

\hfill hep-th/0612235\\
\hfill DESY-THESIS-2006-013

\vskip 0.5cm

{\LARGE \bf Massive Kaluza-Klein Theories and their Spontaneously
      Broken Symmetries\footnote{Based on the author's Ph.D. thesis, defended at
the II. Institute for Theoretical Physics, University of Hamburg,
July 2006.}
\\[0.2cm]}

\vskip 1.5cm

{\bf Olaf Hohm} \\
\vskip 20pt

{\em Spinoza Institute and Institute for Theoretical Physics\\
Utrecht University \\
Leuvenlaan 4, 3584 CE Utrecht,\\
The Netherlands \vskip 5pt }

{email: {\tt o.hohm@phys.uu.nl}} \\

\vskip 0.8cm

\end{center}

\vskip 0.2cm

\begin{center} {\bf ABSTRACT}\\[3ex]

\begin{minipage}{13cm}
\small In this thesis we investigate the effective actions for
massive Kaluza-Klein states, focusing on the massive modes of
spin-3/2 and spin-2 fields. To this end we determine the
spontaneously broken gauge symmetries associated to these
`higher-spin' states and construct the unbroken phase of the
Kaluza-Klein theory. We show that for the particular background
$AdS_3\times S^3\times S^3$ a consistent coupling of the first
massive spin-3/2 multiplet requires an enhancement of local
supersymmetry, which in turn will be partially broken in the
Kaluza-Klein vacuum. The corresponding action is constructed as a
gauged maximal supergravity in $D=3$. Subsequently, the symmetries
underlying an infinite tower of massive spin-2 states are analyzed
in case of a Kaluza-Klein compactification of four-dimensional
gravity to $D=3$. It is shown that the resulting gravity--spin-2
theory is given by a Chern-Simons action of an affine algebra. 
The global symmetry group
is determined, which contains an affine extension of the Ehlers
group. We show that the broken phase can in turn be constructed
via gauging a certain subgroup of the global symmetry group.
Finally, deformations of the Kaluza-Klein theory on $AdS_3\times
S^3\times S^3$ and the corresponding symmetry breakings are
analyzed as possible applications for the AdS/CFT correspondence.

\end{minipage}

\end{center}

\noindent



December 2006

\end{titlepage}

\newpage

\vspace*{15cm}

\begin{center}
 {\bf \large Acknowledgments}
\end{center}
I am greatly indebted to Henning Samtleben for his excellent
supervision. I would also like to thank Jan Louis for supporting
this PhD thesis in many ways.

For scientific discussions and/or social interactions I would like
to thank Iman Benmachiche, Christian Becker, Marcus Berg, Jens
Fjelstad, Thomas Grimm, Hans Jockers, Paolo Merlatti, Falk
Neugebohrn, Thorsten Pr\"ustel, Sakura Sch\"afer-Nameki, Bastiaan
Spanjaard, Silvia Vaula, Martin Weidner and Mattias Wohlfarth.

This work has been supported by the EU contracts
MRTN-CT-2004-503369 and MRTN-CT-2004-512194, the DFG grant SA
1336/1-1 and DAAD -- The German Academic Exchange Service.

\tableofcontents

\newpage

\pagenumbering{arabic}

\chapter{Introduction}

\section[String theory as a candidate for quantum gravity]
{String theory as a candidate for quantum \\gravity}\label{strings}
By now it is common wisdom that one of the most important and
also most challenging problems of theoretical physics is the
reconciliation of gravity, i.e. of general relativity, with
quantum theory. Any `naive' approach to the problem of quantizing
gravity along the lines of conventional field theories has failed,
in that the resulting theories are non-renormalizable, so that
they can be viewed at best only as an effective description.

It is generally appreciated that the
conceptually different nature of general relativity lies at
the heart of the problem of formulating a consistent theory of
quantum gravity.
As a so-called background-independent theory it is invariant
under the full diffeomorphism group of the underlying manifold.
In accordance with that there is no preferred reference frame and
no fixed background space-time. Instead, space-time is completely
determined dynamically. Since any conventional formulation of
quantum field theory relies in contrast on a background space-time
(as Minkowski space) with its preferred notion of distance, causality
and time, it is not surprising that the quantization of gravity
is a complicated problem.
Therefore, quantum gravity struggles with a number
of conceptually severe problems, like the famous so-called
`problem of time' (for a pedagogical introduction see, e.g.,
\cite{Isham:1991mm}.)

Among the approaches to quantize gravity are those which try to
maintain the conventional ideas of quantum field theory, like
the notion of particles, S-matrix etc., while the others
aim to take seriously the lessons which general relativity has taught us
by maintaining general covariance also in the
quantum theory. For instance,
`loop quantum gravity' belongs to the latter
\cite{Rovelli:2004tv,Thiemann:2001yy,Gaul:1999ys}.
While these theories are truly
background-independent, they suffer from the humbling drawback that
it is not clear how to get a semi-classical limit or even how to
identify the observables \cite{Nicolai:2005mc}.
String theory on the other hand is a
perfectly well-defined \textit{perturbative} quantum theory of
gravity \cite{Green:1987sp,Polchinski:1998rq}.
Even though it extends the standard concepts of quantum
field theory in the sense that it replaces the notion of a point-particle
by a one-dimensional string, it still yields a conventional
formulation in that it basically predicts certain S-matrix elements.
It provides a framework to compute finite results for,
say, graviton-graviton-scattering on a Minkowski background
to arbitrary accuracy.
Accordingly, string theory in its original
formulation is not background-independent, since it quantizes
the propagation of strings on a fixed 10-dimensional background
space-time (which is usually taken to be Minkowski space).
Among the excitations of the string is the massless spin-2 mode,
which is interpreted as the graviton, i.e.~as the (quantized)
fluctuations of the metric around the fixed background
space-time.\footnote{From a particle physicist's point of view
the diffeomorphism symmetry of general relativity is accordingly
the gauge symmetry that guarantees consistency of the graviton's
self interactions.}
Even though string theory thus provides a consistent theory
of quantum gravity, it is unable to answer the conceptually important
problems mentioned above, simply because of its intrinsically perturbative
formulation. In particular, string theory incorporates
the symmetry principles known from classical general relativity at
best only in a quite intricate way.

Apart from the massless spin-2 mode, the string spectrum contains
massless~spin-0 and spin-1 states together with their
fermionic superpartners, as well as an infinite tower of massive
higher-spin states. The former support the picture of unification,
namely that all known interactions and matter fields should
combine into one common theory. The latter in turn have been
argued to result from a spontaneous symmetry breaking of an
infinite-dimensional symmetry \cite{Gross:1988ue,Evans:1989cs}.
But, what kind of symmetry this
might be and how the `unbroken phase' of string theory might look
like has always been unclear. Again we are faced with the problem
that the (symmetry) principles underlying
string theory remain unknown.
It would be clearly desirable to uncover these principles,
since they would presumably enrich our understanding in the same
sense as the notion of general covariance and the equivalence
principle have done for our understanding of gravity \cite{Witten:1988sy}.

It should be noted that string theory is not unique,
instead there are five consistent (super-) string theories in 10 dimensions:
Type IIA and type IIB, heterotic string theory with gauge groups
$SO(32)$ and $E_8\times E_8$ and type I superstring theory.
In their low-energy description, i.e.~in the limit that the
strings look effectively like particles, they are described by
field theories, which are the associated supergravity theories.
Beyond these supergravities related to string theories there
exists a unique 11-dimensional supergravity theory, whose
origin was unclear from the point of view of `conventional'
string theory.

However, this picture of string theory changed dramatically
during the second string revolution in 1995
\cite{Witten:1995ex} (for reviews see \cite{Townsend:1996xj,Duff:1996aw}).
Since then various dualities have been argued to exist,
which relate the different string theories together with 11-dimensional
supergravity.
As the result of this change of perspective it is now expected
that there should exist a unifying framework for string theory,
which is called `M-theory', that presumably combines all
known string theories into one common theory and whose low-energy
limit is given by 11-dimensional supergravity.
The different string theories will
then appear as certain limits of M-theory.
This theory is believed to provide a background-independent
description of quantum gravity
(in which not even the space-time dimension is given
a priori), which at the same time unifies all
known interactions and matter fields. In particular, space-time
should appear as an emergent phenomenon \cite{Horowitz:2004rn}.
Moreover, such a theory should also be able to answer the
conceptual questions about quantum gravity mentioned above.
Thus, the question of the `unbroken phase' of string theory translates
in modern parlance into the question of the underlying symmetries
of M-theory, whose identification would presumably be the first
step in identifying this theory.
Even though we are far away from an actual formulation of
M-theory, the first glimpses of such a deeper understanding
have already shown up, e.g.~in the AdS/CFT correspondence
or the holographic principle, which will be important in this
thesis.

\section{Holography and Kaluza-Klein theories}
It was first proposed by 't Hooft and Susskind that a quantum
theory of gravity should pervade a so-called
holographic principle \cite{'tHooft:1993gx,Susskind:1994vu}.
This principle states that quantum gravity ought to have
some description in terms of degrees of freedom that are defined
in one dimension less. The reason for such an expectation results
from the Bekenstein-Hawking formula for the entropy of a black hole,
according to which the entropy is given not by the volume,
but instead by the area of the black hole's event horizon.
This in turn yields a bound for the entropy of a
given region of space-time.
In fact, if the entropy for such a region would be
greater than expected from the area of its boundary
one could violate the second law of thermodynamics by throwing
matter into this region until a black hole forms.
The latter would then have an entropy bounded by its area,
thus implying an effective decrease of entropy.
Since the entropy counts the number of microstates in that region,
the entire information about the degrees of freedom therefore
seems to be contained already in the boundary,
i.e. effectively in one dimension less.

The first concrete realization of this principle was given by
Maldacena's conjecture or the so-called AdS/CFT correspondence
\cite{Maldacena:1997re,Aharony:1999ti}.
It claims an equivalence between certain string theories on
backgrounds containing Anti-de Sitter spaces (AdS) on the one hand
and conformal field theories on its boundary on the other hand.
The most prominent form of the
correspondence is the one between type IIB string theory on
$AdS_5\times S^5$ and ${\cal N}=4$ super-Yang-Mills theory
on four-dimensional Minkowski space (which may be viewed as the
boundary of $AdS_5$). Another form of the correspondence, which we will
mainly focus on in this thesis, is between type IIB string theory
on $AdS_3\times S^3\times K$ (where the compact manifold $K$ can be
$S^3\times S^1$, $T^4$ or $K3$) and certain two-dimensional
conformal field theories.

More technically, the AdS/CFT correspondence associates to
any field in the supergravity theory on AdS
a source $\phi_0$ in the CFT.\footnote{In the following we will
restrict ourselves mainly to the case, where the supergravity
approximation of string theory is valid.} Any of these sources gives
rise to a unique solution of the supergravity equations of motion
which coincides with the CFT data on the boundary of AdS.
The AdS/CFT duality in turn claims a precise relation
between the correlation functions in the CFT and the supergravity
action. Schematically, the correlators can be computed
according to the formula
 \bea
  \vev{\exp \int_{S^d}\phi_0\mathcal{O}}_{\text{CFT}}
  =\exp (-S(\phi))\;,
 \eea
where the supergravity action $S$ has to be evaluated on the appropriate
solutions of its equations of motion, and $\mathcal{O}$ denotes an
operator in the CFT. Thus the knowledge of
the full supergravity action allows the computation of all
correlation functions, and vice versa.
To be more precise, for the evaluation of higher-order
correlators, all \textit{non-linear} couplings in the
supergravity action have to be known.
For instance, if one wants to follow an RG flow in a
broken phase of the field theory, one has to analyze
arbitrary finite movements of the supergravity scalars in their
potential. Therefore the precise shape of the scalar potential
has to be known beyond quadratic order.

Concerning the problem of determining the non-linear
couplings, it should be stressed that the AdS/CFT correspondence
involves the full ten-dimensional string or supergravity theory,
and not just the AdS factor. In fact, consistent geometric string
backgrounds have to be ten-dimensional and should solve the supergravity
equations of motion. Accordingly, the internal manifolds in
$AdS_5\times S^5$ or $AdS_3\times S^3\times K$ enter the
supergravity action in a crucial way through the appearance
of so-called Kaluza-Klein harmonics and their non-linear couplings.

Kaluza-Klein theories were originally introduced as an attempt
to unify gravity with Yang-Mills gauge theories
through the introduction
of higher-dimensional space-times \cite{Appelquist:1987nr}.
They are based on the assumption that the ground
state of a gravity theory might not be given by a maximally symmetric
space (i.e.~Minkowski or de Sitter spaces), but instead by a
product of a lower-dimensional space with a compact manifold
('spontaneous compactification').
The ground state manifold then may read
 \bea
  M_D = \mathbb{R}^{1,d-1}\times K_{D-d}\;,
 \eea
where $K_{D-d}$ denotes the compact internal manifold.
The fields of the $D$-dimensional gravity theory can then be
expanded in harmonics $Y_a$, satisfying
 \bea
  \left(\square_{K_{D-d}}+m_a^2\right)Y_a=0\;,
 \eea
after which one integrates the original action
over the internal manifold. The effectively $d$-dimensional action
then looks like a gravity theory composed of a finite
number of massless fields (resulting from the zero-modes $m_a=0$),
coupled to an infinite tower of massive modes.
Among the massless fields one finds Yang-Mills gauge fields,
whose gauge group is given by the isometry group of the internal
manifold, thus unifying space-time and internal symmetries.
The massive states in turn have a mass scale
$1/R$, where $R$ is some characteristic length scale
of the internal manifold (as, e.g., the radius of a circle).
In most phenomenological considerations one assumes the compact
manifold to be very small, such that the higher Kaluza-Klein modes
are heavy and can in fact be integrated out.
However, the internal manifolds
appearing in compactifications on AdS are typically of the
same size as the AdS space itself, i.e. of `cosmological' scales, in order
to be a solution of the supergravity equations of motion.\footnote{This
will be shown explicitly in chapter~\ref{spin32chapter}
for the $AdS_3\times S^3\times S^3\times S^1$ background.}
The Kaluza-Klein
harmonics have actually masses of order one, in units of the AdS length
scale, while in contrast stringy excitations are much heavier.
Thus, the inclusion of massive Kaluza-Klein states in the
supergravity description is of crucial importance, while so far
the focus was mainly on the effective description of the
lowest modes. In this thesis we therefore aim to incorporate
also Kaluza-Klein states of arbitrary mass.
Since the direct construction of effective Kaluza-Klein
actions in case of a generic compact
manifold is a technically cumbersome problem, we will follow here instead a
different strategy of constructing these theories directly in the
lower-dimensional spacetime by identifying the underlying
symmetries and the allowed couplings.

This strategy has also been followed in the original constructions of
the effective supergravity on $AdS_5\times S^5$
\cite{Gunaydin:1984fk,Pernici:1985ju,Gunaydin:1984qu} (and also of
11-dimensional supergravity on $AdS_4\times S^7$
and others \cite{deWit:1981eq,deWit:1982ig,Duff:1983gq}).
The accomplishment of such a program is based on so-called
gauged supergravities, which incorporate non-abelian
gauge vectors into locally supersymmetric theories.
Since the sphere $S^5$ leads according to the general Kaluza-Klein
recipe to a gauge group $SO(6)$ and preserves moreover all supercharges,
the theory can be constructed directly as a gauged maximal supergravity
with this particular gauge group.
This theory is unique, and it is
therefore natural to identify it with a truncation to the 5-dimensional
maximal supergravity multiplet of the full Kaluza-Klein theory.
Even though this cannot be viewed as an effective description
due to the low mass scale of the higher modes, it is
believed to provide a consistent truncation.\footnote{We will
make this more precise in sec. \ref{truncation}, and will also
try to illuminate why this statement is far from being self-evident.
Nevertheless, throughout the thesis we will use the term `effective'
for the resulting Kaluza-Klein actions, even though it is
in general not an effective description in the sense of
`integrating out' degrees of freedom in field theory.}
More recently, a tower of massive spin-1 multiplets appearing on
$AdS_3\times S^3$ has been described as a unique gauging
of three-dimensional supergravity \cite{Nicolai:2003ux}.
(Effective supergravities for massive Kaluza-Klein states
have also been considered in different contexts
in \cite{Giveon:1990mw,Giveon:1990er,Duff:1989cr}
and more recently in \cite{Grana:2005ny,Benmachiche:2006df}.)

As long as only scalar fields and their superpartner
are considered, this strategy of constructing the effective
supergravities applies.
However, these theories
require already the introduction of spin-1 fields as gauge fields.
Moreover, as supergravities they contain spin-3/2 fields
(gravitinos) and a spin-2 field (the metric).
These `higher-spin' fields are associated to local symmetries in the
AdS bulk.\footnote{These local symmetries do not appear on the CFT side,
but correspond to conserved currents.}
This in turn is the main obstacle for the direct construction of
supergravity theories containing also higher Kaluza-Klein modes.
In fact, the construction of interacting higher-spin theories seems
to be prohibited in general by various no-go theorems which in principle
apply to all fields with $s>1$.
The appearing consistency
problems are precisely due to the fact that the local higher-spin
symmetries apparently cannot be maintained at the interacting
level \cite{Sorokin:2004ie}.
Supergravity theories handle these no-go theorems
in that they are able to couple the graviton self-consistently
(as in general relativity) together with a finite number of gravitinos
(due to the existence of local supersymmetry).
However, the consistency problems seem to reappear once all
Kaluza-Klein modes are taken into account.
The higher Kaluza-Klein modes of the metric and the gravitino
will in fact show up as infinite towers of massive spin-2 and
spin-3/2 fields coupled to gravity.
While a finite number of
massive spin-3/2 fields can be described within the framework of
spontaneously broken supersymmetry (limited by the maximal number
of real supercharges $\leq 32$), an arbitrary number of them cannot
be coupled consistently to gravity -- not to speak of the spin-2 fields.
On the other hand, we know that the
infinite Kaluza-Klein towers of massive spin-2 and spin-3/2 fields
have to be consistent, simply because their higher-dimensional
ancestors are.
Therefore the question appears, how are the no-go results mentioned
above circumvented?
Moreover, is it possible not only to describe
these massive higher-spin states in a consistent way, but to
realize them as the spontaneously broken phase of a
theory possessing an enhanced local symmetry?

These questions were in part analyzed by Dolan and Duff
in \cite{Dolan:1983aa}, where they showed that the higher-dimensional
diffeomorphism group shows up as a spontaneously broken
infinite-dimensional gauge symmetry in the lower-dimensional
Kaluza-Klein theory. More precisely, the diffeomorphism group
of the internal manifold will appear as a Yang-Mills-like gauge
group, while an infinite-dimensional spin-2 symmetry is supposed to
ensure the consistency of the gravity--spin-2 couplings.
In subsequent investigations the focus was on the realization
of the diffeomorphism algebra as a gauge symmetry
\cite{Aulakh:1985un,Cho:1992xv,Cho:1991xk}.
In this thesis we will concentrate instead on the consistency of the
infinite-dimensional spin-2 symmetries.
Moreover, we will also discuss the similar phenomenon for
spin-3/2 symmetries (for the early literature see \cite{Dolan:1984fm}).

The strategy in this thesis will be to focus on compactifications
to three space-time dimensions, and
leave possible generalizations to arbitrary dimensions
for a second step. Among other things
this is motivated by the fact that the existence
of so-called Chern-Simons gauge theories allows an investigation of local
symmetries, which treats internal and space-time symmetries
on an equal footing. For instance, general relativity in three dimensions
has an interpretation as a Chern-Simons theory, in which the diffeomorphisms
are realized as Yang-Mills gauge transformations \cite{Witten:1988hc}.
Accordingly, all higher-spin fields -- starting with spin-1 vectors --
can be described in a similar fashion \cite{Blencowe:1988gj}.
Thus Chern-Simons theories provide the natural arena for an analysis of the
spin-2 and other local (super-)symmetries, which are
expected to appear in Kaluza-Klein theories.
Finally, gauged supergravities in three dimensions represent a natural
extension of these Chern-Simons theories (which are contained as subsectors)
and, correspondingly, are well adapted to an analysis by
symmetry arguments.
Irrespective of the meaning as a toy-model for higher-dimensional cases,
these theories are of significance by themselves, as they afford
the necessary background for an investigation of the dual two-dimensional
CFT's.

Apart from the relevance for the AdS/CFT correspondence
and the conceptual understanding of Kaluza-Klein theories
in general, these questions have a
striking similarity with the analogous search for the
underlying symmetries in string or M-theory mentioned in \ref{strings},
and one may even hope to get some new insights into these subjects.


\section{Outline of the thesis}

This thesis aims to analyze massive Kaluza-Klein theories through
their spontaneously broken symmetries, with a focus on the consistency
of massive spin-3/2 and spin-2 couplings. The organization is
as follows.

In chapter 2 we give a brief review of the consistency problems
related to higher-spin couplings in general, and how they are partially
resolved in supergravity. Then we give a short introduction into
gauged supergravity in three dimensions, which will be applied in
forthcoming chapters.

In chapter 3 we turn to the problem of coupling massive spin-3/2
fields to gravity. For this we consider the example of
Kaluza-Klein supergravity on $AdS_3\times S^3\times S^3\times S^1$.
This background is half-maximally supersymmetric and contains
in its Kaluza-Klein spectrum massive supermultiplets with maximal
spin 3/2. We construct their respective effective actions
as gauged maximal supergravities in $D=3$, whose supersymmetry
is partially broken in the vacuum, thus giving rise to massive
spin-3/2 fields via a super-Higgs mechanism.

The analogous problem for spin-2 fields is discussed in chapter 4.
We show that an `unbroken phase' exists, in which the spin-2 fields
appear to be massless and therefore possess an enhanced
(infinite-dimensional) gauge symmetry, thus circumventing
the no-go theorems. It will be shown that a
geometrical interpretation exists, which is analogous to the one
for general relativity, and consists of a notion of
`algebra-valued' differential geometry developed by Wald.
Moreover we will see that the `broken phase' and the affiliated
Higgs mechanism originate from a gauging of certain global symmetries,
which is to a certain extent similar to the gauging of supergravity
introduced in chapter 2.
In particular, the rigid invariance group is enhanced and the
gauge fields will show up together with the spin-2 fields
in a Chern-Simons form.

In chapter 5 we discuss potential applications of the
results from chapter 3 for the AdS/CFT correspondence.
We consider so-called marginal deformations, which leave the
AdS background intact, but break some supercharges and part of the
gauge group spontaneously. More specifically, we discuss a breaking
of ${\cal N}=(4,4)$ to ${\cal N}=(4,0)$ and ${\cal N}=(3,3)$,
respectively, together with the resulting spectrum and the
reorganization into supermultiplets.

Chapter 6 closes with an outlook and discussions, while the
appendices contain the required technical background.
These include a review on $E_{8(8)}$, an overview of Kac-Moody and
Virasoro algebras and the technicalities of an explicit
Kaluza-Klein reduction containing all Kaluza-Klein modes
(appendices A -- C).
An extension of the spin-2 theories analyzed in chapter 4 to
generic compactification manifolds will be given in appendix D.

\chapter{Higher-spin fields and supergravity}
To set the stage for later examinations we give in
this chapter a short introduction into the interaction
problem for higher-spin fields and its partial resolution
within supergravity. Moreover, we briefly review gauged supergravity
with special emphasis on in its three-dimensional version,
since this will be applied in chapter \ref{spin32chapter} and
generalized in chapter \ref{spin2chapter}.

\section{Consistency problems of higher-spin theories}\label{higherspin}
As mentioned in the introduction the massive Kaluza-Klein
states appearing in supergravity are of significance for the
AdS/CFT correspondence. Among these massive fields are an
infinite tower of spin-3/2 and spin-2 states. The effective
Kaluza-Klein supergravity will describe the coupling of these
infinite towers to gravity (or more precisely, to the supergravity
multiplet). Even more optimistically we would like to couple
massless spin-3/2 and spin-2 fields to gravity that would
exhibit an enhanced gauge symmetry and then establish
a novel version of the Higgs mechanism, such that they can
become massive via breaking the symmetry spontaneously.
However, there exist various no-go theorems that forbid
the existence of interacting higher-spin theories (where
higher-spin means $s>1$), implying in particular that couplings of spin-3/2
and spin-2 fields to gravity -- in the massive and even more
severely in the massless case -- are
impossible.\footnote{There is however a vast
literature on the problem of consistent higher-spin theories.
(For early papers see
\cite{Fierz:1939ix,Fronsdal:1978rb,deWit:1979pe} and for recent
reviews \cite{Bekaert:2005vh,Sorokin:2004ie}.)
Even though there was tremendous progress within the last few years, it is
probably fair to say, that we are still far away from a
conclusive picture.}
In the following we will briefly discuss the consistency problem
related to these higher-spin couplings and how they can be
partially solved within supergravity.

Let us start with a non-interacting spin-2 field
$h_{\mu\nu}$ on a Minkowski background. Its action has been
determined by Pauli and Fierz \cite{Fierz:1939ix} and is given by
 \begin{equation}\label{free}
  \begin{split}
   S_{\text{PF}}=\int d^D x \big[ & \frac{1}{2}\partial_{\mu}h_{\nu\rho}
   \partial^{\mu}h^{\nu\rho}-\partial_{\mu}h^{\mu\nu}\partial^{\rho}
   h_{\rho\nu}+\partial_{\mu}h^{\mu\nu}\partial_{\nu}\hat{h}
   -\frac{1}{2}\partial_{\mu}\hat{h}\partial^{\mu}\hat{h} \\
   &-\frac{1}{2}m^2(h_{\mu\nu}h^{\mu\nu}-\hat{h}^2)\big]\;,
  \end{split}
 \end{equation}
where $\hat{h}=\eta^{\mu\nu}h_{\mu\nu}$ denotes the trace evaluated
in the Minkowski metric.
The equations of motion derived from this action read
 \bea
  \begin{split}
   \square h_{\mu\nu}-\partial_{\mu}\partial^{\rho}h_{\rho\nu}
   -\partial^{\rho}\partial_{\nu}h_{\rho\nu}+\eta_{\mu\nu}\partial_{\rho}
   \partial_{\sigma}h^{\rho\sigma}&-\eta_{\mu\nu}\square\hat{h}
   +\partial_{\mu}\partial_{\nu}\hat{h}\\
   &+ m^2 \left(h_{\mu\nu}-\eta_{\mu\nu}\hat{h}\right)=0\;.
  \end{split}
 \eea
If one takes the divergence and the trace of this equation one gets
 \bea
  \begin{split}
   m^2\partial^{\mu}(h_{\mu\nu}-\eta_{\mu\nu}\hat{h})&=0\;, \\
   (D-2)\partial^{\mu}\partial^{\nu}(h_{\mu\nu}-\eta_{\mu\nu})
   -m^2 (D-1)\hat{h} &=0\;.
  \end{split}
 \eea
>From this we conclude
 \bea\label{spincon}
  \partial^{\mu}h_{\mu\nu}=0\;, \qquad \hat{h}=0\;.
 \eea
Reinserting these relations into the equations of motion,
one obtains
 \bea\label{masscon}
  (\square +m^2)h_{\mu\nu}=0\;.
 \eea
The equations (\ref{spincon}) and (\ref{masscon}) guarantee that
the dynamical content of (\ref{free}) is given by an irreducible
representation of the Poincar\'e group. In four-dimensional
language (\ref{spincon}) states that $h_{\mu\nu}$ has spin 2,
while (\ref{masscon}) ensures that $h_{\mu\nu}$ is an eigenstate
of the mass operator $P^2$. Accordingly, the number of propagating
degrees of freedom can be counted as follows. In $D$ dimensions
the symmetric $h_{\mu\nu}$ has $\frac{D(D+1)}{2}$ components, and
in case that only (\ref{masscon}) is present, all of them would
propagate. In turn one would have to specify $D(D+1)$ Cauchy data
on an initial-value surface, namely the $h_{\mu\nu}$ and
$\partial_t h_{\mu\nu}$. However, in the present case the
conditions (\ref{spincon}) have to be taken into account, i.e.
they have to be imposed as constraints on the initial data.
$\hat{h}=0$ then implies two constraints, namely that $\hat{h}$
and $\partial_t\hat{h}$ are initially zero, while
$\partial^{\mu}h_{\mu\nu}=0$ yields $2D$ further
constraints.\footnote{That $\partial_t\partial^{\mu}h_{\mu\nu}=0$
is a constraint, even though it contains a second-order time
derivative, can be seen by rewriting it with the help of the
Klein-Gordon equation \cite{Hindawi:1995an}.} In total one ends up
with $D^2-D-2$ initial conditions. In other words, there are
$\frac{(D-1)D}{2}-1$ propagating degrees of freedom, which match
exactly the number of components of a symmetric traceless 2-tensor
under the massive little group $SO(D-1)$.

For the massless case $m^2=0$ the conditions (\ref{spincon}) no
longer follow from the equations of motion. Instead, the action
develops a local symmetry
$\delta h_{\mu\nu}=\partial_{\mu}\xi_{\nu}+\partial_{\nu}\xi_{\mu}$,
which decouples additional degrees of freedom. With a similar
analysis as above one finds $\frac{1}{2}D(D-3)$ propagating
degrees of freedom.

What is now the problem of promoting this spin-2 theory
to an interacting theory, e.g. via coupling to gravity?
First of all, the massless $h_{\mu\nu}$ can of course be elevated
to a self-interacting field, namely to the graviton of
general relativity, since (\ref{free}) for $m^2=0$ describes
nothing else than the linearization of the Einstein-Hilbert
action. (Correspondingly, the free spin-2 gauge symmetry
is the linearization of the diffeomorphism symmetry of general
relativity.)
In fact, the existence of this theory has a deep geometrical
reason, namely the existence of Riemannian geometry.
If instead one wants to couple a number of spin-2 fields to gravity,
one would have to replace all
partial derivatives in (\ref{free}) by covariant ones (with
respect to the space-time metric).
However, this would violate either the conditions (\ref{spincon})
in the massive case or the invariance of the massless theory
under any obvious covariantisation of the spin-2 transformation.
This happens because of the non-commutativity of
covariant derivatives, $[\nabla_{\mu},\nabla_{\nu}]\sim R_{\mu\nu}$.
In turn, the number of degrees of freedom would be
larger than compared to the free case, or in other words, some
of the propagating modes would disappear in the free limit,
which clearly indicates an inconsistency.
(For the no-go theorems in case of spin-2 fields see
\cite{Aragone:1979bm,Boulanger:2000rq,Hindawi:1995an,
Cutler:1986dv,Wald:1986dw}, and also
\cite{Boulware:1973my} and \cite{Buchbinder:1999ar,Buchbinder:2000fy}.)

Similar consistency problems appear for all higher-spin fields,
in particular for spin-3/2 fields. However, for the latter we know
already how to circumvent the aforementioned no-go theorem,
namely by introducing local supersymmetry, to which we will
turn now.

\subsubsection{The supergravity miracle}

A free massless spin-3/2 field $\psi_{\mu}$ on a Minkowski background
can be described by the Rarita-Schwinger equation, which reads
 \bea
  \gamma^{\mu\nu\rho}\partial_{\nu}\psi_{\rho}=0\;,
 \eea
where $\gamma^{\mu\nu\rho}=\gamma^{[\mu}\gamma^{\nu}\gamma^{\rho]}$, and
we have suppressed the spinor index on $\psi_{\mu}$.
These equations of motion are clearly invariant under the spin-3/2
gauge symmetry $\delta\psi_{\mu}=\partial_{\mu}\epsilon$,
which in turn guarantees consistency. If one wants to couple
this spin-3/2 field to gravity one encounters the same problems
as mentioned in the last section. Namely, due to the non-commutativity
of covariant derivatives the Rarita-Schwinger equation
transforms under the covariantisation of the spin-3/2 symmetry
into
 \bea\label{vary}
  \delta(\gamma^{\mu\nu\rho}\nabla_{\nu}\psi_{\rho})=
  \frac{1}{2}\gamma^{\mu\nu\rho}[\nabla_{\nu},\nabla_{\rho}]\epsilon
  \sim R^{\mu}_{\hspace{0.3em}\nu}\gamma^{\nu}\epsilon\;.
 \eea
Since the Ricci tensor does not vanish in an interacting theory,
but is instead determined by the energy-momentum tensor
$T_{\mu\nu}(\psi)$ of the spin-3/2 field, the gauge symmetry is
explicitly broken and the theory becomes inconsistent.
Therefore we find a no-go theorem for
spin-3/2 fields which is analogous
to the one for gravity/spin-2 couplings mentioned above.

But, as the discovery of supergravity has shown, there is a
loophole in the no-go theorem: It was assumed in the
variation (\ref{vary}) that only the spin-3/2 field transforms
under the symmetry. If one allows instead for a variation also of
the metric, which is schematically of the form
 \bea
  \delta g_{\mu\nu}=\frac{1}{2}(\bar{\epsilon}\gamma_{\mu}\psi_{\nu}
  +\bar{\epsilon}\gamma_{\nu}\psi_{\mu})\;,
 \eea
the Rarita-Schwinger equation transforms into
 \bea
  \delta(\gamma^{\mu\nu\rho}\nabla_{\nu}\psi_{\rho})\sim
  (R_{\mu\nu}-T_{\mu\nu}+\frac{1}{2} g_{\mu\nu}
  T^{\rho}_{\rho})\gamma^{\nu}\epsilon \; .
 \eea
This means that a non-trivial transformation for the metric
can be introduced in such a way that the Rarita-Schwinger equation
rotates exactly into the Einstein equation containing the energy-momentum
tensor for the spin-3/2 fields. Correspondingly, the Einstein
equation transforms into the Rarita-Schwinger equation.
The theory is therefore consistent also at the interacting level
and propagates only massless spin-2 and spin-3/2 modes
(`the supergravity miracle', \cite{Aragone:1979bm}).

More precisely, $(g_{\mu\nu},\psi_{\mu})$ build a multiplet for a
locally realized  ${\cal N}=1$ supersymmetry.
This means that the introduction of an extended space-time symmetry
allowed consistent couplings of
`higher-spin' fields. In fact, the spin-3/2 fields are realized
as gauge fields for supersymmetry.
If one extends the symmetry further by
introducing additional supercharges $Q^I$, where $I=1,...,{\cal N}$,
consistent couplings of the same number of spin-3/2 fields
(gravitinos) are possible. But, the number of spin-3/2 fields that
can be coupled in this way is bounded (thus being only of
limited use for the required Kaluza-Klein theories).
This can be seen by inspecting
the representation theory of the superalgebra \cite{Nahm:1977tg}.
Since the supercharges raise and lower
the spins in a given multiplet, the maximal number of real supercharges
consistent with spin $s\leq 2$ is $32$. In other words, the number
of gravitinos is bounded by this number, if not
at the same time additional higher-spin fields are introduced.
However, the latter seems to be still impossible, even if
supersymmetry is used. For instance, one may consider the
${\cal N}=1$ multiplet which contains besides the metric
not a massless spin-3/2 field, but a massless spin-5/2 field.
This would be equally sensible from the point of view of the
representation theory of the superalgebra, but still a consistent
field theory (`hypergravity') does not exist \cite{Aragone:1979hx}.

Subsequently we will concentrate on three-dimensional
supergravities, which are special for the following reasons.
First of all, in $D=3$ an arbitrary number of spin-3/2 fields
\textit{can} be coupled to gravity. This is due to the fact,
that they are so-called topological fields, and hence they can
be added to an action without affecting the number of
local degrees of freedom. Even though this flexibility is lost
once the spin-3/2 fields are coupled to matter, one might
hope that the necessary avoidance of the no-go theorems
can be studied more directly.
This topological character actually extends to all massless
higher-spin fields in $D=3$, including spin-2
fields. Thus, also for them an analysis in a three-dimensional
framework seems to be more promising.


\newpage

\section{Gauged supergravity}\label{gaugedsugra}

\subsection{Generalities}
In this section we review the construction of gauged supergravities.
To begin with, we recall that pure supergravity in $D=3$ consists
of a metric (described by the vielbein $e_{\mu}^a$) and Rarita-Schwinger
gravitinos $\psi_{\mu}^I$, $I=1,...,{\cal N}$,
which together build a supermultiplet
for ${\cal N}$-extended supersymmetry. According to the counting of
degrees of freedom in \ref{higherspin} the metric possesses no
propagating degrees of freedom in $D=3$. In this sense
it is purely topological. Similarly, also the gravitino
is topological, as it should be in order to match bosonic
and fermionic degrees of freedom.
Due to this topological nature consistent
field theories exist for arbitrary ${\cal N}$.
Namely, the action
 \bea\label{puresugra}
  {\cal L}_{\text{s.g.}}=-\frac{1}{4}\varepsilon^{\mu\nu\rho}\left(e_{\mu}^a
  R_{\nu\rho a}(\omega)+\bar{\psi_{\mu}}^I\nabla_{\nu}\psi_{\rho}^I
  \right)
 \eea
stays invariant under
 \bea
  \delta_{\epsilon}\hspace{0.1em}e_{\mu}^a
  =\frac{1}{2}\bar{\epsilon}^I\gamma^a
  \psi_{\mu}^I\;, \qquad \delta_{\epsilon}\psi_{\mu}^I=
  \nabla_{\mu}\epsilon^I\;.
 \eea
In fact, the
no-go theorem excluding couplings of an arbitrary number of spin-3/2
fields mentioned in \ref{higherspin} does not apply, since the
multiplet structure refers only to propagating degrees of freedom.
In view of the fact that the gravitational fields are topological,
an arbitrary number of spin-3/2 fields can in turn be coupled.

\subsubsection{Chern-Simons theories}\label{CSgravity}
Another way of providing invariant supergravity actions for
an arbitrary number of topological fields is given by the so-called
Chern-Simons supergravities. To introduce them let us first
discuss the description of pure gravity theories in the framework of
Chern-Simons theories.

The Chern-Simons action for a gauge field ${\cal A}$,
taking values in the Lie algebra of a certain gauge group $G$, reads
 \begin{equation}\label{CSbare}
  S_{\rm CS}=\int \text{Tr}\big({\cal A}\wedge d{\cal A}+\frac{2}{3}
  {\cal A}\wedge {\cal A}\wedge {\cal A}\big)\;.
 \end{equation}
Here the trace refers symbolically to an invariant and non-degenerate
quadratic form on the Lie algebra. It has been shown in a classic
paper by Witten \cite{Witten:1988hc} that three-dimensional
gravity (with or without cosmological constant) can be interpreted
as a Chern-Simons theory with a particular gauge group. If
the cosmological constant $\Lambda$ is positive, negative or zero, the
gauge group is given by the de Sitter, anti-de Sitter of Poincar\'e
group, respectively. We will illustrate this for AdS gravity, in which
case the isometry group decomposes as
 \bea\label{so22}
  SO(2,2)=SL(2,\mathbb{R})_L \times SL(2,\mathbb{R})_R\;.
 \eea
To make contact with the conventional formulation of AdS gravity
one rewrites the respective gauge fields corresponding to the
left or right $SL(2,\mathbb{R})$ factors according to
 \bea
  {\cal A}_{\mu}^{a\hspace{0.2em}L,R}=\omega_{\mu}^a\pm\sqrt{-\Lambda}
  \hspace{0.2em}e_{\mu}^a\;.
 \eea
Then $e_{\mu}^a$ will be interpreted as the vielbein and
$\omega_{\mu}^a$ as the spin connection, which is treated
as an independent field in the first order Palatini formalism.
It turns out that the difference of the two corresponding
Chern-Simons terms
 \bea
   {\cal L}_{\text{CS}} =
   {\rm Tr}
   \big({\cal A}_{L} \wedge d{\cal A}_{L} + \frac{2}{3} {\cal A}_{L}\wedge
   {\cal A}_{L}\wedge {\cal A}_{L} \big)
   -{\rm Tr}
   \big({\cal A}_{R} \wedge d{\cal A}_{R} + \frac{2}{3} {\cal A}_{R}\wedge
   {\cal A}_{R}\wedge {\cal A}_{R}\big)
   \label{CS}
 \eea
coincides exactly with the Einstein-Hilbert term with cosmological
constant $\Lambda$ \cite{Witten:1988hc}.
In this sense pure AdS-gravity
can be interpreted as Chern-Simons gauge theory.

Along these lines AdS supergravity theories can in turn be constructed
as Chern-Simons theories for appropriate superextensions of
the AdS group (\ref{so22}). This has first been observed by
Achucarro and Townsend \cite{Achucarro:1987vz}, which considered
the supergroup $OSp(p|2)\times OSp(q|2)$. Here each factor contains
the $SL(2,\mathbb{R})$ factor in (\ref{so22}) together with
supercharges and internal symmetry generators.
More precisely, the entire supergroup carries ${\cal N}=p+q$
supersymmetry and yields a $SO(p)\times SO(q)$ gauge
symmetry. The resulting Chern-Simons action takes the form
(\ref{puresugra}) for the corresponding number of supercharges,
augmented by additional Chern-Simons terms for the $SO(p)\times SO(q)$
gauge fields together with gravitino `mass' terms and
a cosmological constant. Since these supergroups exist for
arbitrary $p,q$, an arbitrary number of gravitinos can be
consistently coupled to gravity.

\subsubsection{Matter couplings in supergravity}

In the last section we have seen that three-dimensional supergravities
are special in the sense that the purely gravitational fields
do not carry local degrees of freedom. This in turn allowed
the existence of theories with an arbitrary number ${\cal N}$ of
supercharges. There is also another respect in which three-dimensional
theories are special: For massless fields (on Minkowski
backgrounds) there is no notion of spin or helicity.\footnote{In
the following we will nevertheless refer to spin-$s$ fields,
if they are symmetric tensors in $s$ vector indices and carry
an associated gauge symmetry (and analogously for spinors).}
This implies that the standard argument, restricting the number of
supercharges to be less or equal to $32$ in order to arrange for
maximal spin $2$, does not apply,
also if propagating matter fields are taken into
account. Thus one might also hope to get matter-coupled
supergravities for arbitrary ${\cal N}$.
However, it turns out that
upon coupling to propagating matter the no-go theorem
reappears. Adding a globally supersymmetric $\sigma$-model action for
scalar multiplets $(\phi^i,\chi^i)$ of the form
 \bea\label{supersigma}
  {\cal L}_{\text{matter}}=\frac{1}{2}g_{ij}(\phi)\left(\partial_{\mu}\phi^i
  \partial^{\mu}\phi^j-\frac{i}{2}\bar{\chi}^i\gamma^{\mu}D_{\mu}\chi^j
  \right)+{\cal L}_{\chi^4}
 \eea
to the supergravity action (\ref{puresugra}) implies severe
conditions on the geometry of the target space.
For instance, ${\cal N}=8$ supersymmetry restricts the scalar
manifold to be a coset space of the form
 \bea\label{so8coset}
  G/H=\frac{SO(8,n)}{SO(8)\times SO(n)}\;,
 \eea
where $n$ indicates the number of scalar multiplets.
For ${\cal N}>8$ the target spaces are uniquely determined,
and ${\cal N}=16$ implies the exceptional coset space $E_{8(8)}/SO(16)$.
Moreover, one finds that there exist no consistent theories with
${\cal N}>16$ \cite{deWit:1992up}.
Thus the bound ${\cal N}\leq 16$ (implying $\leq 32$ real supercharges)
is still satisfied.

Since the homogeneous spaces appearing as the target spaces for
${\cal N}=8$ and {${\cal N}=16$} supergravity play an important
role in this thesis, we will briefly review
their description \cite{deWit:2002vz}.
For a coset space $G/H$ a convenient parameterization for the scalar
fields is given in terms of a $G$-valued matrix $L(x)$. This matrix
is then subject to local $H$ transformations which reduce the number
of physical degrees of freedom to the required
$\text{dim}\hspace{0.1em}G-\text{dim}\hspace{0.1em}H$
of the coset space. Moreover, $L$ transforms under global $G$
transformations, i.e. in total
 \bea\label{cosetsymmetry}
  L(x) \longrightarrow  g L(x) h^{-1}(x)\;, \qquad
  g\in G ,\;  h(x)\in H\;.
 \eea
The $\sigma$-model action can then be defined through the
Lie algebra--valued
current $L^{-1}\partial_{\mu}L={\cal Q}_{\mu}+{\cal P}_{\mu}$, which we
have decomposed according to ${\cal Q}_{\mu}\in \frak{h}$ and
${\cal P}_{\mu}\in \frak{g}\backslash\frak{h}$, where $\frak{h}$ and $\frak{g}$
denote the Lie algebras of $G$ and $H$, respectively.
Then the action reads
 \bea\label{cosetlagrangian}
  {\cal L}=\frac{1}{2}g^{\mu\nu}\text{Tr}\left(
  {\cal P}_{\mu}{\cal P}_{\nu}\right)\;.
 \eea
For explicit computations one usually takes a gauged-fixed parametrisation
for $L$, e.g.~a gauge, where $L$ takes only values in the non-compact
part of $G$. Writing $L=\exp(\phi^i t_i)$, where
$i=1,...,\text{dim}\hspace{0.1em}G-\text{dim}\hspace{0.1em}H$
and $t_i$ are the corresponding generators, and
then inserting into (\ref{cosetlagrangian})
yields a bosonic $\sigma$-model action of the form (\ref{supersigma}).
The isometry group $G$ as the rigid invariance group of
(\ref{cosetlagrangian}) is then realized via a left multiplication
on $L$ as in (\ref{cosetsymmetry}) together with a compensating
$H$ transformation in order to restore the chosen gauge for $L$.
In a given parametrisation this results in a non-linear action on
the $\phi^i$, generated by the corresponding Killing vectors.

\subsubsection{Gauging of supergravity}\label{sugragauging}
So far we have described ungauged supergravities, since no gauge
fields were involved. Let us now turn to the problem of gauging some
of the global symmetries, i.e.~to the problem of promoting a
certain subgroup of the isometry group $G$ to a local symmetry.

As usual this comprises the introduction of gauge fields and minimal
coupling to charged matter.
However, in supersymmetric theories one immediately encounters
the severe problem that the introduction of additional bosonic
degrees of freedom is inconsistent.
In generic dimensions one evades this problem via starting
from an ungauged supergravity, which already contains a number of
vector multiplets (with abelian gauge fields).
The gauging then deforms the theory in such a way that the
vectors become gauge fields for a non-abelian gauge group.
In contrast, the three-dimensional supergravities mentioned
above contain only scalar fields, which in turn is no restriction
of generality since abelian vectors can in $D=3$ always be
dualized into scalars. But, in order for the theory to express
the maximal rigid symmetry in a coset space structure
(as the $E_{8(8)}$ for ${\cal N}=16$) all bosonic degrees of
freedom have to reside in scalars.
Thus it seems to be impossible to gauge three-dimensional supergravity,
while maintaining at the same time the large rigid symmetry.
It has been shown in \cite{Nicolai:2000sc,Nicolai:2001sv} (for a review
see \cite{deWit:2004yr})
that it is possible, however, to elude this conflict to a certain
extent via the introduction of a Chern-Simons term for the gauge vectors
instead of a Yang-Mills term. Since the former yields only
topological gauge fields, bosonic and fermionic degrees of freedom
still match, and accordingly the bosonic degrees of freedom
are, also in the gauged theory, all carried by scalars.

Let us now consider the gauging in more detail.
To begin with, each partial derivative
in the ungauged theory has to be replaced by a covariant one:
 \bea\label{minimal}
  D_{\mu}= \partial_{\mu}+g\Theta_{\cM \cN}t^\cM A_{\mu}^\cN\; .
 \eea
Here $g$ denotes the gauge coupling constant which measures the
deformation of the ungauged theory into a gauged one. The
$A_{\mu}^{\cM}$ are the gauge fields and $\Theta_{\cM \cN}$ is a symmetric
tensor, where the indices $\cM,\cN,...$ label the
adjoint representation of the global symmetry (as, e.g., $E_{8(8)}$
in the ${\cal N}=16$ case).
$\Theta_{\cM \cN}$ is the so-called embedding tensor.
It describes the embedding of the
gauge group $G_0$ into the rigid symmetry group $G$ in the
sense that the Lie algebra
$\mathfrak{g}_0$ of $G_0$ is spanned by the generators
$\Theta_{\cM \cN}t^{\cN}$, where $t^{\cM}$ denote the generators
of $G$. In particular, the
rank of $\Theta$ is given by the dimension of the gauge group.
To be more precise, the embedding tensor is an element in the
symmetric tensor product, i.e.
 \bea
  \Theta = \Theta_{\cM \cN}t^{\cM}\otimes t^{\cN} \in \text{Sym}
  (\mathfrak{g}\otimes\mathfrak{g})\;.
 \eea
The introduction of $\Theta$ formally preserves covariance with respect
to the full global symmetry group, even though in the gauged theory
the latter is broken to the gauge group (since $\Theta_{\cM \cN}$
is constant and does not transform under $G$).
We will see below that this formalism nevertheless substantially
simplifies the analysis of gauged supergravities.

Generically, the form of gauged supergravities is highly restricted
due to the fact that the minimal substitution (\ref{minimal})
spoils the invariance under supersymmetry. This has to be
compensated by the introduction of additional couplings, and
only in special cases this can be done in a consistent way.
More specifically, the bosonic terms have to be supplemented by
a scalar potential $V$, such that the matter action for them reads
 \bea \label{Lag}
  \cal{L}_{{\rm matter}} &=& e \,{\rm Tr} \Big\langle
  [{\cal V}^{-1}D_{\mu}{\cal V}]_{\mathfrak{k}}\,
  [{\cal V}^{-1}D^{\mu}{\cal V}]_{\mathfrak{k}}
  \Big\rangle + e \,V(\cal V) +{\rm fermions } \; ,
 \eea
where $[\cdot]_{\mathfrak{k}}$
denotes the projection of the Lie algebra $\mathfrak{g}$
associated to $G$ onto its noncompact part.
Similarly, Yukawa type couplings between scalars and fermions are
necessary in the gauged theory, and their explicit form
will be given in sec.~\ref{N8gauging}.
Finally the Chern-Simons term for the gauge fields has to be added,
which can be written as
 \bea\label{LCS}
  \mathcal{L}_{\text{CS}}=\frac{1}{4}\varepsilon^{\mu\nu\rho}
  \Theta_{\cM \cN}A_{\mu}^{\cM}\big(\partial_{\nu}A_{\rho}^{\cN}+\frac{1}{3}
  f^{\cN \cP}_{\hspace{1.3em}\cL}\Theta_{\cP\cK}A_{\nu}^{\cK}
  A_{\rho}^{\cL}\big)\; .
 \eea
This coincides with (\ref{CSbare}), where $\Theta_{\cM \cN}$ serves
as a non-degenerate quadratic form on the subalgebra that will be
gauged. Note that the Chern-Simons term for the compact gauge vectors
combines together with the Einstein-Hilbert term and the kinetic terms
for the gravitinos into a Chern-Simons theory
based on an $AdS_3$-supergroup, as discussed in sec.~\ref{CSgravity}.
Thus the gauged supergravities are the natural matter-coupled
extensions of the topological theories of \ref{CSgravity}.

Depending on the amount of required supersymmetry the constructed
theory is not automatically supersymmetric, but still several
constraints have to be satisfied.
However, the advantage of the given formalism based on the
embedding tensor consists of the fact that all conditions
implied by supersymmetry translate into purely algebraic constraints
on the embedding tensor.
First of all $\Theta$ has to fulfill some purely group-theoretical
constraints. In order for the subset $G_0$ to be a closed algebra,
$\Theta$ has to be invariant under the gauge group. In terms of the
structure constants $f^{\cM \cN}_{\hspace{1.2em}\cK}$ of
$G$ this invariance,
i.e. the fact that $\Theta$ commutes with all gauge group
generators $\Theta_{\cM \cN}t^{\cN}$, implies the quadratic condition
 \bea\label{quad}
  \Theta_{\cK \cP}\Theta_{\cL(\cM}f^{\cK \cL}_{\hspace{1.2em}\cN)}=0\;.
 \eea
Second, supersymmetry requires an algebraic constraint, whose
explicit form we will discuss in the next subsection for
${\cal N}=8$ and ${\cal N}=16$ supergravity, respectively.
The full supergravity action is entirely determined by the
embedding tensor.

We have seen that consistent gaugings of supergravity are
possible in $D=3$ through the introduction of Chern-Simons
gauge fields. On the other hand, we know that gauged supergravities
exist which carry Yang-Mills vectors as gauge fields. The latter
can, for instance, be constructed by Kaluza-Klein compactification of
higher-dimensional supergravities.
Therefore one might be tempted to conclude that
the gauged supergravities mentioned
here are not the most general ones. But this is not the case:
All supergravities in $D=3$ with Yang-Mills type
gauging are on-shell dual to a Chern-Simons gauged supergravity
with an enlarged number of scalar fields, as has been shown in
\cite{Nicolai:2003bp,deWit:2003ja}.
In the following we will shortly review this equivalence.

We start from the generic form of a Yang-Mills gauged supergravity
 \bea\label{YMgauged}
  {\cal L}=-\frac{1}{4}eR -\frac{1}{4}e M_{ab}(\phi)F_{\mu\nu}^a
           F^{\mu\nu b} + {\cal L}^{\prime}(A,\phi)\;,
 \eea
where $A_{\mu}^a$ and $F_{\mu\nu}^a$ denote the non-abelian
gauge field and field strength for a certain gauge group $G$.
Moreover, ${\cal L}^{\prime}$ indicates some additional matter
couplings and fermionic terms.
We will assume that they are separately
gauge-invariant, which in turn implies that $A_{\mu}^a$
enters only through a covariant derivative or maybe an additional
Chern-Simons term. It is exactly this explicit dependence on
$A_{\mu}^a$ that forbids a standard dualization into scalars.
But we will show that a dualization is still possible in which
the gauge fields survive as topological fields, and where
the bosonic degrees of freedom are instead carried by new scalars.
To see this we have to introduce for each of the former Yang-Mills
fields a scalar $\varphi_a$ and also an additional
gauge field $B_{\mu \hspace{0.2em}a}$.
An improved duality relation can then be written as follows
 \bea\label{impdual}
  \frac{1}{2}e\varepsilon_{\mu\nu\rho}F^{a\hspace{0.2em}\nu\rho}
  =M^{ab}(\phi)(D_{\mu}\varphi_b-B_{\mu\hspace{0.2em}b})\;,
 \eea
where $D_{\mu}$ denotes the gauge covariant derivative with
respect to $G$.
Moreover we have assumed that the scalar field matrix $M_{ab}(\phi)$
is invertible, such that $M^{ac}M_{cb}=\delta^a_{\hspace{0.2em}b}$.
The structure of (\ref{impdual}) suggests that $\varphi_a$ transforms
under a shift symmetry which is gauged by the $B_{\mu\hspace{0.2em}a}$.
More specifically, if we define the transformations
 \bea
  \delta B_{\mu\hspace{0.2em}a}=D_{\mu}\Lambda_a \;, \qquad
  \delta\varphi_a =  \Lambda_a\;,
 \eea
the duality relation (\ref{impdual}) stays invariant.
Equivalently, the right hand side of (\ref{impdual}) defines
the covariant derivative $\tilde{D}_{\mu}\varphi_a$ for $\varphi_a$
with respect to the shift gauge symmetry.
In the next step we have to give a new Lagrangian that reproduces the
duality relation (\ref{impdual}), and whose equations of motion
are equivalent to those of the original Lagrangian (\ref{YMgauged}):
It can be written as
 \bea\label{dualaction}
  {\cal L}=-\frac{1}{4}eR + \frac{1}{2}eM^{ab}(\phi)\tilde{D}_{\mu}\varphi_a
  \tilde{D}^{\mu}\varphi_b + \frac{1}{2}\varepsilon^{\mu\nu\rho}
  B_{\mu\hspace{0.2em}a}F_{\nu\rho}^a +{\cal L}^{\prime}(A,\phi)\;.
 \eea
Indeed, we observe the appearance of a Chern-Simons-like term
$B\wedge F$. Moreover, varying (\ref{dualaction}) with
respect to $B_{\mu\hspace{0.2em}a}$
yields the duality relation (\ref{impdual}). That (\ref{dualaction})
is on-shell equivalent to (\ref{YMgauged}) can be most easily seen
by choosing the gauge $\varphi_a =0$
(i.e. fixing the gauge symmetry, which is absent in (\ref{YMgauged})
anyway), such that $\tilde{D}_{\mu}\varphi_a=-B_{\mu\hspace{0.2em}a}$,
and then integrating out $B_{\mu\hspace{0.2em}a}$.
The latter coincides with the Yang-Mills gauged action (\ref{YMgauged}).

In total we have seen that any Yang-Mills gauged supergravity
with gauge group $G$ is on-shell equivalent to a Chern-Simons
gauged theory, in which the gauge group has been enhanced
by $\text{dim}\,G$ additional shift gauge fields. To be more precise,
the gauge algebra is extended by $\text{dim}\,G$ nilpotent generators
(i.e.~commuting among themselves), which transform under the
adjoint action of $G$ (see eq.~(\ref{semidirect}) below).
The Chern-Simons term for this extended gauge group
coincides indeed with the Chern-Simons term in (\ref{dualaction}).
Moreover, additional Chern-Simons terms can be coupled,
which will act as gauge invariant mass terms for the vectors.
The latter corresponds to a further enhancement of the gauge
algebra, see the second line of eq.~(\ref{semidirect}) below.
These nilpotent symmetries will be broken in the vacuum and
give rise to massive vector fields.
Later on the massive spin-1 fields will therefore be identified
through their Goldstone scalars \cite{Fischbacher:2002fx,Hohm:2004rc}.

\subsection{Gauged ${\cal N}=8$ and ${\cal N}=16$ supergravity
in $D=3$}\label{N8gauging}
In this subsection we will describe the gauged supergravities
required for our analysis in more detail.

\subsubsection{The ${\cal N}=8$ supergravity}
In the ${\cal N}=8$ supergravity the scalar fields are, in accordance
with the coset structure discussed above, described by a $SO(8,n)$
valued matrix $L$ and are subject to local and global
transformations (\ref{cosetsymmetry}) \cite{Nicolai:2001ac}.
The current covariantized with respect to the gauge group
reads
 \bea
  L^{-1}(\partial_{\mu}+g\Theta_{{\cal MN}}B_{\mu}^{\hspace{0.3em}{\cal M}}
  t^{\cal N})L = \frac{1}{2}Q_{\mu}^{IJ}X^{IJ}+\frac{1}{2}Q_{\mu}^{rs}X^{rs}
  +{\cal P}_{\mu}^{Ir}Y^{Ir}\;,
 \eea
where $(X^{IJ},X^{rs})$ and $Y^{Ir}$ denote the compact and noncompact
generators of $SO(8,n)$, respectively. In this section
$I,J=1,...,8$ and $r,s=1,...,n$ are $SO(8)$ and $SO(n)$ vector indices.
The gravitinos $\psi_{\mu}^A$ and the matter fermions $\chi^{\dot{A}r}$
carry spinor and conjugate spinor indices under $SO(8)$,
respectively.\footnote{In all other parts of this thesis these
indices will instead refer to $SO(16)$ quantities. Since the
${\cal N}=8$ notation enters only in this section, this cannot
be a source of confusion.}
The action reads
 \bea
  \begin{split}
   {\cal L}=&-\frac{1}{4}eR+\frac{1}{2}\varepsilon^{\mu\nu\rho}
   \overline{\psi}_{\mu}^A D_{\nu}\psi_{\rho}^A+\frac{1}{4}e
   {\cal P}_{\mu}^{Ir}{\cal P}^{\mu \hspace{0.2em}Ir}
   -\frac{1}{2}ie\overline{\chi}^{\dot{A}r}\gamma^{\mu}
   D_{\mu}\chi^{\dot{A}r} + {\cal L}_{\text{CS}}\\
   &+\frac{1}{2}geA_1^{AB}\overline{\psi}_{\mu}^A\gamma^{\mu\nu}\psi_{\nu}^B
   +igeA_2^{A\dot{A}r}\bar{\chi}^{\dot{A}r}\gamma^{\mu}\psi_{\mu}^A
   +\frac{1}{2}geA_3^{\dot{A}r\dot{B}s}\bar{\chi}^{\dot{A}r}\chi^{\dot{B}s}
   +eV\;,
  \end{split}
 \eea
with the Chern-Simons term ${\cal L}_{\text{CS}}$ defined in
(\ref{LCS}).
The explicit expressions for $A_1$, $A_2$, $A_3$ and the potential
$V$ in terms of $\Theta$ can be found in \cite{Nicolai:2001ac}.
The supersymmetry conditions consist basically of some
symmetry requirements on $\Theta$ \cite{Nicolai:2001ac}.

\subsubsection{The ${\cal N}=16$ supergravity}
We have already seen that for ${\cal N}=16$ the supergravity
multiplet contains 16 Majorana gravitino fields $\psi_{\mu}^I$,
$I=1,...,16$, and we may view $I$ as a vector index of
$SO(16)$. The bosonic matter content consists of the $128$ scalar fields
that parametrize the coset space $E_{8(8)}/SO(16)$.
This means they can be represented by an $E_{8(8)}$
valued matrix $\mathcal{V}(x)$, which transforms under
global $E_{8(8)}$ and local $SO(16)$ transformations as follows:
 \bea\label{coset}
  \mathcal{V}(x)\rightarrow g\mathcal{V}(x)h^{-1}(x)\;, \qquad
  g\in E_{8(8)}\;,\quad h(x)\in SO(16)\;.
 \eea
Their fermionic partners are given by $128$ Majorana fermions
$\chi^{\dot{A}}$, $\dot{A}=1,...,128$, where $\dot{A}$ indicates
the conjugate spinor index of $SO(16)$.
Specifying~(\ref{Lag}) to the ${\cal N}=16$ case,
the maximal supergravity Lagrangian
up to quartic couplings in the fermions is given by
 \bea
 \label{action}
  \mathcal{L}&=&-\frac{1}{4}eR+\frac{1}{4}e\,{\cal P}^{\mu A}{\cal P}_{\mu}^A+
  \frac{1}{2}\varepsilon^{\mu\nu\rho}\overline{\psi}_{\mu}^I {\cal D}_{\nu}\psi_{\rho}^I-
  \frac{i}{2}e\,\bar{\chi}^{\dot{A}}\gamma^{\mu}{\cal D}_{\mu} \chi^{\dot{A}}
  -\ft12 e \bar{\chi}^{\dot{A}}\gamma^{\mu}\gamma^{\nu}\psi_{\mu}^I
  \Gamma^{I}_{A\dot{A}}\,{\cal P}^{A}_{\nu}
  \nonumber\\[1ex]
  &&{}
+\frac{1}{2}eg\,A_1^{IJ}\,\bar{\psi_{\mu}}^I
  \gamma^{\mu\nu}\psi_{\nu}^J + ieg\,A_2^{I\dot{A}}\,\bar{\chi}^{\dot{A}}
  \gamma^{\mu}\psi_{\mu}^I
  + \frac{1}{2}eg\,A_3^{\dot{A}\dot{B}}\, \bar{\chi}^{\dot{A}}\chi^{\dot{B}}
    \nonumber\\[1ex]
  &&
{}
  +
\ft{1}{4}\, g\,\varepsilon^{\mu\nu\rho} \,
A_\mu^{\cal{M}} \,\GTh_{\cal{M}\cal{N}} \Big( \partial_{\nu} A_{\rho}^{\cal{N}} - \ft13 g\, \GTh_{\cal{K}\cal{L}} f^{\cal{N}\cal{L}}{}_{\cal{P}} \,
A_{\nu}^{\cal{K}} A_{\rho}^{\cal{P}} \Big)
-e\,V(\cal V)
\;.
\eea
As discussed above, the theory is entirely encoded
in the symmetric constant matrix $\Theta_{{\cal{M}\cal{N}}}$.
The minimal coupling of vector fields to scalars
reads in the given case
 \bea\label{current}
  \mathcal{V}^{-1}D_{\mu}\mathcal{V}~\equiv~
  \mathcal{V}^{-1} \partial_{\mu}{\cal V} +
  g A^{\cal{M}}_{\mu}\,\Theta_{{\cal{M}\cal{N}}}\,
  (\mathcal{V}^{-1} t^{\cal{N}}\,{\cal V})~\equiv~
  \ft{1}{2}{\cal Q}_{\mu}^{IJ}X^{IJ}
   +{\cal P}_{\mu}^AY^A\;,
 \eea
with $X^{IJ}$ and $Y^{A}$ labeling the 120 compact and 128 noncompact
generators of $E_{8(8)}$, respectively.\footnote{See appendix \ref{A1} for
our $E_{8(8)}$ and $SO(16)$ conventions.}
The Yukawa couplings or fermionic mass terms in~(\ref{action})
are defined as linear functions
of $\Theta_{\cal{M}\cal{N}}$ via
 \bea\label{A123}
   A_1^{IJ}&=&\frac{8}{7} \theta \delta_{IJ} +\frac{1}{7}T_{IK|JK}\;,
   \quad
   A_2^{I\dot{A}}=-\frac{1}{7}\Gamma_{A\dot{A}}^JT_{IJ|A}\;, \nonumber\\
   A_3^{\dot{A}\dot{B}}&=&2\theta\delta_{\dot{A}\dot{B}}+\frac{1}{48}
   \Gamma_{\dot{A}\dot{B}}^{IJKL}T_{IJ|KL},
\eea
with $SO(16)$ gamma matrices $\Gamma^I_{A\dot{B}}$,
and the so-called $T$-tensor
\bea\label{T}
  T_{\cal M|\cal N}  &=&\mathcal{V}^{\cal K}{}_{\cal M}
  \mathcal{V}^{\cal L}{}_{\cal N}
  \Theta_{\cal K\cal L}\;.
\eea
The scalar potential $V(\cal V)$ is given by
\bea
\label{pot}
  V=-\frac{g^2}{8}\,\big(A_1^{KL}A_1^{KL}-\frac{1}{2}A_2^{K\dot{A}}
  A_2^{K\dot{A}}\big)\;.
\eea
For later use we give the condition of stationarity of this potential,
as it has been shown in \cite{Nicolai:2000sc}
 \bea\label{stationary}
  \delta\, V =0 \qquad
  \Longleftrightarrow \qquad
  3 A_1^{IM}A_2^{M\dot{A}}=A_3^{\dot{A}\dot{B}}A_2^{I\dot{B}}\;.
 \eea
The quartic fermionic couplings and the supersymmetry transformations
of~(\ref{action}) can be found in~\cite{Nicolai:2000sc}.
For consistency of the theory, the embedding tensor $\Theta_{\cal M\cal N}$
needs to satisfy two algebraic constraints.
First, it has to satisfy the quadratic constraint (\ref{quad}).
Second, $\Theta$ as an element of the symmetric $E_{8(8)}$ tensor product
\bea
  (\f{248}\otimes \f{248})_{\text{sym}}&=&
  \f{1}\oplus \f{3875} \oplus \f{27000}
 \;,
\eea
is required to only live in the ${\bf 1}\oplus{\bf 3875}$ representation,
i.e. to satisfy the projection constraint
 \bea\label{27000}
  (\mathbb{P}_{\f{27000}})_{\cM\cN}^{\hspace{1.6em}\cP\cQ}
  \hspace{0.3em}\Theta_{\cP\cQ}=0\;.
 \eea
Explicitly, this constraint takes the form~\cite{Koepsell:1999uj}
\bea
\Theta_{\cal M\cal N}
+\frac{1}{62}\,\eta_{\cal M\cal N}\,\eta^{\cal K\cal L}\,
\Theta_{\cal K\cal L}
+\frac{1}{12}\, \eta_{\cal P\cal Q} f^{\cal K\cal P}{}_{\cal M}
f^{\cal L\cal Q}{}_{\cal N}\,\Theta_{\cal K\cal L}&=& 0 \;,
\label{conex}
\eea
with the $E_{8(8)}$ structure constants $f^{\cal M\cal N}{}_{\cal K}$
and Cartan-Killing form
 \bea
  \eta^{\cal M\cal N}=\frac{1}{60}\,
  f^{\cM\cK}_{\hspace{1.5em}\cL}\,f^{\cN\cL}_{\hspace{1.5em}\cK}\;.
 \eea
Any solution of (\ref{conex}) and (\ref{quad})
defines a consistent maximally supersymmetric theory~(\ref{action})
in three dimensions.

\chapter{Massive spin-3/2-multiplets in supergravity}\label{spin32chapter}

In this chapter we will construct the effective supergravities for
massive supermultiplets on $AdS_3\times S^3\times S^3\times S^1$.
First, in sec.~\ref{adssol}, we review the solution of type IIB supergravity
leading to this background and discuss in sec.~\ref{kkspec}
the resulting Kaluza-Klein spectrum. In sec.~\ref{lowspin} we
discuss the effective supergravities for the lowest multiplets,
i.e. spin-1/2 and spin-1 multiplets, and in sec.~\ref{spin32}
we finally turn to the first massive spin-3/2 multiplet.

\section{IIB supergravity on
$AdS_3\times S^3\times S^3\times S^1$}\label{adssol}
\subsection{The 10-dimensional supergravity solution}
First we are going to discuss the required solution of type
IIB supergravity, which is the effective low-energy theory for
the corresponding string theory.
The bosonic spectrum consists of the metric $g_{MN}$, two 2-forms
$B_{MN}$, $C_{MN}$, two scalars $\phi$ (the dilaton) and $C_0$ and
finally a self-dual 4-form $C_4$. Its action reads \cite{Gukov:2004ym}
(with ${\alpha^{\prime}=1/(2\pi)^2}$)
 \bea\label{IIB}
  \begin{split}
   S_{\text{IIB}}=\hspace{0.3em}&\frac{2\pi}{g_B^2}
   \int d^{10}x\sqrt{-g}e^{-2\phi}
   (R+4(\nabla\phi)^2)-\frac{\pi}{g_B^2}\int e^{-2\phi}H\wedge *H \\
   &-\pi\int R_1\wedge *R_1-\pi\int R_3\wedge *R_3-\frac{\pi}{2}
   \int R_5\wedge *R_5 + \pi\int C_4\wedge H \wedge F_3\;,
  \end{split}
 \eea
where $H_3=dB_2$, $R_1=d C_0$, $R_3=dC_2-C_0 H_3$,
$R_5=dC_4-H_3\wedge C_2$ and
$F_3=dC_2$. This action is actually not a true off-shell
formulation of type IIB supergravity, since the self-duality of the
4-form has to be imposed by hand,
 \bea
  *R_5 =  R_5\;.
 \eea

We are  now looking for solutions of the
equations of motion which contain an AdS factor, i.e.~which possess
in its space-time part a maximally symmetric manifold. Since the
type IIB action (\ref{IIB}) does not allow for a cosmological
constant term, AdS spaces of generic dimension without matter
sources cannot be a solution of (\ref{IIB}). Thus we have to
give vacuum expectation values to certain matter fields in order
to provide for the right source for the gravitational field.
For the pure NS solution we are going to construct we set
$R_5=F_3=\phi=0$ and assume for the background metric the
direct product $AdS_3\times S^{3+}\times S^{3-}\times S^1$, i.e.
 \bea
  ds^2 = ds^2(AdS_3) + R_+^2 ds^2(S_+^3) + R_-^2 ds^2(S_-^3)
         + L^2(d\theta)^2\;.
 \eea
The source for the metric will be given by a non-trivial
vev for the 3-form flux $H_3$:
 \bea\label{hflux}
  H=\lambda_0\omega_0 + \lambda_+\omega_+ + \lambda_-\omega_-\;,
 \eea
while all other fields are set to zero.
Here $\omega_0=(\frac{L_0}{x_2})^3 dt\wedge dx^1\wedge dx^2$
is the volume form on $AdS_3$ and
$\omega_{\pm}=\text{vol}(S_{\pm}^3)$ are the volume forms of the
spheres. The Bianchi identity $dH_3=0$ is trivially satisfied
since all 3-forms are closed on their respective 3-manifolds.
The curvature for the AdS part is given in terms of
the AdS length scale $L_0$ by
 \bea
  \begin{split}
   R_{\mu\nu\lambda\rho}&=-L_0^{-2}(g_{\mu\lambda}g_{\nu\rho}-g_{\mu\rho}
   g_{\nu\lambda})\;, \\
   R_{\mu\nu}&=-2L_0^{-2}g_{\mu\nu}\;, \\
   R &=-6L_0^{-2}\;.
  \end{split}
 \eea
Similarly the curvature for the two spheres is given in terms of
their radii by
 \bea
  \begin{split}
   R_{mnpq} &= R_+^{-2}(g_{mp}g_{nq}-g_{mq}g_{np})\;, \\
   R_{mn} &= 2R_+^{-2}g_{mn} \;, \\
   R &= 6R_+^{-2} \; ,
  \end{split}
 \eea
and analogously for $S_-^3$ (with indices $\bar{m}, \bar{n},...$).

Next we have to verify the equations of motion, in particular
we have to compare the Ricci tensor with the energy-momentum
tensor induced by the 3-form flux.
First of all, among the equations of motion
for the fields vanishing in the background we also have to check the
one for $\phi$, since it couples to the non-vanishing $H_3$.
Its equation of motion derived from (\ref{IIB})
 \bea
  \nabla^{M}(e^{-2\phi}\partial_M\phi)\sim H_{MNP}H^{MNP}\;,
 \eea
implies for $\phi=0$ the relation $H_{MNP}H^{MNP}=0$.
The latter in turn yield with (\ref{hflux})
 \bea\label{hrel}
  \frac{1}{6}H_{MNP}H^{MNP}=-\lambda_0^2 + \lambda_+^2 + \lambda_-^2=0\;.
 \eea
Moreover, it follows from (\ref{hrel}) that the energy-momentum
tensor simplifies and one has $R=0$. Because of the direct product
structure the Einstein equations can in turn be solved,
if we set
 \bea
  L_0^{-2}=\frac{1}{4}\lambda_0^2\;, \quad
  R_+^{-2}=\frac{1}{4}\lambda_+^2\;, \quad R_-^{-2}=\frac{1}{4}\lambda_-^2\;,
 \eea
which are, however, subject to the constraint (\ref{hrel}).
Therefore the equations of motion for the matter fields -- in this case
$\phi$ -- imply a relation between the curvature scales of AdS on
the one side and the internal manifolds on the other side.
Thus the size of the
internal manifolds is necessarily of the same order as the
AdS space, as mentioned in the introduction.

We have seen that type IIB supergravity admits a
spontaneous compactification to a three-dimensional Anti-de Sitter
space. In the next section we will discuss the resulting
three-dimensional spectrum.

\subsection{The Kaluza-Klein spectrum}\label{kkspec}
Now we will turn to the Kaluza-Klein spectrum of type IIB supergravity
on $AdS_3\times S^3\times S^3\times S^1$. Usually for this one would have
to expand all fields in type IIB (or rather of the nine-dimensional theory
constructed by reducing on the $S^1$) into spherical harmonics of
$S^3\times S^3$ and linearize the resulting couplings. Masses and
spin\footnote{To be more precise, one should talk about the
representations of the $AdS_3$ isometry group.}
of the various fields could then be extracted.
This is a technically cumbersome procedure and has been done
explicitly, e.g., for the case $AdS_3\times S^3$ \cite{Deger:1998nm}.
There is however
a more convenient way, based on a group-theoretical analysis
due to Salam and Strathdee \cite{Salam:1981xd}.
The latter is applicable to all coset spaces (like the sphere
$S^3=SO(4)/SO(3)$), and can be explained as follows.

Suppose first for simplicity that we would compactify on a group
manifold $G$. A (scalar) field $\phi$ would then be expanded
according to
 \bea\label{modeexpansion}
  \phi(x,g) = \sum_n\sum_{p,q} \phi_{qp}^n(x)\hspace{0.3em} D_{pq}^n(g)\;,
 \eea
where $g\in G$. The sum runs over all irreducible representations
of $G$, labeled by~$n$. Due to the Peter-Weyl theorem,
in case of a group manifold the harmonics are determined by the
matrix elements $D_{pq}^n(g)$
of these irreducible representations.\footnote{One
may think of the illustrative example of
a compactification on $S^1$, i.e. on the group manifold $U(1)$.
Here the irreducible representations are labeled by an integer $n$
according to $\theta\rightarrow e^{in\theta}$, which are on the
other hand the Fourier modes in which one would expand.}
This expansion reflects the fact that the Kaluza-Klein modes can
be organized into representations of the isometry group.
(In contrast to generic coset spaces $G/H$, where the isometry group
is $G$, in the special case of a group manifold the isometry group
is actually enhanced to $G\times G$.)
If one compactifies instead on a coset space $G/H$, the fields
generically transform in some representation of $H$ (which will later
be identified with the local Lorentz group). Specifically,
 \bea
  \phi^i(hg)=D^i_{\hspace{0.4em}j}(h)\phi^j(g)\;.
 \eea
This is only consistent with an expansion of the type
(\ref{modeexpansion}), if the irreducible representations $D^n(g)$
fulfill the condition
 \bea
  D^n(hg)=D(h)D^n(g)\;,
 \eea
where $D(h)$ is the representation under which $\phi$ transforms.
In other words, for a coset space only those representations of
the isometry group will appear in the mode expansion that include the
given representation $D(h)$ upon restricting to $H$.

Applied to a compactification on $S^3$, this gives us a criterion
to determine which Kaluza-Klein modes actually appear \cite{deBoer:1998ip}.
Namely, any field transforms in a
representation $R_3$ of the local Lorentz group $SO(3)$ of $S^3$.
In the Kaluza-Klein tower then only those
representations $R_4$ of the isometry group $SO(4)$ will appear,
that contain $R_3$ in a decomposition of $R_4$ into $SO(3)$
representations. For illustration let us consider a scalar,
i.e.~a singlet under the Lorentz group. Denoting the
representations by spin quantum numbers of $SO(3)$ and
$SO(4)\cong SO(3)\times SO(3)$ (which contains the Lorentz group as
the diagonal subgroup), we see that the $SO(4)$ representations contained
in the Kaluza-Klein tower have to be of the form $(j,j)$,
since only these contain a singlet in the decomposition
$(j,j)\rightarrow 0\oplus 1\oplus ... \oplus 2j$.
Similarly, the vector-tower contains also states of
the form $(j,j+1)$ and $(j+1,j)$, and thus the Kaluza-Klein towers
on top of scalars and vectors read
 \bea\label{s3towers}
  0\longrightarrow \bigoplus_{j\geq 0} \hspace{0.2em}(j,j)\;, \qquad
  1\longrightarrow \bigoplus_{j\geq 1/2} \hspace{0.2em}(j,j)
  \Oplus \bigoplus_{j\geq 0}\hspace{0.2em}(j,j+1)
  \Oplus \bigoplus_{j\geq 0}\hspace{0.2em}(j+1,j)\;,
 \eea
and analogously for all other states.

Next we are going to apply this procedure to the compactification of type IIB
supergravity on $AdS_3\times S^3\times S^3\times S^1$.
The spectrum in $D=10$ is organized into representations of the
little group $SO(8)$ and reads \cite{Green:1987sp}
 \bea\label{IIBspec}
  ({\bf 8}_V - {\bf 8}_S)^2= ({\bf 8}_V\otimes {\bf 8}_V \oplus
  {\bf 8}_S\otimes {\bf 8}_S)
  -({\bf 8}_S\otimes {\bf 8}_V \oplus {\bf 8}_V\otimes {\bf 8}_S)\;,
 \eea
where the negative sign indicates fermionic states.
Since we compactify also on a circle, but take only the zero-modes
here, we start effectively from the nine-dimensional theory, i.e.
the little group is $SO(7)$. The relevant $SO(8)$ representations
decompose according to
 \bea
  {\bf 8}_V\rightarrow {\bf 1} \oplus {\bf 7}_V ,  \qquad
  {\bf 8}_S\rightarrow {\bf 8}\;,
 \eea
where ${\bf 8}$ denotes the spinor representation of $SO(7)$.
The $SO(7)$ gets actually further reduced to the little group of
the local Lorentz group of $AdS_3\times S^{3+}\times S^{3-}$.
As the little group of $AdS_3$ becomes trivial, this means
that the $SO(7)$ gets reduced to $SO(3)^+\times SO(3)^-$, so
that the representations decompose as
 \bea\label{so3reduc}
  {\bf 7}_V\rightarrow (0,0)\oplus (1,0)\oplus (0,1)\;, \qquad
  {\bf 8}\rightarrow (\ft12,\ft12)\oplus (\ft12,\ft12)\;.
 \eea
According to the criterion we have to associate to each of the
representations of the Lorentz group $SO(3)^+\times SO(3)^-$ appearing
in the IIB spectrum the tower of those representations of the isometry group
$SO(4)^+\times SO(4)^-$, which contains in the decomposition in
representations of $SO(3)^+\times SO(3)^-$
the given representation of the Lorentz group.
For instance, if we start from a field which is
a scalar $(0,0)$, we get the Kaluza-Klein tower
 \bea\label{kktower}
  (0,0)\hspace{0.5em} \longrightarrow
  \bigoplus_{j_1\geq 0,j_2\geq 0}(j_1,j_2;j_1,j_2) =:T_{KK}\;,
 \eea
which extends (\ref{s3towers}). Here we have denoted the representations
of the isometry groups
$SO(4)^{\pm}\equiv SO(3)_{L}^{\pm} \times SO(3)_{R}^{\pm}$ according to
the decomposition
 \bea\label{so4deco}
  G_{\rm c}&=&
  SO(3)_{L}^{+}\times SO(3)_{L}^{-} \times SO(3)_{R}^{+}\times SO(3)_{R}^{-}
  \;
  \label{so34}
 \eea
as $(\ell_L^+,\ell_L^-;\ell_R^+,\ell_R^-)$.
Specifically, the $SO(3)\times SO(3)$ content derived from the
type IIB spectrum (\ref{IIBspec}) by use of (\ref{so3reduc}) reads
 \begin{equation}\label{so3spec}
  \begin{split}
   T_{\text{IIB}}
   = \hspace{0.3em}&[10\hspace{0.1em}(0,0)\Oplus 9\hspace{0.1em}(1,0)\Oplus
   9\hspace{0.1em}(0,1)
   \Oplus 6\hspace{0.1em}(1,1)
   \Oplus (2,0) \Oplus (0,2)]_{\text{B}} \Oplus \\
   &[16\hspace{0.1em}(\ft12,\ft12)\Oplus 4\hspace{0.1em}(\ft32,\ft12)
   \Oplus 4\hspace{0.1em}(\ft12,\ft32)]_{\text{F}}\;,
  \end{split}
 \end{equation}
where the first line contains bosonic and the second line
fermionic states.
The lowest $SO(4)^+\times SO(4)^-$ states appearing on top of these
states are summarized in tab.~\ref{statesIIb}.
Here $h_{mn}$, $b_{\mu m}$, etc. denote the fluctuations of
the metric, the scalars, the 2-forms and the 4-form, respectively.
We have omitted all components which do not give rise to
propagating degrees of freedom on $AdS_3$, and have also suppressed
the components of the 4-form in the $S^1$ direction denoted by 9,
in accordance with the self-duality constraint.
In addition, the spectrum contains the 2-form $C_2$, whose
Kaluza-Klein tower is identical to the one for $B_2$ given
in tab.~\ref{statesIIb}.
Note that the fields of the first column in tab.~\ref{statesIIb}
represent the bosonic $SO(3)^+\times SO(3)^-$ states
in (\ref{so3spec}).\footnote{For instance, the singlets in
tab.~\ref{statesIIb} are given by $h^m_m$, $h^{\bar{m}}_{\bar{m}}$,
$h_{\mu 9}$, $h_{99}$, $\phi$, $C_0$, $b_{\mu 9}$, $c_{\mu 9}$,
$a_{\mu mnk}$ and $a_{\mu \bar{m}\bar{n}\bar{k}}$, in accordance
with the $10$ singlets in (\ref{so3spec}).}

If one now associates to each of the $SO(3)\times SO(3)$ states in
(\ref{so3spec}) a tower of $SO(4)\times SO(4)$ representations as in
(\ref{kktower}), one gets the Kaluza-Klein tower, which can be
organized into supermultiplets, as has been shown
in \cite{deBoer:1999rh}.
In the present case there exists, however, a more concise way to describe
the Kaluza-Klein tower. Instead of reducing each $SO(7)$
representation to $SO(3)\times SO(3)$ and then lifting to a tower
of $SO(4)\times SO(4)$ representations as explained above, the
$SO(7)$ representation can be lifted to a $SO(8)$ representation,
 \bea
  {\bf 8}\rightarrow {\bf 8}_S\;, \qquad {\bf 7}_V\rightarrow
  {\bf 8}_V-{\bf 1}\;,
 \eea
and then directly reduced to $SO(4)\times SO(4)$ according to
 \bea\label{8reduc}
  {\bf 8}_V\rightarrow (\ft12,0;\ft12,0)\oplus (0,\ft12;0,\ft12)\;,\qquad
  {\bf 8}_S\rightarrow (\ft12,0;0,\ft12)\oplus (0,\ft12;\ft12,0)\;.
 \eea
The claim is that if one associates to each $SO(8)$ representation
$R_{SO(8)}$ and the resulting $SO(4)\times SO(4)$ representation
$R_{SO(4)\times SO(4)}$ an entire tower of those representations according
to
 \bea\label{towerrule}
  R_{SO(8)}\longrightarrow  R_{SO(4)\times SO(4)}\otimes T_{KK}\;,
 \eea
where $T_{KK}$ denotes the Kaluza-Klein tower corresponding to the singlet
defined in (\ref{kktower}),
then this gives the same $SO(4)\times SO(4)$ content for
the Kaluza-Klein tower as the procedure introduced in (\ref{kktower})
above.

\begin{landscape}

\begin{table}[bt]
 \centering
{\footnotesize
  \begin{tabular}{|c||c|c|c|c|c|c|c|c|c|} \hline
    &
   $(0,0;0,0)$ & $(\ft12,0;\ft12,0)$ & $(0,\ft12;0,\ft12)$ &
   $(\ft12,\ft12;\ft12,\ft12)$ & $(1,0;1,0)$
   & $(0,1;0,1)$ & $(\ft12,1;\ft12,1)$ & $(1,\ft12;1,\ft12)$ & $(1,1;1,1)$
   \rule[-2ex]{0pt}{5.5ex}\\
    \hline\hline
   $h_{mn}$ & & & & & + & & & + & + \rule[-1.5ex]{0pt}{4ex}\\
   $h^m_m$ & + & + & + & + & + & + & + & + & +
   \rule[-1.5ex]{0pt}{3.5ex}\\
   $h_{\bar{m}\bar{n}}$ & & & & & & + & + & & + \rule[-1.5ex]{0pt}{3.5ex}\\
   $h^{\bar{m}}_{\bar{m}}$ & + & + & + & + & + & + & + & + & +
   \rule[-1.5ex]{0pt}{4ex}\\
   $h_{\mu m}$ & & + & & + & + & & + & + & + \rule[-1.5ex]{0pt}{3.5ex}\\
   $h_{\mu\bar{m}}$ & & & + & + & & + & + &
   + & +  \rule[-1.5ex]{0pt}{3.5ex}\\
   $h_{m\bar{n}}$ & & & &  + & & & + & + & + \rule[-1.5ex]{0pt}{3.5ex}\\
   $h_{\mu 9}$ & + & + & + & + & + & + & + & + & +
    \rule[-1.5ex]{0pt}{3.5ex}\\
   $h_{m 9}$ & & + & & + & + & & + & + & + \rule[-1.5ex]{0pt}{4ex}\\
   $h_{\bar{m} 9}$ & & & + & + & & + & + & + & + \rule[-1.5ex]{0pt}{3.5ex}\\
   $h_{99}$ & + & + & + & + & + & + & + & + & + \rule[-1.5ex]{0pt}{3.5ex}\\
    \hline
   $\phi, C_0$ & + & + & + & + & + & + & + & + & +
   \rule[-1.5ex]{0pt}{4ex}\\ \hline
   $b_{\mu m}$ & & + & & + & + & & + & + & +
   \rule[-1.5ex]{0pt}{3.5ex}\\
   $b_{\mu\bar{m}}$ & & & + & + & & + & + & + & + \rule[-1.5ex]{0pt}{3.5ex}\\
   $b_{mn}$ & & + & & + & + & & + & + & + \\
   $b_{\bar{m}\bar{n}}$ & & & + & + & & + & + & + & + \\
   $b_{m\bar{n}}$ & & & & + & & & + & + & + \\
   $b_{\mu 9}$ & + & + & + & + & + & + & + & + & + \\
   $b_{m 9}$ & & + & & + & + & & + & + & + \\
   $b_{\bar{m}9}$ & & & + & + & & + & + & + & + \\ \hline
   $a_{\mu mnk}$ & + & + & + & + & + & + & + & + & + \\
   $a_{\mu\bar{m}\bar{n}\bar{k}}$ & + & + & + & + & + & + & + & + & + \\
   $a_{\mu mn\bar{k}}$ & & & & + & & & + & + & + \\
   $a_{\mu m\bar{n}\bar{k}}$ & & & & + & & & + & + & + \\
   $a_{mnk\bar{l}}$ & & & + & + & & + & + & + & + \\
   $a_{mn\bar{k}\bar{l}}$ & & & & + & & & + & + & + \\
   $a_{m\bar{n}\bar{k}\bar{l}}$ & & + & & + & + & & + & + & +  \\
    \hline
   $\Sigma_{scal}$ & 5 & 10 & 10 & 18 & 11 & 11 & 19 & 19 & 20 \\
   $\Sigma_{vect}$ & 3 & 6 & 6 & 9 & 6 & 6 & 9 & 9 & 9 \\
   $\Sigma_{form}$ & $2$ & $3$ & $3$ & $7$ & $3$ &
   $3$ & $7$ & $7$ & $7$ \\
    \hline
  \end{tabular}
  }
 \caption{\small The lowest KK states on top of the IIB fields.}
\label{statesIIb}
\end{table}

\end{landscape}
This is trivially the case for the scalars. To confirm this statement
in general one has to check explicitly that both
procedures coincide for the $SO(7)$ representations ${\bf 7}_V$
and ${\bf 8}$.
Let us show this for ${\bf 7}_V$. Reducing it
directly to $SO(3)^+\times SO(3)^-$ according to (\ref{so3reduc})
and then lifting as in (\ref{kktower}) results in
 \begin{eqnarray}\label{7vreduc}
  {\bf 7}_V &\longrightarrow& (0,0)\Oplus (1,0)\Oplus (0,1) \\ \nonumber
            &\longrightarrow& \bigoplus_{j_1,j_2 \leq 0} (j_1,j_2;j_1,j_2)
  \oplus \bigoplus_{j_1\leq 1/2, j_2 \leq 0} (j_1,j_2;j_1,j_2)
  \oplus \bigoplus_{j_1\leq 0, j_2 \leq 1/2} (j_1,j_2;j_1,j_2) \\ \nonumber
  &\oplus& \bigoplus_{j_1,j_2 \leq 0} (j_1+1,j_2;j_1,j_2)
  \oplus \bigoplus_{j_1,j_2 \leq 0} (j_1,j_2;j_1+1,j_2)
  \oplus \bigoplus_{j_1,j_2 \leq 0} (j_1,j_2+1;j_1,j_2) \\ \nonumber
  &\oplus& \bigoplus_{j_1,j_2 \leq 0} (j_1,j_2;j_1,j_2+1)   \;.
 \end{eqnarray}
For the second procedure we first have to lift to $SO(8)$, then reducing
to $SO(4)\times SO(4)$ according to (\ref{8reduc}) and finally
tensoring with the $T_{KK}$ in (\ref{kktower}). This yields
 \begin{equation}
  \left( (\ft12,0;\ft12,0) \Oplus (0,\ft12;0,\ft12) - (0,0;0,0)\right)
  \hspace{0.3em} \otimes \bigoplus_{j_1,j_2\geq 0}(j_1,j_2;j_1,j_2)\;,
 \end{equation}
and one may check explicitly that this tensor product coincides
with (\ref{7vreduc}).

Up to now we have defined an algorithm to determine the
Kaluza-Klein spectrum in terms of representations of the
$SO(4)\times SO(4)$ isometry group. Next we are going to use
these results in order to write the entire Kaluza-Klein tower
in a very compact form in terms of supermultiplets (see (\ref{tower})
below, \cite{deBoer:1999rh}).
However, since this group-theoretical
procedure cannot reveal the masses of the various Kaluza-Klein
states, we first have to comment on
the required uniqueness of the supermultiplet structure.
Standard sphere compactifications preserve maximal supersymmetry,
so that for them massive long multiplets
would contain states with spin up to 4. Therefore the Kaluza-Klein
spectra can only be organized into (short) massive BPS multiplets, which
are consistent with maximum spin 2. In turn, these BPS multiplets
determine the masses of the states completely, such that the entire
spectrum of supermultiplets is fixed. Even though the
$AdS_3\times S^3\times S^3$ background is only half-maximally supersymmetric,
we will, following \cite{deBoer:1998ip,deBoer:1999rh}, nevertheless assume
that the spectrum is organized into short multiplets.
This is a reasonable assumption, since it
naturally reflects the bound $m^2=0$ for the massless fields in
the higher-dimensional theory we started with.

Let us first review the structure of supermultiplets on $AdS_3$.
As the background is half-maximally supersymmetric, it preserves
${\cal N}=8$ in $D=3$, which corresponds to $16$ real supercharges.
Thus the spectrum is organized under some background
(super-) isometry group,  which in the given case turns out to be
a direct product of two ${\cal N}=4$ supergroups \cite{Sevrin:1988ew}
 \bea
  D^1(2,1;\alpha)_{L} \times D^1(2,1;\alpha)_{R} \;.
  \label{supergroup}
 \eea
This is the required supersymmetrisation of the AdS group (\ref{so22}),
in which each factor combines a bosonic
$SO(3)\times SO(3)\times SL(2,\mathbb{R})$
with eight real supercharges~(see~\cite{Sevrin:1988ew}
for the exact definitions).
More precisely, the noncompact factors
$SL(2,\mathbb{R})_{L}\times SL(2,\mathbb{R})_{R} = SO(2,2)$
join into the isometry group of $AdS_3$ while the compact factors
build the isometry groups
$SO(4)^{\pm}\equiv SO(3)_{L}^{\pm} \times SO(3)_{R}^{\pm}$
of the two spheres $S^{3+}\times S^{3-}$ in (\ref{so4deco}).
Accordingly, this group will show up as the gauge group of the
effectively three-dimensional supergravity action.
The parameter $\alpha$ of~(\ref{supergroup})
corresponds to the ratio of the radii of the two spheres $S^3$.

The group~(\ref{supergroup})
possesses short and long multiplets (analogously to massless and
massive supermultiplets for Poincar\'e supersymmetry).
A short $D^1(2,1;\alpha)$ supermultiplet is defined by its highest
weight state $(\ell^+,\ell^-)^{h_{0}}$, where $\ell^{\pm}$ label spins
of $SO(3)^{\pm}$ and
$h_{0}=\frac1{1+\alpha}\,\ell^{+}+ \frac\alpha{1+\alpha}\,\ell^{-}$
is the charge under the Cartan subgroup ~$SO(1,1)\subset SL(2,\mathbb{R})$.
The corresponding supermultiplet will be denoted by
$(\ell^+,\ell^-)_{\rm S}$. It is
generated from the highest weight state by the action of
three out of the four supercharges $G^a_{-1/2}$ ($a=1,...,4$) and
carries $8(\ell_{+}+\ell_{-}+4\ell_{+}\ell_{-})$ degrees of freedom.
Its $SO(3)^{\pm}$ representation content is summarized in table~\ref{short}.

\begin{table}[bt]
\centering
  \begin{tabular}{c | c c c}
  $h$ & \\
  \hline
   $h_{0}$ & & $(\ell^+,\ell^-)$ & \\[.5ex]
   $h_{0}+\frac{1}{2}$ & $(\ell^+-\frac{1}{2},\ell^- -\frac{1}{2})$ &
   $(\ell^+ +\ft12,\ell^- -\ft12)$ & $(\ell^+ -\ft12,\ell^- +\ft12)$ \\[.5ex]
   $h_{0}+1$ & $(\ell^+,\ell^- -1)$ &
   $(\ell^+ -1,\ell^-)$ & $(\ell^+,\ell^-)$ \\[.5ex]
   $h_{0}+\ft32$ & & $(\ell^+ -\ft12,\ell^- -\ft12)$ &
  \end{tabular}
  \caption{\small The generic
  short supermultiplet $(\ell^+,\ell^-)_{\rm S}$ of $D^1(2,1;\alpha)$,
with  $h_{0}=\frac1{1+\alpha}\,\ell^{+}+ \frac\alpha{1+\alpha}\,\ell^{-}$.}
  \label{short}
\end{table}

The generic long multiplet $(\ell^+,\ell^-)_{\rm long}$
instead is built from the action
of all four supercharges $G^a_{-1/2}$
on the highest weight state and carries
$16\,(2\ell_{+}\!+\!1)(2\ell_{-}\!+\!1)$ degrees of freedom.
Its highest weight state satisfies the unitarity bound
$h\ge\frac1{1+\alpha}\,\ell^{+}+ \frac\alpha{1+\alpha}\,\ell^{-}$.
In case this bound is saturated, the long multiplet decomposes
into two short multiplets (table \ref{short}) according to
 \bea
  (\ell^+,\ell^-)_{\rm long} &=&
  (\ell^+,\ell^-)_{\rm S} \Oplus (\ell^+\!+\ft12,\ell^-\!+\ft12)_{\rm S}
  \;.
  \label{long}
 \eea
The lowest short supermultiplets $(0,\ft12)_{\rm S}$,
$(0,1)_{\rm S}$, and $(\ft12,\ft12)_{\rm S}$ of
$D^1(2,1;\alpha)$ are further degenerate and collected in table~\ref{lower},
and similar for $\ell^+\leftrightarrow\ell^-$, $\Ga \leftrightarrow1/\Ga$.
The long multiplet $(0,0)_{\text{long}}$ is given in table~\ref{lang}.
Moreover, an unphysical short multiplet $(0,0)_S$ can be defined in such
a way that (\ref{long}) applies also to $(0,0)_{\rm long}=(0,0)_{\rm S}
\Oplus (\ft12,\ft12)_{\rm S}$ (see table~\ref{lang}).
Here, negative states are understood as constraints that
eliminate physical degrees of freedom.

\begin{table}[bt]
\centering
  \begin{tabular}{c|c}
   $h$ & $(\ell^+,\ell^-)$ \\ \hline
   $\frac{\Ga}{2(1+\Ga)}$ & $(0,\ft12)$ \\
   $\frac{2\Ga+1}{2(1+\Ga)}$ & $(\ft12,0)$ \\ \vl \\ \vl
  \end{tabular}
 \qquad\quad
  \begin{tabular}{c|c}
   $h$ & $(\ell^+,\ell^-)$ \\ \hline
   $\frac{\Ga}{1+\Ga}$ & $(0,1)$ \\
   $\frac{3\Ga+1}{2(1+\Ga)}$ & $(\ft12,\ft12)$ \\
   $\frac{2\Ga+1}{1+\Ga}$ & $(0,0)$  \\ \vl
  \end{tabular}
 \qquad\quad
  \begin{tabular}{c|c}
   $h$ & $(\ell^+,\ell^-)$ \\ \hline
   $\ft12$ & $(\ft12,\ft12)$ \\
   $1$ & $(0,0)+(0,1)+(1,0)$ \\
   $\ft32$ & $(\ft12,\ft12)$ \\
   $2$ & $(0,0)$
  \end{tabular}
    \caption{\small The lowest short supermultiplets $(0,\ft12)_{\rm S}$,
$(0,1)_{\rm S}$, and $(\ft12,\ft12)_{\rm S}$ of  $D^1(2,1;\alpha)$.}
\label{lower}
\end{table}

\begin{table}[b]
\centering
  \begin{tabular}{c|c}
   $h$ & $(\ell^+,\ell^-)$ \\ \hline
   $0$ & $(0,0)$ \\
   $\ft12$ & $(\ft12,\ft12)$ \\
   $1$ & $(0,1)$ + $(1,0)$ \\
   $\ft32$ & $(\ft12,\ft12)$ \\
   $2$ & $(0,0)$
  \end{tabular}
  \qquad\qquad\quad
  \begin{tabular}{c|c}
   $h$ & $(\ell^+,\ell^-)$ \\ \hline
   $0$ & $(0,0)$ \\
   $\ft12$ & $0$ \\
   $1$ & $-(0,0)$
  \end{tabular}
    \caption{\small The long multiplet $(0,0)_{\text{long}}$ and the
   unphysical short multiplet $(0,0)_{\rm S}$ }
   \label{lang}
\end{table}

Short representations of the full supergroup~(\ref{supergroup}) are
constructed as tensor products of the supermultiplets (tab.~\ref{short})
for the left and right factors, and correspondingly
will be denoted by $(\ell_L^+,\ell_L^-;\ell_R^+,\ell_R^-)_{\rm S}$.
The quantum numbers which denote the
representations of the $AdS_3$ group $SO(2,2)$
are labeled by numbers $s$ and $\GD$,
which encode the $AdS$ analogue of spin and mass, respectively.
They are related to the values of $h_{R}$ and $h_{L}$
by $s=h_R-h_L, \GD=h_L+h_R$.

The massive Kaluza-Klein spectrum of nine-dimensional supergravity on
$AdS_3\times S^3\times S^3$ can now be written in terms
of supermultiplets.
For this we use that the
IIB spectrum $({\bf 8}_V-{\bf 8}_S)^2$ yields via the reduction
in (\ref{8reduc}) a
bosonic and fermionic $SO(4)\times SO(4)$ spectrum which coincides
exactly with the content of the supermultiplet
$(0,0;0,0)_{\rm long}$, which can be easily checked
with tab.~\ref{lang}. Put differently, to associate to
each $SO(8)$ representation of type IIB a Kaluza-Klein tower according to
(\ref{towerrule}) is equivalent to take the tensor product of
$(0,0;0,0)_{\rm long}$ (to which we will therefore refer as the
fundamental multiplet) with the Kaluza-Klein tower in  (\ref{kktower}).
Thus the full resulting Kaluza-Klein spectrum is described by
 \bea\label{longspec}
  (0,0 ; 0,0)_{\rm long}\hspace{0.3em}\otimes
  \bigoplus_{\ell^{\pm}\geq 0}(\ell^+,\ell^-;\ell^+,\ell^-)
  \hspace{0.2em}= \hspace{0.2em}
  \bigoplus_{\ell^{\pm}\geq 0}(\ell^+,\ell^-;\ell^+,\ell^-)_{\rm long}\;,
 \eea
where the last equation follows from the fact that the long multiplets
are given by tensor products of $(0,0)_{\rm long}$ with the highest-weight
state $(\ell^+,\ell^-)$,
 \bea
  (\ell^+,\ell^-)_{\rm long}=(0,0)_{\rm long}\otimes (\ell^+,\ell^-)\;.
 \eea
This can be easily checked by use of (\ref{long}) and table~\ref{lang}.
Equivalently, as the spectrum is decomposable into short multiplets,
it can be rewritten with (\ref{long}) as
 \begin{eqnarray}\label{tower}
  && \bigoplus_{\ell^+\ge0,\ell^-\ge1/2}
  (\ell^+,\ell^-;\ell^+,\ell^-)_{\rm S}~~\oplus
  \bigoplus_{\ell^+\ge 1/2,\ell^-\ge 0}
  (\ell^+,\ell^-;\ell^+,\ell^-)_{\rm S} \nonumber\\[1ex]
  &&
  \qquad\qquad \oplus\bigoplus_{\ell^+,\ell^-\ge 0}
  \big( (\ell^+,\ell^-;\ell^+\!+\ft12,\ell^-\!+\ft12)_{\rm S}
  \oplus (\ell^+\!+\ft12,\ell^-\!+\ft12;\ell^+,\ell^-)_{\rm S} \big)
 \end{eqnarray}
where we have omitted the unphysical multiplet $(0,0;0,0)_{\rm S}$
according to tab.~\ref{lang}.
Note that the multiplets $(\ell^+,\ell^-;\ell^+,\ell^-)_{\rm S}$ generically
contain massive fields with spin running from 0 to $\ft32$, whereas
multiplets of the type
$(\ell^+,\ell^-;\ell^+\!+\!\ft12,\ell^-\!+\!\ft12)_{\rm S}$
represent massive spin-2 multiplets.

In addition, the tower (\ref{tower})
contains $(\ft12,\ft12;0,0)_{\rm S}\Oplus (0,0;\ft12,\ft12)_{\rm S}$,
which is the massless supergravity multiplet. It consists of the
vielbein, eight gravitinos, transforming as
 \bea
  \psi_{\mu}^{I} &:&\;\;
  (\ft12,\ft12;0,0)\oplus (0,0;\ft12,\ft12) \;,
  \label{supercharges}
 \eea
under (\ref{so34}),
and topological gauge vectors, corresponding to the
$SO(4)_{L}\times SO(4)_{R}$
gauge group. In accordance with the counting of~table~\ref{short} it
does not contain any physical degrees of freedom.

In total we have seen that generically the short multiplets
appearing in the Kaluza-Klein spectrum~(\ref{tower})
may combine into long multiplets~(\ref{long})~\cite{deBoer:1999rh}.
This holds for all multiplets, except for the supergravity
multiplet and one of the lowest massive spin-$3/2$ multiplets
$(\ft12,\ft12;\ft12,\ft12)_{\rm S}$, since we have omitted the
unphysical short multiplet $(0,0;0,0)_{\rm S}$ in the step
from (\ref{longspec}) to (\ref{tower}).
This is in contrast to other sphere compactifications, which
preserve maximal supersymmetry.
The conformal weight of the long representations appearing in
(\ref{tower}) is not protected
by anything and may vary throughout the moduli space.
This gives a distinguished role to the
supermultiplet~$(\ft12,\ft12;\ft12,\ft12)_{\rm S}$
that we shall analyze here.

\section[Effective supergravities for spin-1/2 and spin-1 multiplets]
{Effective supergravities for spin-1/2 and \\spin-1 multiplets}\label{lowspin}
Next we are going to construct the effective supergravities for
the lowest Kaluza-Klein multiplets \cite{Hohm:2005ui}.
As the background preserves half of
the supersymmetries, i.e. in total 16 real supercharges, the theories
will have ${\cal N}=8$ in $D=3$.

The lowest multiplet is the massless supergravity multiplet
$(0,0;\ft12,\ft12)\oplus (\ft12,\ft12;0,0)$. We have already seen
in sec.~\ref{CSgravity} that it is described by a Chern-Simons theory for
the appropriate supergroup.

Furthermore, the lowest massive multiplets in the Kaluza-Klein
tower (\ref{tower}) are the degenerate multiplets
$(0,\ft12;0,\ft12)_{\rm S}$ and $(0,1;0,1)_{\rm S}$ (together with
$(\ft12,0;\ft12,0)_{\rm S}$ and $(1,0;1,0)_{\rm S}$), to which
we will refer as the spin-$\ft12$ and spin-$1$ multiplet, respectively,
in accordance with their states of maximal spin.
Their precise representation content is collected in table~\ref{spinspin}.

\begin{table}[bt]
\centering
  \begin{tabular}{c||c|c|}
   \raisebox{-1.25ex}{$h_L$} \raisebox{1.25ex}{$h_R$} &
   $\frac{\alpha}{2(1+\alpha)}$ & $\frac{1+2\alpha}{2(1+\alpha)}$
   \rule[-2ex]{0pt}{5.5ex}\\
    \hline\hline
   $\frac{\alpha}{2(1+\alpha)}$ & $(0,\ft12;0,\ft12)$ & $(0,\ft12;\ft12,0)$
   \rule[-1.5ex]{0pt}{4ex}\\
    \hline
   $\frac{1+2\alpha}{2(1+\alpha)}$ &  $(\ft12,0;0,\ft12)$ &
   $(\ft12,0;\ft12,0)$  \rule[-1.5ex]{0pt}{4ex} \\
   \hline
  \multicolumn{3}{c}{\vphantom{I}}\\[.9ex]
  \end{tabular}
\qquad\;\;
  \begin{tabular}{c||c|c|c|}
   \raisebox{-1.25ex}{$h_L$} \raisebox{1.25ex}{$h_R$} &
   $\frac{\alpha}{1+\alpha}$ & $\frac{3\alpha+1}{2(1+\alpha)}$
   & $\frac{2\alpha+1}{1+\alpha}$
   \rule[-2ex]{0pt}{5.5ex}\\
    \hline\hline
   $\frac{\alpha}{1+\alpha}$ & $(0,1;0,1)$  & $(0,1;\ft12,\ft12)$ &
   $(0,1;0,0)$
   \rule[-1.5ex]{0pt}{4ex}\\
    \hline
   $\frac{3\alpha+1}{2(1+\alpha)}$ &  $(\ft12,\ft12;0,1)$
   & $(\ft12,\ft12;\ft12,\ft12)$ &
   $(\ft12,\ft12;0,0)$  \rule[-1.5ex]{0pt}{4ex} \\
     \hline
   $\frac{2\alpha+1}{1+\alpha}$ &  $(0,0;0,1)$ & $(0,0;\ft12,\ft12)$
   & $(0,0,0,0)$ \rule[-1.5ex]{0pt}{4ex} \\
  \hline
  \end{tabular}
      \caption{\small The spin-$\ft12$ multiplet $(0,\ft12;0,\ft12)_{\rm S}$,
      and the massive spin-$1$ multiplet~$(0,1;0,1)_{\rm S}$.}
\label{spinspin}
\end{table}

To construct their respective gauged supergravities one first of
all has to identify the corresponding ungauged theory.
Since it should be an ${\cal N}=8$ supersymmetric theory, we know
already from sec. \ref{CSgravity} the general form of the target spaces in
(\ref{so8coset}). It remains to determine the actual dimension
of this coset space, i.e. the number $n$ in
(\ref{so8coset}). The two spin-$\ft12$ multiplets of table~\ref{spinspin}
contain together 16 bosonic degrees of freedom.
This suggests that together they are effectively described by
a gauging of the ${\cal N}=8$ theory
with target space $SO(8,2)/(SO(8)\times SO(2))$.
To confirm this we have to show that the gauge group $SO(4)\times SO(4)$
can be embedded into the rigid symmetry group $SO(8,2)$ in such a
way that the induced spectrum coincides with the representations
expected from table \ref{spinspin}.
As reviewed in sec. \ref{N8gauging},
the propagating degrees of freedom of the ${\cal N}=8$ theory
are carried entirely by the scalars $P_{\mu}^{Ir}$ and the fermions
$\chi^{\dot{A}r}$. Thus they transform under $SO(8)$ as
${\bf 8}_V\oplus{\bf 8}_C$.
And indeed, one verifies that the field content of
$(0,\ft12;0,\ft12)_{\rm S}$
can be lifted from a representation of the gauge group~(\ref{so34})
to an ${\bf 8}_{V}\oplus {\bf 8}_{C}$  with the embedding
 \bea
  {\bf 8}_{V} \rightarrow (0,\ft12;0,\ft12)\Oplus (\ft12,0;\ft12,0) \;,
  \qquad
  {\bf 8}_{C} \rightarrow (0,\ft12;\ft12,0)\Oplus (\ft12,0;0,\ft12) \;,
  \label{embedding12}
 \eea
while the supercharges~(\ref{supercharges}) lift to the
spinor representation
${\bf 8}_{S}$ of $SO(8)$. This corresponds to the
canonical embedding of the $SO(4)\times SO(4)$ gauge group (see (\ref{so34}))
into $SO(8)$ according to
${\bf 8}_{V}\rightarrow {\bf 4}_V\hspace{0.1em}\oplus\hspace{0.1em}{\bf 4}_V$,
etc. Hence, the two spin-$1/2$ multiplets reproduce the field content
$({\bf 8}_{V}\oplus{\bf 8}_{C},{\bf 2})$ of the ungauged
$SO(8,2)/(SO(8)\times SO(2))$ theory. It remains to
verify that the embedding~(\ref{embedding12})
of the gauge group into $SO(8,2)$ is compatible with the
constraints imposed by supersymmetry on the embedding tensor
$\Theta_{{\cM\cN}}$.
Along the lines of \cite{Nicolai:2001ac} it can be shown that
these requirements determine $\Theta_{{\cM\cN}}$
completely up to a free parameter
corresponding to the ratio $\alpha$ of the two
sphere radii.
The effective theory is then completely determined.
Its scalar potential will be further investigated in
chapter \ref{adschapter} (see also~\cite{Berg:2001ty})
and indeed reproduces the correct scalar masses
predicted by table~\ref{spinspin}.

The coupling of the spin-$1$ multiplets
$(0,1;0,1)_{\rm S}\oplus (1,0;1,0)_{\rm S}$
is slightly more involved due to the presence of
massive vector fields but can be achieved by a
straightforward generalization of the case of
a single $S^{3}$ compactification~\cite{Nicolai:2003bp,Nicolai:2003ux}.
Here, the effective theory for $128$ degrees of freedom is a gauging
of the ${\cal N}=8$ theory
with coset space $SO(8,8)/(SO(8)\times SO(8))$.
Again, the first thing to verify in this case is
that the field content of
$(0,1;0,1)_{\rm S}\oplus (1,0;1,0)_{\rm S}$
(table~\ref{spinspin})
can be lifted from a representation of the gauge group~(\ref{so34})
as above to an
$({\bf 8}_{V}\oplus{\bf 8}_{C},{\bf 8}_{V})$ of $SO(8)\times SO(8)$
via the embedding
 \bea
  &&({\bf 8}_{V},{\bf 1}) \rightarrow (0,\ft12;0,\ft12)\oplus
  (\ft12,0;\ft12,0) \;,
  \quad
  ({\bf 8}_{C},{\bf 1}) \rightarrow (0,\ft12;\ft12,0)\oplus
  (\ft12,0;0,\ft12) \;,
  \nonumber\\[1ex]
  &&({\bf 1},{\bf 8}_{V}) \rightarrow (0,\ft12;0,\ft12)\oplus
  (\ft12,0;\ft12,0) \;,
  \quad
  ({\bf 1},{\bf 8}_{C}) \rightarrow
  (0,\ft12;\ft12,0)\oplus (\ft12,0;0,\ft12)
  \label{embedding22}
 \eea
This corresponds to the embedding of groups
$SO(8)\times SO(8)\supset SO(8)_{D}\supset SO(4)\times SO(4)$,
where $SO(8)_D$ denotes the diagonal subgroup
of the two $SO(8)$ factors. For instance, (\ref{embedding22})
implies that the bosonic part decomposes as
 \bea
  ({\bf 8}_{V},{\bf 8}_{V}) &\rightarrow&
  \Big((0,\ft12;0,\ft12)\oplus (\ft12,0;\ft12,0)\Big)\, \otimes \,
  \Big((0,\ft12;0,\ft12)\oplus (\ft12,0;\ft12,0)\Big)
   \\[1ex]
   &=&
  (0,1;0,1)\oplus(0,1;0,0)\oplus(0,0;0,1)\oplus
  (1,0;1,0)\oplus(1,0;0,0)\oplus(0,0;1,0)
  \nonumber\\
   &&{}
  \oplus 2\cdot(0,0;0,0)\oplus2\cdot
  (\ft12,\ft12;\ft12,\ft12) \;,
  \nonumber
 \eea
in agreement with table~\ref{spinspin}
and its conjugate.
It is important to note that the massive
spin-$1$ fields
show up in this decomposition
through their Goldstone scalars, in accordance with the general
discussion in \ref{gaugedsugra}.

Moreover, we have to keep in mind that
in order to reproduce the correct coupling for these massive vector fields,
the total gauge group $G_{0}\subset SO(8,8)$ for the Chern-Simons
gauged supergravity should not just be the compact
factor $G_{\rm c}$~(\ref{so34}), but is rather extended by some
nilpotent generators. In fact, the algebra takes the form of a
semi-direct product
 \bea
  G_{0}&=& G_{\rm c}\ltimes T_{12} \;,
 \label{GT12}
 \eea
with the abelian $12$-dimensional translation group $T_{12}$
transforming in the adjoint representation of
$G_{\rm c}$~\cite{Nicolai:2003bp}.
In the $AdS_{3}$ vacuum, these translational symmetries are
broken and the corresponding vector fields gain their masses
in a Higgs effect.
The proper embedding of~(\ref{GT12}) into $SO(8,8)$ is again
uniquely fixed by the constraints imposed by supersymmetry
on the embedding tensor $\Theta_{{\cM\cN}}$
up to the free parameter $\alpha$~\cite{Nicolai:2003ux}.

Finally, it is straightforward to construct the effective theory
that couples both the spin-$1/2$ and the spin-$1$ supermultiplets
as a gauging of the theory with coset space $SO(8,10)/(SO(8)\times SO(10))$
which obviously embeds the two target spaces described above.

\section{The spin-3/2 multiplet}\label{spin32}

Finally we turn to the coupling of the spin-3/2 multiplet
$(\ft12,\ft12;\ft12,\ft12)_{\rm S}$, which is contained
twice~in the Kaluza-Klein tower (\ref{tower}).
Its $SO(4)\times SO(4)$ representation content
is summarized in table~\ref{spin32}.
\begin{table}[bt]
 \centering
  \begin{tabular}{|c||c|c|c|c|} \hline
   \raisebox{-1.25ex}{$h_L$} \raisebox{1.25ex}{$h_R$} &
   $\ft{1}{2}$ & $1$ & $\ft32$ & $2$ \rule[-2ex]{0pt}{5.5ex}\\
    \hline\hline
   & & $(\ft12,\ft12;0,0)$ & & \rule[-1.5ex]{0pt}{4ex}\\
   $\ft12$ & $(\ft12,\ft12;\ft12,\ft12)$ & $(\ft12,\ft12;0,1)$ &
   $(\ft12,\ft12;\ft12,\ft12)$ & $(\ft12,\ft12;0,0)$
   \rule[-1.5ex]{0pt}{3.5ex}\\
   & & $(\ft12,\ft12;1,0)$ & & \rule[-1.5ex]{0pt}{3.5ex}\\
    \hline
   & &  $(0,0;0,0)$ & &  \rule[-1.5ex]{0pt}{4ex}\\
   & $(1,0;\ft12,\ft12)$ & $(0,1;0,0),(0,0;0,1)$ & $(0,1;\ft12,\ft12)$ &
     $(0,0;0,0)$ \rule[-1.5ex]{0pt}{3.5ex}\\
   $1$ & $(0,0;\ft12,\ft12)$ & $(1,0;0,0),(0,0;1,0)$ & $(0,0;\ft12,\ft12)$ &
     $(0,1;0,0)$ \rule[-1.5ex]{0pt}{3.5ex}\\
   & $(0,1;\ft12,\ft12)$ & $(0,1;0,1),(1,0;1,0)$ & $(1,0;\ft12,\ft12)$ &
     $(1,0;0,0)$ \rule[-1.5ex]{0pt}{3.5ex}\\
   & &  $(0,1;1,0),(1,0;0,1)$ & & \rule[-1.5ex]{0pt}{3.5ex}\\
    \hline
   & & $(\ft12,\ft12;0,0)$ & & \rule[-1.5ex]{0pt}{4ex}\\
   $\ft32$ & $(\ft12,\ft12;\ft12,\ft12)$ & $(\ft12,\ft12;0,1)$ &
   $(\ft12,\ft12;\ft12,\ft12)$ & $(\ft12,\ft12;0,0)$
   \rule[-1.5ex]{0pt}{3.5ex}\\
   & & $(\ft12,\ft12;1,0)$ & & \rule[-1.5ex]{0pt}{3.5ex}\\
    \hline
   & & $(0,0;0,0)$ & & \rule[-1.5ex]{0pt}{4ex}\\
   $2$ & $(0,0;\ft12,\ft12)$ & $(0,0;0,1)$ & $(0,0;\ft12,\ft12)$
   & $(0,0;0,0)$ \rule[-1.5ex]{0pt}{3.5ex}\\
   & & $(0,0;1,0)$ & & \rule[-1.5ex]{0pt}{3.5ex}\\
    \hline
  \end{tabular}
 \caption{\small The massive spin-$3/2$ supermultiplet
 $(\ft12,\ft12;\ft12,\ft12)_{\rm S}$.}
\label{spin32con}
\end{table}
In analogy to the aforementioned couplings of the spin-$\ft12$ and
spin-$1$ multiplet to ${\cal N}=8$ supergravity,
a natural candidate for the effective theory
might be an ${\cal N}=8$ gauging of the theory
with coset space $SO(8,16)/(SO(8)\times SO(16))$,
reproducing the correct number of 128 bosonic degrees of freedom.
(The appearance of massive spin-$\ft32$
fields would then require some analogue of the
dualization taking place in the scalar/vector sector.)
Let us check the representation content of table~\ref{spin32con}.
It is straightforward to verify that
the states of this multiplet may be lifted from a representation
of the gauge group~(\ref{so34}) to a representation
$({\bf 8}_{V}\oplus {\bf 8}_{C},{\bf 8}_{V}\oplus {\bf 8}_{C})$
of an $SO(8)_{L}\times SO(8)_{R}$ according to
 \bea
  &&({\bf 8}_{V},{\bf 1}) \rightarrow (0,0;0,0)\oplus
  (0,0;0,0)\oplus (1,0;0,0)\oplus (0,1;0,0)\;,
  \nonumber\\
  &&({\bf 8}_{C},{\bf 1}) \rightarrow (\ft12,\ft12;0,0)\oplus
  (\ft12,\ft12;0,0) \;,
  \quad
  ({\bf 8}_{S},{\bf 1})
  \rightarrow (\ft12,\ft12;0,0)\oplus (\ft12,\ft12;0,0) \;,
  \nonumber\\
  &&({\bf 1},{\bf 8}_{V}) \rightarrow(0,0;0,0)\oplus (0,0;0,0)\oplus
  (0,0;1,0)\oplus (0,0;0,1) \;,
  \quad
  \nonumber\\
  &&({\bf 1},{\bf 8}_{C}) \rightarrow (0,0;\ft12,\ft12)\oplus
  (0,0;\ft12,\ft12)\;,
  \quad
  ({\bf 1},{\bf 8}_{S}) \rightarrow
  (0,0;\ft12,\ft12)\oplus (0,0;\ft12,\ft12)
  \label{embedding33}
 \eea
This corresponds to an embedding of groups according to
 \bea\label{so4embed}
  SO(4)_{L} = {\rm diag}[SO(4)\times SO(4)]
  \subset SO(4)\times SO(4) \subset SO(8)_{L}\;,
 \eea
and similarly for $SO(4)_{R}$.
In order to be described as a gauging of the ${\cal N}=8$
theory, the field content would have to be further lifted
to the $({\bf 8}_{V}\oplus {\bf 8}_{C},{\bf 16})$
of $SO(8)\times SO(16)$.
This is only possible, if $SO(8)_{R}$ is entirely embedded
into the $SO(16)$.
On the other hand, we know from sec.~\ref{N8gauging} that the
gravitino of the ${\cal N}=8$ theory transform in
the $({\bf 8}_{S},{\bf 1})$.
This in turn implies with the embedding (\ref{embedding33})
that they would decompose
as $(\ft12,\ft12;0,0)\oplus (\ft12,\ft12;0,0)$, in contrast to the
gravitinos~(\ref{supercharges})
of the Kaluza-Klein spectrum.
We conclude that the massive spin-$3/2$ multiplet
cannot be described as a gauging of the $SO(8,16)/(SO(8)\times SO(16))$
theory.

Rather we will find that the effective theory describing
this multiplet is a maximally supersymmetric
gauging of the ${\cal N}=16$ theory in its broken phase.
Half of the supersymmetry is broken down to ${\cal N}=8$ and
correspondingly eight gravitinos acquire mass via a super-Higgs
mechanism.
As a first check
we observe that the total number of degrees of freedom
collected in table~\ref{spin32con} indeed equals the $16^2=256$
of the maximal theory.
This is moreover in agreement with our general considerations in
sec.~\ref{higherspin}, where we have argued that a consistent coupling of
massive spin-3/2 fields requires the existence of spontaneously
broken supercharges.

More specifically, we have to check again whether an embedding
of the gauge group can be found, that reproduces the right spectrum.
In the ${\cal N}=16$ theory we have seen that the fields are described
by $P_{\mu}^A$ and $\chi^{\dot{A}}$, i.e. transforming in the
spinor and conjugate spinor representation of $SO(16)$.
Thus the required spectrum has to be lifted to
${\bf 128}_{S}\oplus {\bf 128}_{C}$.
First we note that according to~(\ref{embedding33}),
the total spectrum can be lifted to an
$({\bf 8}_{V}\oplus {\bf 8}_{S},{\bf 8}_{V}\oplus {\bf 8}_{C})$
of $SO(8)_{L}\times SO(8)_{R}$
and thus further to $SO(16)$
according to
 \bea\label{so16deco}
  && \f{16}\rightarrow (\f{8}_C,\f{1})\oplus (\f{1},\f{8}_S)\;, \nonumber\\
  && \f{128}_S\rightarrow (\f{8}_V,\f{8}_V)\oplus (\f{8}_S,\f{8}_C)\;, \quad
  \f{128}_C\rightarrow (\f{8}_S,\f{8}_V)\oplus (\f{8}_V,\f{8}_C)\;.
 \eea
This corresponds to the
canonical embedding $SO(16)\supset SO(8)_{L}\times SO(8)_{R}$
and an additional triality rotation.
Finally this lifts the spectrum
precisely to the ${\bf 128}_{S}\oplus {\bf 128}_{C}$
field content of the maximal ${\cal N}=16$
theory with scalar target space $G/H=E_{8(8)}/SO(16)$.

\subsection{Gauge group and spectrum}
\label{embed}

In this section, we will identify the full gauge group
of the effective three-dimensional theory and determine
its embedding into the global $E_{8(8)}$ symmetry group
of the ungauged theory.
We have already seen in sec.~\ref{gaugedsugra} that in
order to describe a certain
number of massive vectors fields, which are on-shell
dual to Yang-Mills fields, the gauge algebra has to be enlarged.
Its general form has been determined in \cite{Nicolai:2003bp}
and will be denoted by
 \bea
  G_{0}&=&
  {G}_{\rm c} \ltimes (\hat{T}_p, {T}_\nu)
  \;.
  \label{gengau}
 \eea
In our case, ${G}_{\rm c}$ denotes the compact
gauge group~(\ref{so34}) which from the Kaluza-Klein
origin of the theory is
expected to be realized by propagating vector fields.
In the Chern-Simons formulation given above,
this compact factor needs to be amended by the
nilpotent translation group
${T}_\nu$ whose $\nu={\rm dim}\,{G}_{\rm c}$
generators transform in the adjoint representation
of ${G}_{\rm c}$. This allowed an alternative formulation
of the theory~(\ref{action}) in which
part of the scalar sector is redualized
into propagating vector fields gauging the group ${G}_{\rm c}$,
which accordingly appear with a conventional
Yang-Mills term.
The third factor $\hat{T}_p$ in~(\ref{gengau})
is spanned by $p$ nilpotent translations
transforming in some representation of ${G}_{\rm c}$ and
closing into ${T}_\nu$.
This part of the gauge group is completely broken
in the vacuum and gives rise to $p$ massive vector fields.
Specifically, the algebra underlying~(\ref{gengau}) reads
 \bea\label{semidirect}
&&
[\mathcal{J}^m,\mathcal{J}^n]=f^{mn}_{\hspace{1.2em}k}\,
\mathcal{J}^k\;,
\qquad
  [\mathcal{J}^m,\mathcal{T}^{\underline{n}}]
  =f^{mn}_{\hspace{1.2em}k}\,\mathcal{T}^{\underline{k}}\;,
\qquad
  [\mathcal{T}^{\underline{m}},\mathcal{T}^{\underline{n}}]=0\;,
\nonumber\\[1ex]
&&
  [\mathcal{J}^m,\hat{\mathcal{T}}^{\alpha}]
  =t^{m\alpha}_{\hspace{1.0em}\beta}\,\hat{\mathcal{T}}^{\beta}\;,
  \qquad
  [\hat{\mathcal{T}}^{\alpha},\hat{\mathcal{T}}^{\beta}]
  =t^{\alpha\beta}_{\hspace{1.0em}m}\,\mathcal{T}^{\underline{m}}\;,
  \qquad
  [\mathcal{T}^{\underline{m}},\hat{\mathcal{T}}^{\alpha}]=0\;,
 \eea
with $\mathcal{J}^m$, $\mathcal{T}^{\underline{n}}$, and $\hat{\mathcal{T}}^{\alpha}$
generating ${G}_{\rm c}$, ${T}_\nu$, and $\hat{T}_p$, respectively.
The $f^{mn}_{\hspace{1.2em}k}$ are the structure constants of ${G}_{\rm c}$
while the $t^{m\alpha}_{\hspace{1.0em}\beta}$ denote the
representation matrices for the $\hat{\mathcal{T}}^{\alpha}$.
Indices $m, n, \dots$ are raised/lowered with the Cartan-Killing form of ${G}_{\rm c}$;
raising/lowering of
indices $\alpha$, $\beta$ requires a symmetric ${G}_{\rm c}$
invariant tensor $\kappa^{\alpha\beta}$.

To begin with, we have to reconcile the structure~(\ref{gengau})
with the spectrum collected in table~\ref{spin32con}.
With ${G}_{\rm c}=SO(4)_{L}\times SO(4)_{R}$ from~(\ref{so34}),
${T}_\nu$ transforms in the adjoint representation
$(1,0;0,0)\oplus (0,1;0,0)\oplus(0,0;1,0)\oplus (0,0;0,1)$.
Table~\ref{spin32con} exhibits $34$ additional massive vector
fields, transforming in the
$2\!\cdot\!(\frac12,\frac12;\frac12,\frac12)\oplus2\!\cdot\!(0,0;0,0)$ of
${G}_{\rm c}$.
In total, we thus expect a gauge group
$G_{0}={G}_{\rm c} \ltimes (\hat{T}_{34}, {T}_{12})$
of dimension ${\rm dim}\,G_{0}=12+12+34=58$.
Next, we have to identify this group within $E_{8(8)}$.
To this end, it proves useful to first consider the embedding
of ${G}_{\rm c}$ into $E_{8(8)}$ according to the decompositions
(see appendix \ref{A})
 \begin{eqnarray}
  E_{8(8)} &~\supset~&
  \left\{
  \begin{array}{lcr}
  \supset & SO(16)& \supset  \\[2ex]
  \supset & SO(8,8)& \supset
  \end{array}
   \right\}
   ~\supset~
   SO(8)_{L} \times SO(8)_{R} \\ \nonumber
   &~\supset~& SO(4)_{L} \times SO(4)_{R}
   \label{embbb}
 \end{eqnarray}
with the two embeddings of $SO(8)_{L} \times SO(8)_{R}$ given by
 \bea
SO(16)  &:& \quad  \f{16}\rightarrow (\f{8}_C,\f{1})\oplus (\f{1},\f{8}_S)\;, \quad
 \f{128}_S\rightarrow (\f{8}_V,\f{8}_V)\oplus (\f{8}_S,\f{8}_C)\;,
 \nonumber\\[2ex]
 SO(8,8) &:& \quad  \f{16}\rightarrow (\f{8}_V,\f{1})\oplus (\f{1},\f{8}_V)\;, \quad
 \f{128}_S\rightarrow (\f{8}_C,\f{8}_S)\oplus (\f{8}_S,\f{8}_C)\;.
 \label{16vs88}
  \eea
Accordingly, the group $E_{8(8)}$ decomposes as
\bea
\f{248} ~\rightarrow ~
\Big\{
(\f{28},\f1) \oplus (\f1,\f{28}) \oplus
 (\f{8}_C,\f{8}_S)\Big\}
 ~\oplus~
 \Big\{
  (\f{8}_V,\f{8}_V)\oplus (\f{8}_S,\f{8}_C) \Big\}
  \;,
\eea
and further according to~(\ref{embedding33}).
Here curly brackets indicate the splitting into
its compact and noncompact part and $\f{28}=\f{8}\wedge\f{8}$.
We have already discussed that with this embedding
the noncompact part of $E_{8(8)}$ precisely
reproduces the bosonic
spectrum of table~\ref{spin32con}.

In order to identify the embedding of the full gauge group
$G_{0}={G}_{\rm c} \ltimes (\hat{T}_{34}, {T}_{12})$
we further consider the decomposition
of $E_{8(8)}$ according to
\bea
E_{8(8)} \supset SO(8,8) \supset
SO(6,6)\times SO(2,2)
\supset
SO(6,6)\times SO(1,1)\times SO(1,1)
\;,
\eea
and its grading with respect to these two $SO(1,1)$ factors
which is explicitly given in table~\ref{grading}.
>From this table we can infer that properly identifying
\bea\label{e8grade}
{G}_{\rm c}~\subset~ \f{66}_0^0\;,\quad
{T}_{12} ~=~ \f{12}_{0}^{+1} \;,
\quad
\hat{T}_{34} ~\subset~
\f{32}_{-1/2}^{+1/2} ~\oplus~ \overline{\f{32}}_{+1/2}^{+1/2}
~\oplus~ \f{1}^{+1}_{-1} ~\oplus~ \f{1}^{+1}_{+1}
\;,
\label{GTT}
\eea
precisely reproduces the desired algebra structure~(\ref{semidirect}).
We have thus succeeded in identifying the algebra~$\mathfrak{g}_{0}$
underlying the full gauge group
$G_{0}={G}_{\rm c} \ltimes (\hat{T}_{34}, {T}_{12})$,
which is entirely embedded in the `upper light cone'
of table~\ref{grading}.
In the next section, we will explicitly construct
the embedding tensor $\Theta_{\cM\cN}$
projecting onto this algebra, and show
that it is indeed compatible with the algebraic
constraints~(\ref{conex}), (\ref{quad})
imposed by supersymmetry.

 \begin{table}[bt]
  \begin{center}
   $\begin{array}{cccccccccccc} \rule[-0.4cm]{0cm}{0.4cm}
     \f{1}^{+1}_{-1} &&    \f{12}_{0}^{+1} &&  \f{1}^{+1}_{+1}
     \\ \rule[-0.4cm]{0cm}{0.4cm}
     & \f{32}_{-1/2}^{+1/2} & & \overline{\f{32}}_{+1/2}^{+1/2}
       \\ \rule[-0.4cm]{0cm}{0.4cm}
     \f{12}_{-1}^0 & & \f{66}_0^0 + \f{1}_0^0 + \f{1}_0^0 &&
     \f{12}_{+1}^{0}
   \\ \rule[-0.4cm]{0cm}{0.4cm}
     & \overline{\f{32}}_{-1/2}^{-1/2} & & {\f{32}}_{+1/2}^{-1/2}
   \\ \rule[-0.4cm]{0cm}{0.4cm}
      \f{1}^{-1}_{-1} &&    \f{12}_{0}^{-1} &&  \f{1}^{-1}_{+1}
    \end{array}$
  \end{center}
  \caption{\small Grading of $E_{8(8)}$ according to
$SO(6,6)\times SO(1,1)\times SO(1,1)$.
For later reference, we denote by $SO(1,1)_{a}$ the factor
responsible for the grading from left to right and by
$SO(1,1)_{b}$ the factor
responsible for the grading from top to bottom.}
\label{grading}
 \end{table}

\subsection{The embedding tensor}
\label{constr}

In this section, we will explicitly construct
the embedding tensor $\Theta_{\cM\cN}$
projecting onto the Lie algebra $\mathfrak{g}_{0}$
of the desired gauge group
$G_{0}={G}_{\rm c} \ltimes (\hat{T}_{34}, {T}_{12})$
identified in the previous section.
The embedding tensor then uniquely
defines the effective action~(\ref{action}).
We start from the $SO(4)\times SO(4)$ basis
of $E_{8(8)}$ defined in appendix~\ref{A3}.
In this basis, the grading of table~\ref{grading}
refers to the charges of the
generators $X^{0\hat{0}}$ and $X^{\bar0\hat{\bar0}}$.
We further denote the generators of
$G_{c}$ and ${T}_{12}$ within table~\ref{grading} as
 \begin{equation}
   \f{66}^0_0~\supset~ \f{3}_L^{+}\oplus
   \f{3}_L^{-}\oplus\f{3}_R^{+}\oplus\f{3}_R^{-}\;,
   \qquad \mbox{and}\quad
   \f{12}_{0}^{+1}~=~\f{\hat3}{}_L^{+}\oplus
   \f{\hat3}{}_L^{-}\oplus\f{\hat3}{}_R^{+}\oplus\f{\hat3}{}_R^{-}\;,
 \label{g1}
 \end{equation}
respectively, with the labels $L$, $R$, $\pm$ referring to the four
factors of~(\ref{so34}), i.e.\
$\f{3}_L^{+}=(1,0;0,0)$, $\f{3}_L^{-}=(0,1;0,0)$, etc.
Similarly, we will identify the generators of $\hat T_{34}$ among
 \begin{equation}
   \qquad \f{32}^{+1/2}_{-1/2}\equiv\f{16}_{-}^{(1)} \oplus \f{16}_{-}^{(2)}\;,
   \qquad
   \overline{\f{32}}^{+1/2}_{+1/2} \equiv{\f{16}}_{+}^{(1)}
   \oplus {\f{16}}_{+}^{(2)}\;,
   \qquad
   \f{1}^{+1}_{-1} \equiv \f{1}_{-}\;,\quad
   \f{1}^{+1}_{+1} \equiv \f{1}_{+}\;,
   \label{g2}
 \end{equation}
where $\bf{16}$ denotes the
$(\ft12,\ft12;\ft12,\ft12)$ of $SO(4)\times SO(4)$,
and we use subscripts $(1), (2)$, $\pm$ in order to distinguish
the identical representations.
The split of the $\bf{32}$ representations into two copies of
the $\bf{16}$ is chosen such that the algebra
closes according to
 \bea
   [\f{16}_+^{(1)},\f{16}_-^{(1)}]&\subset& \hat{\f{3}}_L^- \oplus \hat{\f{3}}_R^-\;,
   \qquad
     [\f{16}_+^{(2)},\f{16}_-^{(2)}]~\subset~ \hat{\f{3}}_L^- \oplus \hat{\f{3}}_R^-\;,
    \nonumber\\[1ex]
{}  [\f{16}_+^{(1)},\f{16}_-^{(2)}]&\subset& \hat{\f{3}}_L^+ \oplus \hat{\f{3}}_R^+\;,
   \qquad
  [\f{16}_+^{(2)},\f{16}_-^{(1)}]~\subset~ \hat{\f{3}}_L^+ \oplus \hat{\f{3}}_R^+\;.
 \eea

The embedding tensor $\Theta_{\cM\cN}$ is an object in the
symmetric tensor product of two adjoint representations
of $E_{8(8)}$.
It projects onto the Lie algebra of the gauge group
according to
$\mathfrak{g}_{0}=\langle X_{\cM}\equiv\Theta_{\cM\cN}\,t^{\cN}\rangle$.
We start from the most general ansatz for $\Theta_{\cM\cN}$
that has entries only on the generators (\ref{g1}), (\ref{g2}).
Since the $\Theta_{\cM\cN}$ relevant for our theory moreover
is an $SO(4)\times SO(4)$ invariant tensor,
it can only have non-vanishing entries contracting coinciding
representations, e.g. $\Theta_{\f3_L^+,\f{\hat3}_L^+}$,
$\Theta_{{\f{16}}_{+}^{(1)},\f{16}^{(2)}_-}$, etc.
Using computer algebra (Mathematica), we can
then implement the algebraic constraint~(\ref{conex}).
As one of the main results of this chapter,
we find that this constraint determines the embedding
tensor $\Theta$ with these properties up to five free constants
$\gamma$, $\beta_{1}$, $\beta_{2}$, $\beta_{3}$, $\beta_{4}$,
in terms of which it takes the
form\footnote{Here we have used a somewhat symbolic notation for $\Theta$,
indicating just the multiples of the identity matrix that
$\Theta$ takes in the various blocks.}
 \bea
 \label{theta2}
&&
    \Theta_{\f3_L^+,\f{\hat3}{}_L^+} =\beta_{1}\;, \quad
    \Theta_{\f3_L^-,\f{\hat3}{}_L^-}=\beta_{2}\;, \quad
    \Theta_{\f3_R^+,\f{\hat3}{}_R^+}=\beta_{3}\;, \quad
    \Theta_{\f3_R^-,\f{\hat3}{}_R^-}=\beta_{4}\;, \nonumber\\[2ex]
&&
   \Theta_{\f{1}_+,\f{1}_-}=
    \Theta_{\f{\hat3}{}_R^+,\f{\hat3}{}_R^+}=
    \Theta_{\f{\hat3}{}_R^-,\f{\hat3}{}_R^-}=
    -\Theta_{\f{\hat3}{}_L^+,\f{\hat3}{}_L^+}=
    -\Theta_{\f{\hat3}{}_L^-,\f{\hat3}{}_L^-}=\gamma \;,
    \nonumber\\[2ex]
&&
   \Theta_{{\f{16}}_{+}^{(1)},\f{16}^{(1)}_-}
    ~=~-\frac{1}{32\sqrt{2}}(\beta_2+\beta_4)\;, \qquad
    \Theta_{{\f{16}}_{+}^{(1)},
    \f{16}^{(2)}_-}~=~-\frac{1}{32\sqrt{2}}(\beta_1
    +\beta_3)\;,
\nonumber\\[1ex]
&&
    \Theta_{{\f{16}}^{(2)}_{+},
    \f{16}^{(1)}_-}~=~\frac{1}{32\sqrt{2}}(\beta_1
    -\beta_3)\;, \qquad
    \Theta_{{\f{16}}^{(2)}_{+},\f{16}^{(2)}_-}
    ~=~-\frac{1}{32\sqrt{2}}(\beta_2-\beta_4)\;.
 \eea
A priori, it seems quite surprising that the
constraint~(\ref{conex}) still leaves five free constants in $\Theta$ ---
the ${\bf 27000}$ representation of $E_{8(8)}$ gives
rise to $1552$ different $SO(4)\times SO(4)$ representations
that are separately imposed as constraints on
our general ansatz for $\Theta$.

In order to satisfy the full set of consistency constraints it remains
to impose the quadratic constraint~(\ref{quad}) on the embedding
tensor~$\Theta_{\cM\cN}$. Again using computer algebra, we can
compute the form of this constraint for the embedding tensor~(\ref{theta2})
and find that it reduces to a single condition on the parameters:
\bea\label{quadr}
  \beta_1^2+\beta_2^2 &=& \beta_3^2 + \beta_4^2\;.
\eea
This suggests a parametrisation as
\bea
\beta_{1} = \kappa\sin\alpha_{1}\;,\quad
\beta_{2} = \kappa\cos\alpha_{1}\;,\quad
\beta_{3} = \kappa\sin\alpha_{2}\;,\quad
\beta_{4} = \kappa\cos\alpha_{2}\;.
\label{alpha12}
\eea

Altogether we have shown, that
there is a four parameter family of maximally supersymmetric
theories, described by the embedding tensor~(\ref{theta2}), which satisfies
all the consistency constraints~(\ref{conex}),~(\ref{quad}).

For generic values of the parameters, one verifies that the rank of
the induced gauge group is indeed $58$ as expected.\footnote{Let
us note that the degenerate case $\kappa=0$
induces a theory with
14-dimensional nilpotent abelian gauge group, as can be seen from
(\ref{theta2}). This particular gauge group had already been identified
in~\cite{Fischbacher:2003yw}.}
In particular, (\ref{theta2}), (\ref{quadr}) imply
that on the block of $\f{16}$ representations
one finds
\bea
\Theta_{\cM\cN}\:  t^{\cM}\otimes t^{\cN}\Big|_{\f{16}} &=&
-\frac\kappa{16\sqrt{2}}\,
\Big(
\f{16}_{+}^{(1)}\cos(\ft12(\alpha_{1}\!-\!\alpha_{2}))-
\f{16}_{+}^{(2)}\sin(\ft12(\alpha_{1}\!-\!\alpha_{2}))
\Big)\otimes
\nonumber \\
&&{}
\Big(
 \f{16}_{-}^{(1)}\cos(\ft12(\alpha_{1}\!+\!\alpha_{2}))-
 \f{16}_{-}^{(2)}\sin(\ft12(\alpha_{1}\!+\!\alpha_{2}))
\Big)
\;.
\eea
This implies that out of the 64 generators $\f{16}_{\pm}^{(1)}$, $\f{16}_{\pm}^{(2)}$,
only the 32 combinations
\bea
\f{16}_{+}&\equiv&
\f{16}_{+}^{(1)}\cos(\ft12(\alpha_{1}\!-\!\alpha_{2}))-
\f{16}_{+}^{(2)}\sin(\ft12(\alpha_{1}\!-\!\alpha_{2}))
\;,
\nonumber\\
\f{16}_{-}&\equiv&
\f{16}_{-}^{(1)}\!\cos(\ft12(\alpha_{1}\!+\!\alpha_{2}))-
\!\f{16}_{-}^{(2)}\sin(\ft12(\alpha_{1}\!+\!\alpha_{2}))
\;,
\eea
form part of the gauge group.
These correspond to the $2\!\cdot\!(\ft12,\ft12;\ft12,\ft12)$ generators in $\hat T_{34}$.
The complete gauge algebra spanned by the generators
$X_{\cM}\equiv \Theta_{\cM\cN}\:  t^{\cN}$ is precisely
of the form anticipated in~(\ref{semidirect}).

Let us stress another important property of the embedding
tensor~(\ref{theta2}):
it is a singlet not only under the $SO(4)\times SO(4)$, but also under the
$SO(1,1)_{a}$
generating the grading from left to right in table~\ref{grading}.
In other words, the resulting $\Theta$
contracts only representations for which
these particular charges add up to zero.
As a consequence the gauged supergravity~(\ref{action})
in addition to the local gauge symmetry
$G_{0}={G}_{\rm c} \ltimes (\hat{T}_{34}, {T}_{12})$
is invariant under the action of the global symmetry $SO(1,1)_{a}$.
We will discuss the physical consequences of this extra symmetry
in section~\ref{sec:pot} below.

\subsection{Ground state and isometries}
\label{ground}

In the previous section we have found a four-parameter family
of solutions $\Theta_{\cM\cN}$~(\ref{theta2}) to the algebraic
constraints~(\ref{conex}),~(\ref{quad})
compatible with the gauge
algebra~$G_{0}={G}_{\rm c} \ltimes (\hat{T}_{34}, {T}_{12})$.
We will now show that the four free parameters $\gamma$, $\kappa$,
$\alpha_{1}$, $\alpha_{2}$,
can be adjusted such that the
theory admits an ${\cal N}=(4,4)$ supersymmetric
$AdS$ ground state, leaving only two free parameters that
correspond to the the radii of the two $S^{3}$ spheres.
Furthermore, expanding the action~(\ref{action})
around this ground state precisely reproduces
the spectrum of table~\ref{spin32con}.

In order to show, that the Lagrangian~(\ref{action})
admits an $AdS$ ground state, we first have to check
the condition (\ref{stationary}) equivalent to
the existence of a stationary point of the scalar potential~(\ref{pot}).
For this in turn
we have to compute the tensors $A_1$, $A_2$ and $A_3$
(\ref{A123}) from the $T$-tensor~(\ref{T})
evaluated at the ground state ${\cal V}=I$.
At this point,
the $T$-tensor coincides with the embedding tensor~(\ref{theta2}).
The only technical problem is the translation from
$\Theta$~(\ref{theta2}) in the $SO(8,8)$ basis of appendix~\ref{A2}
into the $SO(16)$ basis of appendix~\ref{A1}, in which the
tensors $A_1$, $A_2$ and $A_3$  are defined.

It follows from (\ref{theta2}) that
$\Theta$ is traceless, $\theta =0$, and moreover that
all components of $\Theta$, which mix bosonic and spinorial parts, like
$\Theta_{ab|\alpha\dot{\beta}}$, vanish.
As a consequence, the tensor
$A_1$ is block-diagonal, and with the index split in appendix \ref{A}
its explicit form turns out to be
 \bea\label{a1}
  A_1^{IJ}=\frac{1}{7}\left(\begin{array}{cc}
  2\Theta_{\dot{\alpha}\gamma|\dot{\beta}\gamma}
  +\overline{\Gamma}^{\hat{a}\hat{b}}_{\dot{\alpha}\dot{\gamma}}
  \overline{\Gamma}^{\hat{c}\hat{d}}_{\dot{\beta}\dot{\gamma}}
  \Theta_{\hat{a}\hat{b}|\hat{c}\hat{d}}  & 0 \\ 0 &
  2\Theta_{\dot{\gamma}\alpha|\dot{\gamma}\beta}
  +\Gamma^{ab}_{\alpha\gamma}\Gamma^{cd}_{\beta\gamma}\Theta_{ab|cd}
  \end{array}\right)\;,
 \eea
with $SO(8)$ $\Gamma$-matrices $\Gamma^{a}_{\alpha\dot{\beta}}$,
see appendix~\ref{A22} for details.
Similarly, $A_2$ and $A_3$ are also block-diagonal and can be written as
 \bea\label{a2}
  A_2^{I\dot{A}}=-\frac{1}{7}\left(\begin{array}{cc}
  2\Gamma^a_{\gamma\dot{\epsilon}}\Theta_{\dot{\alpha}\gamma|
  \beta\dot{\epsilon}}-\Gamma^{\hat{b}}_{\beta\dot{\gamma}}
  \overline{\Gamma}^{\hat{c}\hat{d}}_{\dot{\alpha}\dot{\gamma}}
  \Theta_{\hat{c}\hat{d}|a\hat{b}} & 0 \\
  0 & 2\Gamma^{\hat{a}}_{\delta\dot{\gamma}}\Theta_{\dot{\gamma}\alpha|\delta
  \dot{\beta}}+\Gamma^{cd}_{\alpha\gamma}\Gamma^b_{\gamma\dot{\beta}}
  \Theta_{cd|b\hat{a}} \end{array}\right),
 \eea
and
 \bea\label{a3}
   A_3^{\dot{A}\dot{B}}=\left(\begin{array}{cc} \delta^{\alpha\beta}
   \Theta_{a\hat{c}|b\hat{c}}+2\delta^{ab}\Theta_{\alpha\dot{\gamma}|
   \beta\dot{\gamma}}  & 0 \\
   0 & \delta^{\dot{\alpha}
   \dot{\beta}}\Theta_{c\hat{a}|c\hat{b}}+2\delta^{\hat{a}\hat{b}}
   \Theta_{\gamma\dot{\alpha}|\gamma\dot{\beta}}\end{array}\right).
 \eea
Using these tensors one can now check that the ground state condition
(\ref{stationary}) is fulfilled if the parameters of (\ref{theta2}) satisfy
 \bea\label{groundstate}
  \kappa^{2}&=&16 \gamma^2\;.
 \eea
Moreover, using (\ref{pot}) the value of the scalar potential
at the ground state, i.e.~the cosmological constant, can be computed
and consistently comes out to be negative, $V=-g^{2}/2$,
i.e.~the $AdS$ length is given by $L_{0}=1/g$.

In the following, we will absorb $\kappa$ into the global coupling constant
$g$ and set $\gamma=1/4$ in accordance with~(\ref{groundstate}).
As a result, there remains a two-parameter family of supergravities
admitting an $AdS$ ground state.
Let us now determine the number of unbroken supercharges in the ground state.
It is derived from the Killing spinor equations
 \bea\label{cond1}
  \delta\psi_{\mu}^I = D_{\mu}\epsilon^I+ig A_1^{IJ}\gamma_{\mu}
  \epsilon^J\equiv 0\;,
  \qquad
  \delta\chi^{\dot{A}}=gA_2^{I\dot{A}}\epsilon^I\equiv 0
  \;.
 \eea
As has been shown in~\cite{Nicolai:2000sc},
the number of solutions to~(\ref{cond1}) and thus the
number of preserved supersymmetries is given by the
number of eigenvalues $\alpha_{i}$ of the tensor $A_{1}^{IJ}$
with $|\alpha_{i}|gL_{0}=1/2$.
Computing these eigenvalues from the explicit form of~(\ref{a1}) we find
that the tensor $A_{1}^{IJ}$
may be diagonalized as
\bea
\label{eigenvalues}
A_{1}^{IJ} &=&
{\rm diag}\,
\Big\{
-\ft{3}{2}, -\ft{3}{2}, -\ft{3}{2}, -\ft{3}{2},
-\ft{1}{2},-\ft{1}{2},-\ft{1}{2},-\ft{1}{2},
\ft{1}{2},\ft{1}{2},\ft{1}{2},\ft{1}{2},
\ft{3}{2}, \ft{3}{2}, \ft{3}{2}, \ft{3}{2}
\Big\} \;.
\eea
{}From this, we infer that the $AdS$ ground state of the theory
indeed preserves ${\cal N}=(4,4)$ supersymmetries, as expected.
The other eight gravitinos become
massive through a super-Higgs mechanism~\cite{deWit:2003ja,Hohm:2004rc}.
This implies that due to the broken supersymmetries eight of the
spin-$1/2$ fermions
 \bea
   \eta^I&\equiv& A_2^{I\dot{A}}\chi^{\dot{A}} \;,
   \label{gs}
 \eea
transform by a shift under supersymmetry and
act as Goldstone fermions that get eaten by the gravitino fields which in turn
become massive propagating spin-$3/2$ fields.
With the relation
\bea
 \label{mass}
  |\Delta-1|= |m|\,L_{0} \;,
\eea
between the $AdS$ masses $m$ and conformal dimensions $\Delta$
of fermions and self-dual massive vectors
in three dimensions, (\ref{eigenvalues})
implies that the massive gravitinos correspond to
operators with conformal weights $(\ft12,2)$ and $(2,\ft12)$,
in precise agreement with the spectrum of table~\ref{spin32con}.

To compute the physical masses for the spin-$1/2$ fermions,
we observe from (\ref{action}) that their mass matrix is given by
$gA_3^{\dot{A}\dot{B}}$, except for the eight eigenvalues that
correspond to the Goldstone fermions~(\ref{gs}).
>From the explicit form (\ref{a3})
one computes the spin-1/2
masses and verifies using~(\ref{mass})
that they coincide with those of table~\ref{spin32con}.
Finally, we may check the mass spectrum for the spin-$1$ fields.
Their mass matrix is given
by $\Theta_{AB}$, the projection of the embedding tensor
onto the non-compact part of the algebra~\cite{deWit:2003ja}.
>From (\ref{theta2}) one finds by explicit computation
for these eigenvalues $46$ non-vanishing values in precise accordance with
table~\ref{spin32con}.

Altogether, we have shown the existence of a new family of gauged maximally
supersymmetric theories in $D=3$, which are parametrized by the two
free parameters $\alpha_1$ and $\alpha_2$ and the overall gauge coupling
constant $g$.
These theories admit an ${\cal N}=(4,4)$ supersymmetric
$AdS_{3}$ ground state and linearizing the field equations around this
ground state reproduces the correct spectrum of table~\ref{spin32con}.
In particular, this spectrum does not depend on the particular
values of $\alpha_{1}$ and $\alpha_{2}$.
One may still wonder about the
meaning of these two parameters. From the point of view
of the Kaluza-Klein reduction the only relevant parameter
is the ratio $\alpha$ of the two spheres radii, which enters the
superalgebra~(\ref{supergroup}).
Let us thus compute the background isometry group by expanding
the supersymmetry algebra
\bea
{}\{\,\delta_{\epsilon_{1}},\delta_{\epsilon_{2}} \,\}
&=& (\bar\epsilon_{1}^{I}\epsilon_{2}^{J})
\,{\cal V}^{\cM}{}_{{IJ}}\, \Theta_{{\cM\cN}}\,t^{\cN} + \dots
\;,
\label{bison}
\eea
around the ground state ${\cal V}=I$.
The conserved supercharges $\epsilon^{I}$ are the
eigenvectors of $A_{1}$ from (\ref{eigenvalues})
to the eigenvalues $\pm1/2$ where the different signs
correspond to the split into
left and right supercharges according to~(\ref{supergroup}).
Correspondingly, the algebra~(\ref{bison})
splits into two parts, $L$ and $R$, with anticommutators
\bea
\label{susyclosure}
   \{ G_{-1/2\,L,R}^i,G_{1/2\,L,R}^j \}=
   4\left(\frac{1}{1+\alpha_{L,R}}\,\tau_{kl}^{+ij}\,J^{+kl}_{L,R}
   +\frac{\alpha_{L,R}}{1+\alpha_{L,R}}\,\tau_{kl}^{-ij}\,J^{-kl}_{L,R}\right)+\dots
   \;,
\eea
where $\tau_{kl}^{+ij}\equiv \delta_{kl}^{ij}\pm\ft12\epsilon_{ijkl}$
denote the projectors onto selfdual and anti-selfdual generators of
$SO(4)_{L,R}$ corresponding to the split
$SO(4)=SO(3)^{+}\times SO(3)^{-}\,$.
This coincides with the anticommutators of the superalgebra
$D^{1}(2,1;\alpha_{{L,R}})$~\cite{Sevrin:1988ew}.
Specifically, we find the relation
\bea
   \alpha_{L}= \tan \alpha_1
   \;, \qquad
   \alpha_{R} = \tan \alpha_2\;,
\eea
to the parameters (\ref{alpha12}) of the embedding tensor.
This shows that the three-parameter family of theories constructed
in this section exhibits the background isometry group
\bea
D^1(2,1;\alpha_{L})_{L} \times D^1(2,1;\alpha_{R})_{R}
\;.
\label{supergroup0}
\eea
The theories related to the Kaluza-Klein compactification
on $AdS_{3}\times S^{3}\times S^{3}$ are thus given
by further restricting $\alpha_{L}=\alpha_{R}\equiv\alpha$
where this parameter corresponds to the ratio of radii of the two spheres.

Putting everything together, we have shown that the effective supergravity action
describing the field content of table~\ref{spin32con} is given by the
Lagrangian~(\ref{action}) with the following particular form
of the embedding tensor $\Theta_{\cM\cN}$
 \bea
 \label{thetatotal}
&&
    \Theta_{\f3_L^+,\f{\hat3}{}_L^+} =\Theta_{\f3_R^+,\f{\hat3}{}_R^+}=
    \frac{\alpha}{\sqrt{1\!+\!\alpha^{2}}}\;, \qquad
    \Theta_{\f3_L^-,\f{\hat3}{}_L^-}=
    \Theta_{\f3_R^-,\f{\hat3}{}_R^-}=\frac{1}{\sqrt{1\!+\!\alpha^{2}}}\;, \nonumber\\[1ex]
&&
   \Theta_{\f{1}_+,\f{1}_-}~=~
    \Theta_{\f{\hat3}{}_R^+,\f{\hat3}{}_R^+}~=~
    \Theta_{\f{\hat3}{}_R^-,\f{\hat3}{}_R^-}~=~
    -\Theta_{\f{\hat3}{}_L^+,\f{\hat3}{}_L^+}~=~
    -\Theta_{\f{\hat3}{}_L^-,\f{\hat3}{}_L^-}~=~\frac14 \;,  \nonumber\\[1ex]
&&
   \Theta_{{\f{16}}_{+}^{(1)},\f{16}^{(1)}_-}
    =-\frac{1}{16\sqrt{2}}\,\frac{1}{\sqrt{1\!+\!\alpha^{2}}}\;, \qquad
    \Theta_{{\f{16}}_{+}^{(1)},
    \f{16}^{(2)}_-}=-\frac{1}{16\sqrt{2}}\,\frac{\alpha}{\sqrt{1\!+\!\alpha^{2}}}\;.
 \eea
We have verified that this tensor indeed represents a solution of
the algebraic consistency constraints~(\ref{conex}),~(\ref{quad}).
The resulting theory admits an ${\cal N}=(4,4)$ supersymmetric
$AdS_{3}$ ground state with background isometry group~(\ref{supergroup})
at which half of the 16 supersymmetries are spontaneously broken
and the spectrum of table~\ref{spin32con} is reproduced via a supersymmetric
version of the Higgs effect.

\subsection{The scalar potential for the gauge group singlets}
\label{sec:pot}

We have identified the gauged supergravity theory,
whose broken phase describes the coupling of
the massive spin-$3/2$ multiplet $(\ft12,\ft12;\ft12,\ft12)_{\rm S}$
to the supergravity multiplet.
In particular, the scalar potential~(\ref{pot}) of the
effective three-dimensional theory is completely
determined in terms of the embedding tensor~(\ref{thetatotal}).
In the holographic context this scalar potential
carries essential information about the boundary
conformal field theory, in particular
about higher point correlation functions
and about deformations and renormalization group flows.
Explicit computation of the full potential
is a highly nontrivial task, as it is a function
on the $128$-dimensional target space
$E_{8(8)}/SO(16)$. For concrete applications it is
often sufficient to evaluate this potential on
particular subsectors of the scalar manifold.

As an example, let us in this section
evaluate the potential on the gauge group singlets. From table~\ref{spin32con}
we read off that there are two scalar fields that are singlets
under the $SO(4)_{L}\times SO(4)_{R}$ gauge group.
Let us denote them by $\phi_{1}$ and $\phi_{2}$.
They are dual to a marginal and an irrelevant operator of
conformal dimension $(1,1)$ and $(2,2)$, respectively.
In particular,
the scalar $\phi_{1}$ corresponds to a modulus of the theory.
In order to determine the explicit dependence of the scalar
potential on these fields, we parametrize the scalar $E_{8(8)}$
matrix ${\cal V}$ as
\bea
  \mathcal{V}=\exp\big( \phi_1 X^{0\hat{0}}+\phi_2 X^{\bar0\hat{\bar0}} \big)
  \;,
\label{VV}
\eea
where $X^{0\hat{0}}$ and $X^{\bar0\hat{\bar0}}$ are the
generators of the $SO(1,1)_{a}$ and $SO(1,1)_{b}$
of table~\ref{grading}, respectively.
The potential is obtained by computing with this parametrisation
the $T$-tensor from~(\ref{T}), (\ref{thetatotal}),
splitting it into the tensors $A_1$ and $A_2$ according to~(\ref{A123})
and inserting the result into~(\ref{pot}).

The computation is simplified by first transforming the two singlets
into a basis where their adjoint action is diagonal, such that
their exponentials can be easily computed and afterwards
transforming back to the $SO(16)$ basis of appendix~\ref{A1}.
It becomes now crucial that the embedding tensor $\Theta$ is
invariant under $SO(1,1)_{a}$ and thus under the
adjoint action of $X^{0\hat{0}}$ as we found in
section~\ref{constr} above. This implies, that the $T$-tensor~(\ref{T})
is in fact completely independent of $\phi_{1}$. In turn, neither the fermionic mass
terms nor the scalar potential carries an explicit dependence on $\phi_{1}$.
This scalar thus enters the theory only through its kinetic term and
the dual operator is truly marginal.

\begin{figure}
 \begin{center}
   \mbox{\includegraphics[width=0.7\textwidth]{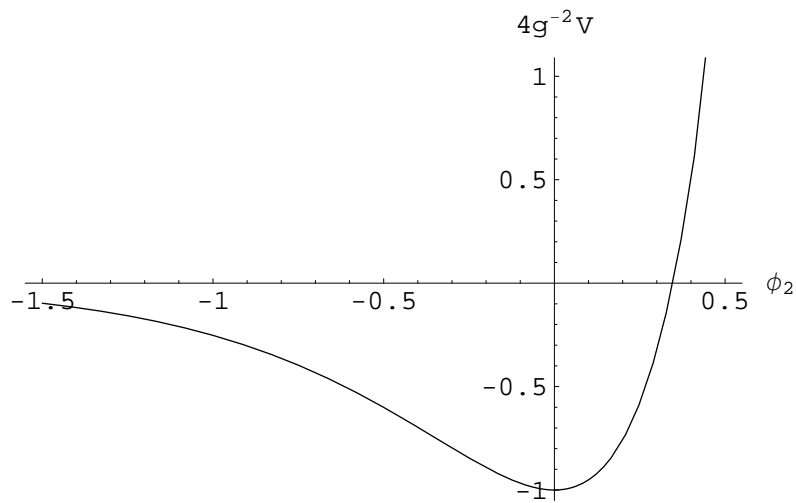}}
 \end{center}
\caption{The scalar potential for the gauge group singlets}\label{graph}
\end{figure}

The scalar potential (\ref{pot}) evaluated on the gauge group singlets
is finally given as a function of $\phi_{2}$ as
 \bea\label{potential}
  V(\phi_1,\phi_2)=\frac{g^2}{4}
  e^{2\phi_2}\big(-2 + e^{2\phi_2}\big)\;.
 \eea
The profile is plotted in Figure~\ref{graph}.
Its explicit form shows that the theory has no other ground state
which preserves the full $SO(4)\times SO(4)$ symmetry.


\chapter{Massive spin-2 fields and their infinite-dimensional symmetries}
\label{spin2chapter}

\section{Is there a spin-2 Higgs effect?}
We have seen in the last chapter that the massive spin-3/2 states
appearing in Kaluza-Klein supergravities have to be described within the
framework of spontaneously broken supersymmetry. Specifically we have
seen that for Kaluza-Klein supergravity on $AdS^3\times S^3\times S^3$
the inclusion of the lowest spin-3/2 multiplet requires already
an enhancement of supersymmetry
from ${\cal N}=8$ to ${\cal N}=16$, which in turn gets spontaneously
broken in the Kaluza-Klein vacuum.

Let us now turn to the similar problem of finding the
effective action for spin-2 fields.
As we have discussed in sec.~\ref{higherspin} their naive
coupling to gravity is inconsistent, both in the massless and in the
massive case. In comparison, the analogous inconsistencies
appearing for spin-3/2 fields were avoided by the introduction of
supersymmetry, which
linked the spin-3/2 fields with the metric in such a way that
the entire theory becomes consistent. In consideration of the fact that
the spin-2 couplings appearing in Kaluza-Klein theories have
to be consistent, one might expect to identify a similar phenomenon
for spin-2 fields. Indeed we will show the appearance
of an infinite-dimensional spin-2 symmetry, that guarantees
consistent couplings in the same sense as supersymmetry does
for spin-3/2 fields.
Even more, we will see that it is possible to parallel the main
steps for the construction of gauged supergravities, where
the spin-3/2 fields get somehow replaced by the
spin-2 fields.
For this we will again focus on compactifications to $D=3$, since
here -- as we have seen -- the gauged supergravities have a more
coherent form due to the topological nature of all `higher-spin'
fields (starting with $s=1$). Thus we expect similar simplifications
for spin-2 theories.

In order to accomplish such a program we have to show the existence
of an unbroken or ungauged phase, where the spin-2 fields appear to be
massless. Furthermore, we expect the bosonic degrees of freedom to be
carried entirely by scalars, i.e. we have to show that even in the
presence of an infinite number of spin-2 fields all appearing
vector fields (including the Kaluza-Klein vectors)
can be dualized into scalars.
After dualization we expect this scalar sector to exhibit an enhanced rigid
symmetry, in analogy to supergravities, in which, e.g.,
the ${\cal N}=16$ theory carries the exceptional symmetry group $E_{8(8)}$.
These enlarged global symmetries will presumably restrict
the possible couplings severely.
In addition, there will be a local spin-2 symmetry for
infinitely many spin-2 fields in much the same way as one has
local supersymmetry in an ungauged supergravity theory.
The broken phase will then be constructed by gauging a certain
subgroup of the global symmetries, or in other words by
switching on a gauge coupling constant. This gauge coupling will later
turn out to be given by the mass scale $M$ characterizing the
inverse radius of the internal manifold (such that the unbroken phase
corresponds to the decompactification limit).
The gauging in turn will modify the spin-2 symmetries by
$M$-dependent terms and will induce a spin-2 mass term.
The latter enables a novel Higgs effect for the spin-2 fields,
which takes place in the same way as the super-Higgs effect
in supergravity.

In practice, we have to address the following questions,
which also determine the organization of this chapter:
 \begin{itemize}
  \item[(i)]
   How can we identify the unbroken phase, and
   how is the spin-2 symmetry realized in this limit?
   In particular, how does this theory fit into the no-go results discussed
   in the literature before?
  \item[(ii)]
   Which global symmetry is realized on the scalar fields in this phase?
  \item[(iii)]
   Which subgroup of the global symmetries has to be
   gauged in order to get the full Kaluza-Klein theory?
   Does a formulation exist also for the gauged phase, where all
   vector and spin-2 fields appear to be topological?
  \item[(iv)]
   How does the spin-2 symmetry get modified due to the gauging?
 \end{itemize}
For answering question (i) we have to identify the ungauged theory
with its symmetries. The symmetries appearing in Kaluza-Klein
theories have in part been analyzed by Dolan and Duff in
\cite{Dolan:1983aa}, where they showed that in the simplest
case of an $S^1$ compactification including all massive modes
a local Virasoro algebra $\hat{v}$
corresponding to the diffeomorphisms on $S^1$
as well as the affine extension $\widehat{iso(1,2)}$
of the Poincar\'e algebra appear. This Kac-Moody algebra
describes the infinite-dimensional spin-2 symmetry, which
in the Kaluza-Klein vacuum will be broken to the lower-dimensional
diffeomorphism group.
Answering question (ii) we will see that the scalar fields span
a generalization of a non-linear $\sigma$-model
(which we are going to make precise later)
with target space $\widehat{SL(2,\mathbb{R})}/\widehat{SO(2)}$.
Thus the global symmetry group contains an enhancement of the
Ehlers group $SL(2,\mathbb{R})$ to its affine extension,
but moreover it will also contain the
Virasoro algebra. By gauging a certain subgroup of the global
symmetries, i.e. after answering question (iii), we will see
that a formulation is still possible in which all degrees of
freedom are carried by scalars. The topological Kaluza-Klein
vectors will in turn combine with the spin-2 fields into a
Chern-Simons theory for an extended algebra. This algebra
structure will also enlighten the deformations of the spin-2
transformations due to the gauging, thus answering in part question (iv).
The strategy underlying this approach can be illustrated
schematically as given below.

\newenvironment{Boxedminipage}%
{\begin{Sbox}\begin{minipage}}%
{\end{minipage}\end{Sbox}\fbox{\TheSbox}}

\begin{equation*}
\begin{CD}
\begin{Boxedminipage}{4.5cm}
 \small{\underline{Unbroken phase}:  \\
  CS theory for $\widehat{iso(1,2)}$ + \\
  $\sigma$-model
  $\widehat{SL(2,\mathbb{R})}/\widehat{SO(2)}$},\\
  spin-2 symmetry
\end{Boxedminipage}
@>\text{gauging } >>
\begin{Boxedminipage}{4.5cm}
 \small{\underline{Broken phase}:  \vspace{0.3em} \\
 CS theory for extended \\algebra
 + gauged $\sigma$-model, \\deformed spin-2 symmetry}
\end{Boxedminipage} \\
@AA{\text{abelian duality}}A
@VV{\text{non-abelian duality}}V\\
\begin{Boxedminipage}{4.7cm}
 \small{\underline{Unbroken phase}:  \vspace{0.3em} \\
  Scalars $\phi^n$, $U(1)$ vectors $A_{\mu}^n$, \\
  reduced global symmetry, \\
  spin-2 symmetry}
\end{Boxedminipage}
@<\text{   }M\rightarrow 0\text{   }<<
\begin{Boxedminipage}{4.5cm}
 \small{\underline{Broken phase}:  \vspace{0.1em}  \\
  4D gravity on $\mathbb{R}^3 \times S^1$, \\
  infinite tower of spin-2 \\fields with mass scale $M$}
\end{Boxedminipage}
\end{CD}
\end{equation*}

In order to get the full Kaluza-Klein theory (given in
the lower right corner) we show how to construct the ungauged theory
with its enlarged symmetry group (given in the upper left corner).
Then we argue that via
gauging part of the global symmetries we obtain a theory
(given in the upper right corner) which is on-shell equivalent to
the original theory.\footnote{This
chapter is based on \cite{Hohm:2005sc}.}

\section[Kac-Moody symmetries in Kaluza-Klein theories]
{Kac-Moody symmetries in Kaluza-Klein\\ theories}
It has been shown by Dolan and Duff \cite{Dolan:1983aa}
that Kaluza-Klein compactification can be analyzed from the
following point of view. The infinite tower of massive modes in the
lower-dimensional Kaluza-Klein spectrum can be viewed as resulting
from a spontaneous symmetry breaking of an infinite-dimensional
Kac-Moody-like algebra down to the Poincar\'e group times the
isometry group of the internal manifold. This infinite dimensional
symmetry group is a remnant of the higher dimensional diffeomorphism group.

To be more specific let us review Dolan and Duff's analysis
applied to the case of a Kaluza-Klein reduction on
$\mathbb{R}^3\times S^1$. We start from pure Einstein gravity in
$D=4$ and split the vielbein $E_M^A$ in $D=4$ as
follows\footnote{$M,N,...=0,1,2,5$ denote $D=4$ space-time indices,
$A,B,...$ are flat $D=4$ indices and
the coordinates are called $x^M=(x^{\mu},\theta /M)$,
where $M$ is a mass scale characterizing the inverse radius of
the compact dimension. Our metric convention is $(+,-,-)$ for
$D=3$ and similar for $D=4$.}:
 \bea\label{metric}
  E_M^A=\left(\begin{array}{cc} \phi^{-1/2}e_{\mu}^a &
  \phi^{1/2} A_{\mu} \\ 0 & \phi^{1/2} \end{array}\right)\;.
 \eea
Here we have chosen a triangular gauge and also performed a Weyl rescaling.
The fields are now expanded in spherical harmonics of the
compact manifold, which for $S^1$ simply reads
 \begin{equation}\label{fieldexp}
  \begin{split}
   e_{\mu}^a(x,\theta)&=\sum_{n=-\infty}^{\infty}
   e_{\mu}^{a(n)}(x)e^{in\theta}\;, \qquad
   A_{\mu}(x,\theta)=\sum_{n=-\infty}^{\infty}A_{\mu}^{n}(x)
   e^{in\theta}\;, \\
   \phi(x,\theta)&=\sum_{n=-\infty}^{\infty}\phi^{n}(x)e^{in\theta},
  \end{split}
 \end{equation}
where we have to impose the reality constraint
$(\phi^*)^n=\phi^{-n}$ and similarly for the other fields.
Truncating to the zero-modes, the effective Lagrangian is given by
 \begin{equation}\label{zero-mode}
  \mathcal{L}=-eR^{(3)}+\frac{1}{2}
  eg^{\mu\nu}\phi^{-2}\partial_{\mu}
  \phi\partial_{\nu}\phi-\frac{1}{4}e\phi^2g^{\mu\rho}g^{\nu\sigma}
  F_{\mu\nu}F_{\rho\sigma}\;,
 \end{equation}
where as usual $F_{\mu\nu}$ denotes the $U(1)$ field strength for
$A_{\mu}$. This action is invariant under three-dimensional
diffeomorphisms and $U(1)$ gauge transformations.

Let us next analyze how the four-dimensional symmetries are
present in the Kaluza-Klein theory without any truncation.
For this we notice that the
diffeomorphisms in $D=4$, which are locally generated by a vector
field $\xi ^M$, are restricted by the topology of the assumed ground
state $\mathbb{R}^3\times S^1$ to be periodic in $\theta$.
Therefore we have to expand similarly
 \begin{equation}\label{transexp}
   \xi^{\mu}(x,\theta)=\sum_{n=-\infty}^{\infty}\xi^{\mu (n)}(x)
   e^{in\theta}\;, \qquad
   \xi^5(x,\theta)=\sum_{n=-\infty}^{\infty}\xi^{5 (n)}(x)
   e^{in\theta}\;.
 \end{equation}
The four-dimensional diffeomorphisms and local Lorentz
transformations act on the vielbein as
 \bea
  \delta_{\xi}E_M^A=\xi^N\partial_NE_M^A + \partial_M\xi^N E_N^A \;, \qquad
  \delta_{\tau}E_M^A = \tau^A_{\hspace{0.4em}B}E_M^B \;.
 \eea
By applying a four-dimensional diffeomorphism to (\ref{metric}) we get
 \begin{equation}\label{diff}
  \begin{split}
   \delta_{\xi}\phi &= \xi^{\rho}\partial_{\rho}\phi+\xi^5\partial_5\phi
   +2\phi\partial_5\xi^{\rho}A_{\rho}+2\phi\partial_5\xi^5\;, \\
   \delta_{\xi}A_{\mu}&=\xi^{\rho}\partial_{\rho}A_{\mu}
   +\xi^5\partial_5A_{\mu} + \partial_{\mu}\xi^{\rho}A_{\rho}+\partial_{\mu}
   \xi^5-A_{\mu}\partial_5\xi^{\rho}A_{\rho}-A_{\mu}\partial_5\xi^5\;, \\
   \delta_{\xi}e_{\mu}^a&=\xi^{\rho}\partial_{\rho}e_{\mu}^a+\xi^5\partial_5
   e_{\mu}^a+\partial_{\mu}\xi^{\rho}e_{\rho}^a+\partial_5\xi^5 e_{\mu}^a
   +\partial_5 \xi^{\rho}A_{\rho}e_{\mu}^a\;.
  \end{split}
 \end{equation}
Moreover we have to add a compensating Lorentz transformation with parameter
$\tau^a_{\hspace{0.3em}5}=-\phi^{-1}\partial_5\xi^{\rho}e_{\rho}^a$
to restore the triangular gauge,
 \bea\label{locomp}
  \delta_{\tau}\phi=0\;, \quad \delta_{\tau}e_{\mu}^a=-A_{\mu}\partial_5
  \xi^{\rho}e_{\rho}^a\;, \quad \delta_{\tau}A_{\mu}=-\phi^{-2}\partial_5
  \xi^{\rho}g_{\rho\mu}\;,
 \eea
where as usual we have written $g_{\mu\nu}=e_{\mu}^ae_{\nu a}$, but now
with $\theta$-dependent vielbein.
Using the mode expansion for the fields in (\ref{fieldexp}) and the
transformation parameter in (\ref{transexp}), one gets an infinite-dimensional
Kaluza-Klein symmetry acting on the fields as\footnote{We will adopt
the Einstein convention also for double indices
$m,n=-\infty,...,\infty$, but indicate summations explicitly, if
the considered indices appear more than twice.}
 \begin{equation}\label{variations}
  \begin{split}
   \delta\phi^n&=\xi_k^{\rho}\partial_{\rho}\phi^{n-k}+iM\sum_k (n+k)\xi_k^5
   \phi^{n-k}+2iM\sum_{k,l}k\xi_k^{\rho}\phi^{n-k-l}A_{\rho}^l\;, \\
   \delta A_{\mu}^n&=
   \partial_{\mu}\xi_n^5+iM\sum_k (n-2k)\xi_k^5A_{\mu}^{n-k}
   +\xi_k^{\rho}\partial_{\rho}A_{\mu}^{n-k}+\partial_{\mu}\xi_k^{\rho}
   A_{\rho}^{n-k}\\ &-iM\sum_{k,l}k\xi_k^{\rho}(\phi^{-2})^{n-k-l}
   g_{\rho\mu}^l-iM\sum_{k,l} k\xi_k^{\rho}A_{\mu}^{n-k-l}A_{\rho}^l \;, \\
   \delta e_{\mu}^{a(n)}&=\xi_k^{\rho}\partial_{\rho}e_{\mu}^{a(n-k)}
   +\partial_{\mu}\xi_k^{\rho}e_{\rho}^{a(n-k)}\\
   &+iMn\xi_k^5e_{\mu}^{a(n-k)}
   +iM\sum_{k,l}k\xi_k^{\rho}\left(e_{\mu}^{a(n-k-l)}A_{\rho}^l
   -e_{\rho}^{a(n-k-l)}A_{\mu}^l\right)\;.
  \end{split}
 \end{equation}
Here $(\phi^{-2})^{n}$ is implicitly
defined by $\phi^{-2}=\sum_{n=-\infty}^{\infty}
(\phi^{-2})^{n}e^{in\theta}$.

We see now that the standard Kaluza-Klein vacuum given by the
vacuum expectation values
 \bea\label{kkvacuum}
  \vev{g_{\mu\nu}}=\eta_{\mu\nu}\;, \qquad \vev{A_{\mu}}=0\;, \qquad
  \vev{\phi}=1\;,
 \eea
is only invariant under rigid $k=0$ transformations, or in other words
the infinite-dimensional symmetry is spontaneously broken to the symmetry
of the zero-modes, i.e. to three-dimensional diffeomorphisms and
a $U(1)$ gauge symmetry.

To explore the group structure which is realized on the whole tower
of Kaluza-Klein modes including its spontaneously broken part,
Dolan and Duff proceeded as follows.
Expanding the generators of the $D=3$ Poincar\'e algebra as well as
those for the diffeomorphisms on $S^1$ into Fourier modes, one gets
 \begin{equation}
  P_a^n=e^{in\theta}\partial_a\;, \quad
  J_{ab}^n=e^{in\theta}(x_b\partial_a-x_a\partial_b)\;, \quad
  Q^n=-Me^{in\theta}\partial_{\theta}\;,
 \end{equation}
where we have used flat Minkowski indices.
This implies after introducing $J^a=\ft12\varepsilon^{abc}J_{bc}$
the following symmetry algebra
 \begin{equation}\label{kacmoody}
  \begin{split}
   [P_a^m,P_b^n]&=0\;, \qquad [J_a^m,J_b^n]=\varepsilon_{abc}J^{c(m+n)}\;,
   \qquad [J_a^m,P_b^n]=\varepsilon_{abc}P^{c(m+n)}\;, \\
   [Q^m,Q^n]&=iM(m-n)Q^{m+n}\;, \\
   [Q^m,P_a^n]&=-iMnP_a^{m+n}\;, \qquad [Q^m,J_a^n]=-iMnJ_a^{m+n}\;,
  \end{split}
 \end{equation}
i.e.~one gets the Kac-Moody algebra associated to the Poincar\'e group
as well as the Virasoro algebra,
both without a central extension. (For definitions see appendix~\ref{km}.)
More precisely, we have a semi-direct product of the Virasoro algebra
$\hat{v}$ with the affine Poincar\'e algebra $\widehat{iso(1,2)}$
in the standard fashion known from the Sugawara
construction \cite{Goddard:1986bp}.
This algebra should be realized as a local symmetry.
However, this latter statement is a little bit contrived
since even the diffeomorphisms (i.e. the $k=0$ transformations)
are known not to be realized in general as gauge
transformations. In fact, even though the idea that general relativity
should have some interpretations as a gauge theory has always been
around \cite{Utiyama:1956sy,Kibble:1961ba}, it is known that
the diffeomorphism in general do not allow an interpretation as
Yang-Mills transformations for a certain Lie algebra, e.g. as
gauge transformations for the $P_a$.
One aim of the present chapter is to clarify this question in the
case of a Kaluza-Klein reduction to $D=3$, and we will see how a modification
of (\ref{kacmoody}) appears as a proper gauge symmetry.

In summary, from this infinite-dimensional symmetry algebra only
$iso(1,2)\times u(1)$ remains unbroken in the Kaluza-Klein vacuum.
This in turn implies that the fields $A_{\mu}^n$ and $\phi^n$ for
$n\neq 0$, which correspond to the spontaneously broken generators
$\xi_n^{\mu}$ and $\xi_n^5$, can be identified
with the Goldstone bosons. They get eaten by the spin-2
fields $e_{\mu}^{a(n)}$, such that the latter become massive.
A massless spin-2 field carries no local degrees of freedom in $D=3$,
while a massless vector as well as a real scalar
each carry one degree of freedom in $D=3$, such that in total
the massive spin-2 fields each carry two degrees of freedom,
as expected.

\section{Unbroken phase of the Kaluza-Klein theory}

\subsection{Infinite-dimensional spin-2 theory}\label{infspin-2}
To construct the
theory containing infinitely many massless spin-2 fields coupled to gravity,
we remember that according to (\ref{kacmoody}) it should have
an interpretation as a gauge theory of the
Kac-Moody algebra $\widehat{iso(1,2)}$.
However, as we already indicated in the discussion at the end of the
previous section,
in general dimensions
this is not a helpful statement, because not even pure gravity has an honest
interpretation as a Yang-Mills-like
gauge theory.
Fortunately, as we reviewed in \ref{CSgravity},
gravity in $D=3$ can in contrast be viewed
as a gauge theory. In case of Poincar\'e gravity the required
gauge theory is a Chern-Simons theory with the Poincar\'e group
$ISO(1,2)$ as (non-compact) gauge group \cite{Witten:1988hc}.
Furthermore, the
symmetries of general relativity, i.e. the diffeomorphisms, are
on-shell realized as the non-abelian gauge transformations.
We are going to show that correspondingly
the Chern-Simons theory of the affine $\widehat{iso(1,2)}$
describes a consistent coupling of
infinitely many spin-2 fields to gravity.

To start with, we recall the Chern-Simons theory for a gauge connection
${\cal A}$, which is given by
 \begin{equation}\label{CS}
  S_{\rm CS}=\int \text{Tr}\big({\cal A}\wedge d{\cal A}+\frac{2}{3}
  {\cal A}\wedge {\cal A}\wedge {\cal A}\big)\;.
 \end{equation}
 The invariance of the quadratic form
$\langle\hspace{0.1em},\rangle$ then implies
that under an arbitrary variation one has
 \begin{equation}\label{var}
  \delta S_{\rm CS}=
  \int \langle \delta {\cal A}_{\mu},{\cal F}_{\nu\rho}\rangle
  dx^{\mu}\wedge dx^{\nu}\wedge dx^{\rho} \;,
 \end{equation}
where ${\cal F}_{\mu\nu}=\partial_{\mu}{\cal A}_{\nu}
-\partial_{\nu}{\cal A}_{\mu}+[{\cal A}_{\mu},{\cal A}_{\nu}]$
denotes the field strength. In particular,
under a gauge transformation $\delta {\cal A}_{\mu}=D_{\mu}u$,
where $D_{\mu}$ denotes the gauge covariant derivative
 \begin{equation}
  D_{\mu}u=\partial_{\mu}u+[{\cal A}_{\mu},u]
 \end{equation}
of an infinitesimal transformation parameter $u$,
the action is invariant due to the Bianchi identity. In addition,
the non-degeneracy of the quadratic form implies the
equations of motion ${\cal F}_{\mu\nu}=0$.

Therefore, to construct the Chern-Simons theory
for $\widehat{iso(1,2)}$, we have to find such a quadratic form.
In contrast to the AdS algebra (\ref{so22}), which is the
direct product of two semi-simple Lie algebras, the
existence of a non-degenerate quadratic form is not self-evident.
However, it turns out that
 \bea\label{quadform}
  \langle P_a^m,J_b^n \rangle = \eta_{ab}\delta^{m,-n}\;, \qquad
  \langle P_a^m,P_b^n \rangle = \langle J_a^m,J_b^n \rangle =0
 \eea
defines an invariant form since the bilinear expression
 \bea
  W:=\sum_{n=-\infty}^{\infty}P^{a(n)}J_a^{(-n)}
 \eea
commutes with all gauge group generators. For instance,
 \bea
  [W,P_b^{k}]=\varepsilon_{abc}\sum_{n=-\infty}^{\infty}
  P^{a(n)}P^{c(k-n)}=0
 \eea
can be seen by performing an index shift
$n\rightarrow n^{\prime}=k-n$, which shows that the sum is symmetric
in $a$ and $c$.\footnote{Upon truncating the quadratic form to the
zero-modes, this reduces to an invariant form of the Poincar\'e algebra,
which was the one used in \cite{Witten:1988hc}
to construct the Chern-Simons action
describing pure Poincar\'e gravity in $D=3$.}

We turn now to the calculation of the action, the equations of motion and
the explicit form of the gauge transformations, which is necessary
to identify the Kaluza-Klein symmetries and fields.
The gauge field takes values in the Kac-Moody algebra,
i.e. it can be written as
 \bea
  {\cal A}_{\mu}=e_{\mu}^{a(n)}P_a^{n}+\omega_{\mu}^{a(n)}J_a^{n}\;.
 \eea
Note, that in the description of ordinary Einstein gravity as
Chern-Simons theory the gauge field $\omega_{\mu}^a$ is
interpreted as the spin-connection,
which like in the Palatini formulation is determined only by the equations
of motion to be the Levi-Civita connection. Here, instead, we have
an infinite number of `connections' and their meaning will be
interpreted later.

With the invariant quadratic form defined in (\ref{quadform}),
the action reads
 \bea\label{spin2CS}
  S_{\rm CS}=\int d^3 x \:\varepsilon^{\mu\nu\rho}e_{\mu a}^{(n)}
  \left(\partial_{\nu}\omega_{\rho}^{a(-n)}-\partial_{\rho}
  \omega_{\nu}^{a(-n)}+\varepsilon^{abc}\omega_{\nu b}^{(m)}
  \omega_{\rho c}^{(-n-m)}\right)\;.
 \eea
If we define `generalized' curvatures
 \bea
  R^{a(n)}=d\omega^{a(n)}+\varepsilon^{abc}\omega^{(m)}_b\wedge
  \omega^{(n-m)}_c\;,
 \eea
the action may be written in a more compact form as
 \bea\label{spin2CS2}
  S_{\rm CS}=\int e^{(n)}_a\wedge R^{a(-n)}\;.
 \eea

The field equations implying vanishing field strength,
${\cal F}_{\mu\nu}=0$,
read in the given case
 \begin{equation}\label{eom}
  \begin{split}
   \partial_{\mu}e_{\nu}^{a(n)}-\partial_{\nu}e_{\mu}^{a(n)}
   +\varepsilon^{abc}e_{\mu b}^{(n-m)}\omega_{\nu c}^{(m)}
   +\varepsilon^{abc}\omega_{\mu b}^{(n-m)}e_{\nu c}^{(m)}&=0\;, \\
   \partial_{\mu}\omega_{\nu}^{a(n)}-\partial_{\nu}\omega_{\mu}^{a(n)}
   +\varepsilon^{abc}\omega_{\mu b}^{(n-m)}\omega_{\nu c}^{(m)}&=0\;.
  \end{split}
 \end{equation}
Due to the mixing of the infinitely many `spin connections', the torsion
defined by $e_{\mu}^{a(0)}$ does
no longer vanish by the equations of motion. This in turn implies
that it is not transparent which part of the Einstein equation
expresses the curvature and which part the energy-momentum tensor
for the higher spin-2 fields. We will clarify this point later.

Next we evaluate the explicit form of the gauge transformations.
Introducing the algebra-valued transformation parameter
$u=\rho^{a(n)}P_a^{n}+\tau^{a(n)}J_a^{n}$, for the
transformations given by $\delta {\cal A}_{\mu}=D_{\mu}u$ one finds
 \begin{equation}\label{kmgauge}
  \begin{split}
   \delta e_{\mu}^{a(n)}&=\partial_{\mu}\rho^{a(n)}+\varepsilon^{abc}
   e_{\mu b}^{(n-m)}\tau_c^{(m)}+\varepsilon^{abc}\omega_{\mu b}^{(n-m)}
   \rho_c^{(m)}\;, \\
   \delta \omega_{\mu}^{a(n)}&=\partial_{\mu}\tau^{a(n)}+\varepsilon^{abc}
   \omega_{\mu b}^{(n-m)}\tau_c^{(m)}\;.
  \end{split}
 \end{equation}
To see that these gauge transformations indeed include the spin-2
Kaluza-Klein transformations (\ref{variations}) for $M=0$, let us define for
a given Kaluza-Klein transformation parameterized by $\xi_k^{\mu}$
the gauge parameters
 \bea\label{kkparam}
  \rho^{a(n)}=\xi_k^{\mu}e_{\mu}^{a(n-k)}\;, \qquad
  \tau^{a(n)}=\xi_k^{\mu}\omega_{\mu}^{a(n-k)}\;.
 \eea
Then the gauge transformation (\ref{kmgauge}) takes the form
 \begin{equation}
  \begin{split}
   \delta e_{\mu}^{a(n)}&=\partial_{\mu}\xi_k^{\rho}e_{\rho}^{a(n-k)}
   +\xi_k^{\rho}\partial_{\rho}e_{\mu}^{a(n-k)} \\
   &+\xi_k^{\rho}\left(\partial_{\mu}e_{\rho}^{a(n-k)}
   -\partial_{\rho}e_{\mu}^{a(n-k)}+\varepsilon^{abc}e_{\mu b}^{(n-k-m)}
   \omega_{\rho c}^{(m)}+\varepsilon^{abc}\omega_{\mu b}^{(n-k-m)}
   e_{\rho c}^{(m)}\right)\;,
  \end{split}
 \end{equation}
where we have again performed an index shift.
We see that the first term reproduces the correct Kaluza-Klein transformation in
(\ref{variations}) with $M=0$,
while the last term vanishes by the equations of
motion (\ref{eom}). On-shell the Kaluza-Klein transformations are therefore
realized as gauge transformations.
That the symmetry is realized only on-shell should not
come as a surprise because this is already the case for the
diffeomorphisms \cite{Witten:1988hc},
which are now part of the Kaluza-Klein-symmetries.

Thus we have determined a theory which is by construction a consistent
coupling of infinitely many spin-2 fields.
One might also ask the question whether the theory can be consistently
truncated to a finite number of spin-2 fields, i.e.~where only
$e_{\mu}^{a(n)}, n=-N,...,N$ for any finite $N$ remain.
These issues will be discussed in sec. \ref{truncation}.

So far we have seen that the action permits
a consistent spin-2 invariance. It remains to be checked that
it can also be viewed as a
deformation of a sum of free Pauli-Fierz
Lagrangians (\ref{free}), in particular that the first-order
theory constructed here is equivalent to a second order action.
To see this we introduce an expansion
parameter $\kappa$ and linearize
the theory by writing
 \bea
  e_{\mu}^{a(0)}=\delta_{\mu}^a+\kappa h_{\mu}^{a(0)}+O(\kappa ^2)\;,
  \qquad e_{\mu}^{a(\pm 1)}=\kappa h_{\mu}^{a(\pm 1)}+O(\kappa ^2)\;.
 \eea
We concentrate for
simplicity reasons only on the case where just
$e_{\mu}^{a(\pm 1)}$ are present.
Even though this will turn out not to be consistent
with the gauge symmetry in general, it yields
correct results up to order $O(\kappa)$ as the corrections
by the full equations are at least of order $O(\kappa^2)$.
Using the equations of motion for $e_{\mu}^{a(\pm 1)}$ we can now
express $\omega_{\mu}^{a(\pm 1)}$ in terms of them.
One finds upon expanding up to $O(\kappa)$
 \bea\label{omega1}
   \omega_{\mu}^{a(1)}&=&\kappa\big[ \varepsilon^{\nu ac}(\partial_{\mu}
   h_{\nu c}^{(1)}-\partial_{\nu}h_{\mu c}^{(1)})
   +\delta_b^{\nu}(h_{\mu}^{b(1)}\omega_{\nu}^{a(0)}-h_{\mu}^{a(1)}
   \omega_{\nu}^{b(0)}\\ \nonumber
   &&+\omega_{\mu}^{b(0)}h_{\nu}^{a(1)}
   -\omega_{\mu}^{a(0)}h_{\nu}^{b(1)})
   -\frac{1}{4}\varepsilon^{\sigma\rho d}(\partial_{\rho}h_{\sigma d}^{(1)}
   -\partial_{\sigma}h_{\rho d}^{(1)})\delta_{\mu}^a \\ \nonumber
   &&-\frac{1}{4}\delta^{\rho c}\delta^{\sigma d}(h_{\rho d}^{(1)}
   \omega_{\sigma c}^{(0)}-h_{\rho c}^{(1)}
   \omega_{\sigma d}^{(0)}+\omega_{\rho d}^{(0)}h_{\sigma c}^{(1)}
   -\omega_{\rho d}^{(0)}h_{\sigma c}^{(1)})\delta_{\mu}^a\big]
   +O(\kappa^2)\;,
 \eea
and analogously for $\omega_{\mu}^{a(-1)}$. The next step would be
to insert these relations into the equation for $e_{\mu}^{a(0)}$ and
solve the resulting expression for the `spin connection'
$\omega_{\mu}^{a(0)}$. However, due to the fact that $e_{\mu}^{a(\pm 1)}$
as well as $\omega_{\mu}^{a(\pm 1)}$ are yet of order $O(\kappa)$,
for the approximation linear in $\kappa$ we just get
 \bea
  0= \varepsilon^{abc}\delta_{\mu b}\omega_{\nu c}^{(0)}+
  \varepsilon^{abc}\delta_{\nu c}\omega_{\mu b}^{(0)}+
  \kappa\left(\partial_{\mu}h_{\nu}^{a(0)}-\partial_{\nu}h_{\mu}^{a(0)}
  +\varepsilon^{abc}(h_{\mu b}^{(0)}\omega_{\nu c}^{(0)}
  +\omega_{\mu b}^{(0)}h_{\nu c}^{(0)})\right)\;.
 \eea
In the limit $\kappa\rightarrow 0$ this is the relation for vanishing
torsion in the flat case and is therefore solved by $\omega_{\mu}^{a(0)}=0$.
Up to order $O(\kappa)$ we have the usual relation of vanishing torsion
for the `metric' $h_{\mu}^{a(0)}$ and it can therefore be solved as
in the standard case. But to solve the equation we have to multiply
with the inverse vielbein and therefore up to order $O(\kappa)$
the solution is just the linearized Levi-Civita connection.
Altogether we have
 \bea
  \omega_{\mu}^{a(0)}=\kappa(\text{lin. Levi-Civita connection})
  +O(\kappa^2)\;.
 \eea
We can now insert this relation into the formulas (\ref{omega1}) for
$\omega_{\mu}^{a(\pm 1)}$ and get
 \bea
  \omega_{\mu}^{a(1)}=\kappa\big[\varepsilon^{\nu ac}(\partial_{\mu}
  h_{\nu c}^{(1)}-\partial_{\nu}h_{\mu c}^{(1)})-\frac{1}{4}
  \varepsilon^{\sigma\rho d}(\partial_{\rho}h_{\sigma d}^{(1)}
  -\partial_{\sigma}h_{\rho d}^{(1)})\delta_{\mu}^a\big]
  +O(\kappa^2)
 \eea
and analogously for $\omega_{\mu}^{a(-1)}$.
Finally inserting this expression into the equation of motion for
$\omega_{\mu}^{a(1)}$ to that order, i.e.
 \bea
  0=\partial_{\mu}\omega_{\nu}^{a(1)}-\partial_{\nu}\omega_{\mu}^{a(1)}\;,
 \eea
and multiplying with $\varepsilon^{\mu\nu\lambda}$ results in
 \bea\label{freespin2}
  0=\partial_{\mu}\partial^{\lambda}h^{\mu\nu}-\partial^{\nu}
  \partial^{\lambda}\hat{h}+\eta^{\lambda\nu}\square\hat{h}-\eta^{\lambda\nu}
  \partial_{\mu}\partial_{\rho}h^{\mu\rho}+\partial^{\nu}\partial_{\mu}
  h^{\mu\lambda}-\square h^{\nu\lambda}\;.
 \eea
This coincides exactly with the equation of motion derived from
the original free spin-2 action (\ref{free}) in the massless case.
Here we have defined $h_{\mu\nu}:=\delta_{\mu}^ah_{\nu a}^{(1)}
+\delta_{\nu}^ah_{\mu a}^{(1)}$,
which also results up to $O(\kappa)$ from the general formula
for the higher spin-2 fields in a metric-like representation:
 \bea\label{spin2}
  g_{\mu\nu}^{(n)}:=e_{\mu}^{a(n-m)}e_{\nu a}^{(m)}\;.
 \eea
Clearly, also for
the linearized Einstein equation we get the free spin-2 equation,
and therefore we can summarize our analysis by saying that the
theory reduces in the linearization up to order $O(\kappa)$
to a sum of Pauli-Fierz terms.
On the other hand, we know that the full theory has the Kaluza-Klein symmetries
(\ref{variations}) which mix fields of different level and
accordingly the full theory (\ref{spin2CS})
has to include non-linear couplings. This in turn implies that
the higher order terms in $\kappa$ cannot vanish and the theory
is therefore a true deformation of a pure sum of Pauli-Fierz
terms. In summary, one can solve the equations of motion for
$\omega_{\mu}^{a(n)}$ at least perturbatively, thus giving a second-order
formulation.
However, it would be much more convenient to have a deeper geometrical
understanding for the $\omega_{\mu}^{a(n)}$. Such a geometrical interpretation
indeed exists and is based on a notion of algebra-valued
differential geometry developed by Wald \cite{Wald:1986dw},
which we are going to discuss in the next section.

\subsection{Geometrical interpretation of the spin-2 symmetry}
Cutler and Wald analyzed in \cite{Cutler:1986dv} the question of
possible consistent extensions of a free spin-2 gauge invariance
to a collection of spin-2 fields.
This may be compared to the similar question of a consistent
gauge symmetry for a collection of spin-1 fields. In this case
one knows that the resulting theories are Yang-Mills theories
determined by a non-abelian Lie algebra.
Analogously it was shown in \cite{Cutler:1986dv} that such a spin-2 theory
is organized by an associative and commutative algebra
$\mathfrak{A}$ (which should not be confused with a Lie
algebra).
Namely, the additional index which indicates the different
spin-2 fields is to be interpreted as an algebra index, and therefore
any collection of spin-2 fields can be viewed as a single spin-2 field,
which takes values in a nontrivial algebra.
An associative and commutative algebra $\mathfrak{A}$
can be characterized by its multiplication law, which is with respect
to a basis given by a tensor $a_{\hspace{0.3em}nm}^{k}$ according to
 \bea
  (v\cdot w)^n = a^n_{\hspace{0.3em}mk}v^mw^k\;,
 \eea
where $v,w \in \mathfrak{A}$.
That the algebra is commutative and associative is encoded in the
relations
 \bea\label{asso}
  a_{\hspace{0.3em}mn}^{k}=a_{\hspace{0.3em}(mn)}^{k}\;, \qquad
  a^k_{\hspace{0.3em}mn}a^n_{\hspace{0.3em}lp}=a^k_{\hspace{0.3em}np}a^n_
  {\hspace{0.3em}ml}\;.
 \eea
With respect to such a given algebra $\mathfrak{A}$, the allowed
gauge transformations can be written according to \cite{Cutler:1986dv} as
 \bea\label{conn}
  \delta g_{\mu\nu}^{(n)}=\partial_{(\mu}\xi_{\nu)}^{(n)}
  -2\Gamma_{\mu\nu\hspace{0.6em}l}^{\sigma \hspace{0.3em}n}\xi_{\sigma}^{(l)}
  =:\nabla_{\mu}\xi_{\nu}^{(n)}+\nabla_{\nu}\xi_{\mu}^{(n)}\;,
 \eea
where the generalized Christoffel symbol is defined by
 \bea\label{christ}
  \Gamma_{\mu\nu\hspace{0.6em}l}^{\sigma \hspace{0.3em}n}=
  \frac{1}{2}g^{\sigma\rho\hspace{0.3em}k}_{\hspace{1.5em}l}
  \left(\partial_{\mu}g_{\rho\nu\hspace{0.3em}k}^{\hspace{0.7em}n}
  +\partial_{\nu}g_{\rho\mu\hspace{0.3em}k}^{\hspace{0.6em}n}
  -\partial_{\rho}g_{\mu\nu\hspace{0.3em}k}^{\hspace{0.6em}n}\right)\;,
 \eea
and
 \bea\label{algmetric}
  g_{\mu\nu\hspace{0.3em}n}^{\hspace{0.6em}k}=
  a_{\hspace{0.3em}nm}^k g_{\mu\nu}^{(m)}\;.
 \eea
We see that $\nabla_{\mu}$ has the formal character of a covariant
derivative.

Moreover, it has been shown in \cite{Wald:1986dw} that beyond this
formal resemblance to an ordinary metric-induced connection,
there exists a geometrical interpretation in the
following sense.
As in pure general relativity, where the symmetry transformations are
given by the diffeomorphisms acting on the fields via a pullback,
the transformation rules (\ref{conn}) are the infinitesimal version
of a diffeomorphism on a generalized manifold.
This new type of manifold introduced in \cite{Wald:1986dw} generalizes the
notion of an ordinary real manifold to `algebra-valued' manifolds,
where the algebra $\mathfrak{A}$ replaces the role of $\mathbb{R}$.
To be more precise, such a manifold is locally
modeled by a $n$-fold cartesian product $\mathfrak{A}^n$ in the
same sense as an ordinary manifold is locally given by $\mathbb{R}^n$.
On these manifolds one can correspondingly define a metric which looks
from the point of view of the underlying real manifold like an ordinary,
but algebra-valued metric. The diffeomorphisms of these generalized
manifolds act infinitesimally on the metric exactly as written above
via an algebra-valued generalization of a Lie-derivative.
This Lie derivative acts, e.g., on algebra-valued $(1,1)$ tensor
fields as
 \bea\label{lieder}
  {\cal L}_{\xi}T^{\mu\hspace{0.2em}(n)}_{\hspace{0.4em}\nu}=
  a^n_{mk}\left(\xi^{\rho (m)}\partial_{\rho}
  T^{\mu\hspace{0.2em}(k)}_{\hspace{0.4em}\nu}-
  T^{\rho\hspace{0.2em}(m)}_{\hspace{0.4em}\nu}\partial_{\rho}\xi^{\mu (k)}
  +T^{\mu\hspace{0.2em}(m)}_{\hspace{0.4em}\rho}
  \partial_{\nu}\xi^{\rho (k)}\right)\;,
 \eea
and in an obvious way on all higher-rank tensor fields.
(For further details see \cite{Wald:1986dw}.)
Moreover, most of the constructions known from Riemannian geometry like
the curvature tensor have their analogue here.

To check whether our theory fits into this general framework we
first have to identify the underlying commutative algebra.
Due to the fact that the theory contains necessarily an infinite
number of spin-2 fields, the algebra has to be infinite-dimensional, too,
and we will assume that the formalism applies also to this case.

We will argue that the algebra is given by the algebra
of smooth functions on $S^1$, on which we had compactified,
together with the point-wise multiplication of functions as
the algebra structure.\footnote{That the spin-2 couplings arising
in Kaluza-Klein compactifications might be related to
Wald's framework in this way
has first been suggested by Reuter in \cite{Reuter:1988ig},
where he analyzed the reduction of a dimensionally continued Euler form
in $D=6$. Namely, the latter has the exceptional
property of inducing an infinite
tower of massless spin-2 fields due to the existence of an
infinite-dimensional symmetry already in the higher-dimensional theory.}
With respect to the complete basis
$\{e^{in\theta},n=-\infty,...,\infty\}$ of functions on $S^1$,
the multiplication is given due to elementary Fourier analysis by
 \bea
  (f\cdot g)^{n}=\sum_{m=-\infty}^{\infty} f^{n-m}\cdot g^{m}
  =\sum_{k,m=-\infty}^{\infty}\delta_{k+m,n}f^{k}g^{m}\;,
 \eea
such that the algebra is characterized by
 \bea\label{alg}
  a_{\hspace{0.3em}km}^{n}=\delta_{k+m,n}\;.
 \eea
This implies that the metric can be written according to (\ref{algmetric})
as
 \bea
  g_{\mu\nu\hspace{0.3em}k}^{\hspace{0.6em}n}=a^n_{\hspace{0.3em}km}
  g_{\mu\nu}^{(m)}=g_{\mu\nu}^{(n-k)}\;.
 \eea
Now it can be easily checked that the Kaluza-Klein
transformations (\ref{variations}) for $M=0$ applied to (\ref{spin2})
can be written as
 \bea
  \delta g_{\mu\nu}^{(n)}=\nabla_{\mu}\xi_{\nu}^{(n)}
  +\nabla_{\nu}\xi_{\mu}^{(n)}\;,
 \eea
i.e. they have exactly the required form.
Here the connection $\nabla_{\mu}$ is calculated as in (\ref{conn})
with respect to the algebra (\ref{alg}). For this we have
assumed that indices are raised and lowered according to
 \bea
  \xi_{\mu}^{(n)}=g_{\mu\nu\hspace{0.3em}k}^{\hspace{0.3em}n}\xi^{(k)\nu}\;,
 \eea
while the inverse metric is defined through the relation
 \bea\label{inverse}
   g^{\mu\rho\hspace{0.3em}n}_{\hspace{1.5em}k}
   g_{\rho\nu\hspace{0.3em}m}^{\hspace{0.6em}k}
   =\delta_{\nu}^{\mu}\delta_m^n\;.
 \eea

With the help of this geometrical interpretation we are now also
able to interpret the existence of an infinite number of
'spin-connections' $\omega_{\mu}^{a(n)}$.
If we assume that the vielbeins are invertible in the sense of
(\ref{inverse}), one can solve the
equations of motion (\ref{eom})
for the connections in terms of $e_{\mu}^{a(n)}$, as we have argued in
sec.~\ref{infspin-2}.
Then one can define a generalized covariant derivative by postulating
the vielbein to be covariantly constant,
 \bea\label{covconst}
  \nabla_{\mu}e_{\nu}^{a(n)}=\partial_{\mu}e_{\nu}^{a(n)}-
  \Gamma_{\mu\nu\hspace{0.6em}m}^{\rho \hspace{0.3em}n}e_{\rho}^{a(m)}
  +\omega_{\mu\hspace{0.3em}b}^{a(n-m)}e_{\nu}^{b(m)}=0\;.
 \eea
Since the antisymmetric part $\nabla_{[\mu}e_{\nu]}^{a(n)}$
vanishes already by the
equations of motion (\ref{eom}), this requirement specifies the
symmetric part of $\nabla_{\mu}e_{\nu}^{a(n)}$.
In turn, the algebra-valued metric (\ref{spin2}) is covariantly constant
with respect to this symmetric connection,
 \bea
  \nabla_{\mu} g_{\nu\rho}^{(n)}=\partial_{\mu}g_{\nu\rho}^{(n)}
  -\Gamma_{\mu\nu\hspace{0.6em}m}^{\sigma \hspace{0.3em}n}g_{\sigma\rho}^{(m)}
  -\Gamma_{\mu\rho\hspace{0.6em}m}^{\sigma \hspace{0.3em}n}
  g_{\sigma\nu}^{(m)}=0\;.
 \eea
But this is on the other hand also the condition which uniquely fixes the
Christoffel connection in (\ref{christ}) as
a function of the algebra-valued metric \cite{Wald:1986dw}.
Altogether, the equations of motion for the Chern-Simons action
(\ref{spin2CS}) together with (\ref{covconst}) determine a
symmetric connection, which is equivalent to the algebra-valued
Christoffel connection (\ref{christ}) compatible with the
metric (\ref{spin2}).

Wald also constructed an algebra-valued generalization of the
Einstein-Hilbert action, whose relation to the Chern-Simons action
(\ref{spin2CS}) we are going to discuss now.
This generalization is (written for three space-times dimensions)
given by
 \bea
  S^m=\int a^m_{nl}R^n\varepsilon_{\mu\nu\rho}^{\hspace{1.3em}l}
  dx^{\mu}\wedge dx^{\nu}\wedge dx^{\rho}\;,
 \eea
where $R^n$ and $\varepsilon_{\mu\nu\rho}^{\hspace{1.3em}l}$
denote the algebra-valued scalar curvature and volume form,
respectively \cite{Wald:1986dw}.
Applied to the algebra (\ref{alg}) it yields
 \bea\label{algaction}
  S^m=\int R^{m-n}\varepsilon_{\mu\nu\rho}^{\hspace{1.3em}n}
  dx^{\mu}\wedge dx^{\nu}\wedge dx^{\rho}\;.
 \eea
Here the volume form is given by
$\varepsilon_{\mu\nu\rho}^{\hspace{1.3em}n}=e^n\epsilon_{\mu\nu\rho}$,
where
 \bea
  e^n=\frac{1}{3!}\epsilon^{\mu\nu\rho}\epsilon_{abc}e_{\mu}^{a(n-m-k)}
  e_{\nu}^{b(m)}e_{\rho}^{c(k)}\;,
 \eea
such that the zero-component of (\ref{algaction})
indeed coincides with the Chern-Simons action
(\ref{spin2CS2}).
One may wonder about the meaning of the other components of the algebra-valued
action (\ref{algaction}), whose equations of motion
cannot be neglected for generic algebras.
However, the quadratic form (\ref{quadform}) which has been used
to construct the Chern-Simons action (\ref{spin2CS}), is actually
not unique, but instead there is an infinite series of
quadratic forms,
 \bea\label{quadform2}
  \langle P_a^m,J_b^n \rangle_k = \eta_{ab}\delta^{m,k-n}\;,
 \eea
each of which is invariant and can therefore be used to define
a Chern-Simons action. These will then be identical
to the corresponding components of the algebra-valued
action (\ref{algaction}).
But, as all of these actions imply the same equations of motion,
namely ${\cal F}_{\mu\nu}=0$, and are separately invariant under gauge
transformations, there is no need to consider the full algebra-valued
metric, but instead the zero-component is sufficient.

Moreover, also matter couplings can be
described in this framework  in a spin-2-covariant way.
For instance, an algebra-valued
scalar field $\phi^n$  can be coupled via
 \bea\label{scalar}
  S^m_{\text{scalar}}=\int a^m_{nk}a^n_{lp}\partial_{\lambda}\phi^l
  \partial^{\lambda}\phi^p\varepsilon_{\mu\nu\rho}^{\hspace{1.3em}k}
  dx^{\mu}\wedge dx^{\nu}\wedge dx^{\rho}\;.
 \eea
Furthermore, (\ref{scalar}) is invariant under algebra-diffeomorphisms.
The latter act via the Lie derivative in (\ref{lieder}),
such that the scalars transform with
respect to the algebra (\ref{alg}) as
 \bea
  \delta_{\xi}\phi^n = a_{km}^n \xi^{\rho (k)}\partial_{\rho}\phi^m
  = \xi^{\rho (k)}\partial_{\rho}\phi^{n-k}\;,
 \eea
i.e. as required by
(\ref{variations}) in the phase $M\rightarrow 0$.
Again, for the algebra (\ref{alg}) considered here,
the zero-component of (\ref{scalar}) is
separately invariant and can be written as
 \begin{equation}
  S_{\text{scalar}}=
  \int d^3x \sqrt{g}^{-n}\partial_{\mu}\phi^l\partial^{\mu}\phi^{n-l}
  =\int d^3xd\theta\sqrt{g}\partial_{\mu}\phi\partial^{\mu}\phi\;.
 \end{equation}
Similarly, the Chern-Simons action can be rewritten by retaining a formal
$\theta$-integration and assuming all fields to be $\theta$-dependent.
For explicit computations it is accordingly often more convenient to work with
$\theta$-dependent expressions and therefore we will give subsequent
formulas in both versions.

Finally let us briefly discuss the resolution of the aforementioned
no-go theorems for consistent gravity/spin-2 couplings.
In \cite{Boulanger:2000rq} it has been shown that
Wald's algebra-valued spin-2 theories for arbitrary algebras
generically contain ghost-like excitations.
Namely, the algebra has to admit a metric specifying the kinetic
Pauli-Fierz terms in the free-field limit and moreover has to be symmetric
in the sense that lowering the upper index in $a_{mn}^k$ by use of
this metric results in a totally symmetric $a_{mnk}=a_{(mnk)}$.
Requiring the absence of ghosts, i.e. assuming the
metric to be positive-definite, restricts the algebra to a
direct sum of one-dimensional ideals (which means $a_{mn}^k=0$
whenever $m\neq n$). The theory reduces in turn to a sum of independent
Einstein-Hilbert terms. For the infinite-dimensional algebra
considered here the metric is given by the $L^2$-norm for
square-integrable functions (see formula (\ref{L2}) in the appendix),
which is clearly positive-definite. The action may instead be viewed as
an integral over Einstein-Hilbert terms and is thus in agreement
with \cite{Boulanger:2000rq}.

\subsection{Non-linear $\sigma$-model and its global symmetries}\label{nonlin}
Apart from the spin-2 sector also the infinite tower of scalar fields
$\phi^n$ will survive in the unbroken limit $M\rightarrow 0$.
We have seen in the last
section how spin-2 invariant couplings for scalar fields can be
constructed. To fix the actual form of these couplings, we have
to identify also the global symmetries in this limit, and
in order to uncover the maximal global symmetry, we will dualize
all degrees of freedom into scalars.

We note from (\ref{variations}) that in the unbroken phase the
Virasoro algebra $\hat{v}$
parameterized by $\xi_k^5$ reduces to an abelian gauge symmetry.
As a general feature of ungauged limits, the full $\hat{v}$
will then turn out to be realized only
as a global symmetry.
More precisely, we expect an invariance under rigid transformations
of the general form
 \bea\label{rep}
  \delta_{\xi^5}\chi^{n}=i \sum_k (n-(1-\Delta)k)\xi_k^5\chi^{n-k}\;,
 \eea
where $\xi_k^5$ is now space-time independent.
One easily checks that these
are representations of $\hat{v}$ (for details see appendix B).
They will be labeled by the conformal dimension $\Delta$.
More precisely, the Kaluza-Klein fields
$e_{\mu}^a$, $A_{\mu}$ and $\phi$ transform as $\Delta=1$, $\Delta=-1$ and
$\Delta=2$, respectively.

We start form the zero-mode action (\ref{zero-mode}) and replace it
by the algebra-valued generalization discussed in the last section.
For the Einstein-Hilbert term we have already seen that
this procedure yields the correct $\widehat{iso(1,2)}$ gauge theory,
and therefore it is sufficient to focus on the scalar kinetic term and
the Yang-Mills term. The action reads
 \bea\label{matteraction}
  S_{\text{matter}}=\int d^3 xd\theta e\left(-\frac{1}{4}
  \phi^2F^{\mu\nu}F_{\mu\nu}+\frac{1}{2}\phi^{-2}
   g^{\mu\nu}\partial_{\mu}\phi
  \partial_{\nu}\phi\right)\;,
 \eea
where all fields are now $\theta$-dependent or, equivalently,
algebra-valued.

To dualize the $U(1)$ gauge fields $A_{\mu}^n$ into new
scalars $\varphi^n$, we define the standard duality relation
 \bea\label{dual}
  \phi^2F_{\mu\nu}=e\varepsilon_{\mu\nu\rho}g^{\rho\sigma}
  \partial_{\sigma}\varphi\;,
 \eea
which is not affected by the $\theta$-dependence of all
fields. Thus, the abelian duality between vectors and scalars
persists also in the algebra-valued case,
and the degrees of freedom can be assigned to
$\phi$ and $\varphi$.
The Lagrangian for the
scalar fields then takes in the unbroken limit $M\rightarrow 0$
the form
 \bea\label{cosetaction}
  \mathcal{L}_{\text{scalar}}=\frac{1}{2}eg^{\mu\nu}\phi^{-2}
  (\partial_{\mu}\phi\partial_{\nu}\phi+\partial_{\mu}\varphi
  \partial_{\nu}\varphi)\;,
 \eea
which coincides formally with the zero-mode action after a
standard dualization, but now with all fields still being
$\theta$-dependent.
>From (\ref{dual}) one determines the transformation properties of
the dual scalar $\varphi$ under $\hat{v}$ and finds
 \bea
  \delta_{\xi^5}\varphi=\xi^5\partial_5\varphi+2\varphi\partial_5\xi^5\;,
 \eea
i.e. it transforms in the same representation as $\phi$ with $\Delta=2$.
(For the computation it is crucial to take into account
that also $e_{\mu}^a$ transforms under $\hat{v}$.)
Now one easily checks that the action is invariant under global
Virasoro transformations.

Moreover, it is well known that the zero-mode scalar fields
span a non-linear $\sigma$-model with coset space $SL(2,\mathbb{R})/SO(2)$
as target space, carrying the `Ehlers group' $SL(2,\mathbb{R})$
as isometry group \cite{Ehlers}.
If one includes all Kaluza-Klein modes at $M=0$,
this symmetry is enhanced to an infinite-dimensional algebra,
which we are going to discuss now.
Defining the complex scalar field $Z=\varphi+i\phi$,
the action (\ref{cosetaction}) can be rewritten as
 \bea
  \mathcal{L}_{\text{scalar}}=\frac{1}{2}eg^{\mu\nu}\frac{\partial_{\mu}Z
  \partial_{\nu}\bar{Z}}{(Z-\bar{Z})^2}\;,
 \eea
which is invariant under the $SL(2,\mathbb{R})$ isometries
acting as
 \bea\label{sl2action}
  Z\rightarrow Z^{\prime}=\frac{aZ+b}{cZ+d}\;,
  \qquad \left(\begin{array}{cc}
  a & b \\ c & d \end{array}\right)\in SL(2,\mathbb{R})\;.
 \eea
This invariance is not spoiled by the fact that $Z$ is still
$\theta$-dependent and so the $SL(2,\mathbb{R})$ acts on the full
tower of Kaluza-Klein modes, as can be seen by expanding (\ref{sl2action})
into Fourier modes. But moreover, also the $SL(2,\mathbb{R})$ group
elements can depend on $\theta$, and therefore an additional
infinite-dimensional symmetry seems to appear.

To determine the algebra structure of this infinite-dimensional
symmetry, let us first introduce a basis for $sl(2,\mathbb{R})$:
 \bea
   h=\left(\begin{array}{cc} 1 & 0 \\ 0 & -1 \end{array}\right), \quad
   e=\left(\begin{array}{cc} 0 & 1 \\ 0 & 0 \end{array}\right), \quad
   f=\left(\begin{array}{cc} 0 & 0 \\ 1 & 0 \end{array}\right).
 \eea
Infinitesimally, with transformation parameter $\alpha=\alpha(\theta)$
they act according to (\ref{sl2action}) as
 \bea
  \delta_{\alpha}(h)Z=-2\alpha Z\;, \qquad
  \delta_{\alpha}(e)Z=-\alpha\;, \qquad
  \delta_{\alpha}(f)Z=\alpha Z^2\;,
 \eea
or, expanded in Fourier components, as
 \begin{equation}
  \begin{split}
   \delta_{\alpha^m}(h)Z^n &= -2\alpha^m Z^{n-m}\;, \qquad
  \delta_{\alpha^m}(e)Z^n=-\delta^{mn}\alpha^m\;, \\
  \delta_{\alpha^m}(f)Z^n &= \alpha^mZ^{n-m-l}Z^l \;.
  \end{split}
 \end{equation}
In particular, the real part of $Z$, i.e. the dual scalar $\varphi$,
transforms as a shift under $e$-transformations, which will later on
be promoted to local shift symmetries in the gauged theory.

We can now compute the closure of these symmetry variations
with the Virasoro variations $\delta_{\xi^m}(Q)$.
One finds
 \begin{equation}
  \begin{split}
   [\delta_{\xi^m}(Q),\delta_{\eta^n}(h)]Z^k &=-in\delta_{(\xi\eta)^{m+n}}
   (h)Z^k\;, \\
   [\delta_{\xi^m}(Q),\delta_{\eta^n}(e)]Z^k &=i(-n-2m)
   \delta_{(\xi\eta)^{m+n}}(e)Z^k\;, \\
   [\delta_{\xi^m}(Q),\delta_{\eta^n}(f)]Z^k &=i(-n+2m)
   \delta_{(\xi\eta)^{m+n}}(f)Z^k\;,
  \end{split}
 \end{equation}
where we have set
 \bea
  (\xi\eta)^{m+n}=\xi^m\eta^n\;.
 \eea
Furthermore, the extended $sl(2,\mathbb{R})$ transformations close
among themselves according to
 \begin{equation}
  \begin{split}
   [\delta_{\alpha^m}(h),\delta_{\beta^n}(e)]Z^k&=
   2\delta_{(\alpha\beta)^{m+n}}(e)Z^k\;, \\
   [\delta_{\alpha^m}(h),\delta_{\beta^n}(f)]Z^k&=
   -2\delta_{(\alpha\beta)^{m+n}}(f)Z^k\;, \\
   [\delta_{\alpha^m}(e),\delta_{\beta^n}(f)]Z^k&=
   \delta_{(\alpha\beta)^{m+n}}(h)Z^k\;, \\
   [\delta_{\alpha^m}(h),\delta_{\alpha^n}(h)]Z^k&=
   [\delta_{\alpha^m}(e),\delta_{\alpha^n}(e)]Z^k=
   [\delta_{\alpha^m}(f),\delta_{\alpha^n}(f)]Z^k=0\;.
  \end{split}
 \end{equation}
Altogether we can conclude that the following Lie algebra is a global
symmetry of the ungauged theory
 \begin{equation}\label{global}
  \begin{split}
   [Q^m,Q^n]&=i(m-n)Q^{m+n}\;, \quad
   [Q^m,e_n]=i(-n-2m)e_{m+n}\;, \\ [Q^m,h_n]&=-inh_{m+n}\;,
   \qquad [Q^m,f_n]=i(-n+2m)f_{m+n}\;, \\
   [h_m,e_n]&=2e_{m+n}\;, \qquad [h_m,f_n]=-2f_{n+m}\;, \\
   [e_m,f_n]&=h_{m+n}\;, \qquad [e_m,e_n]=[h_n,h_m]=[f_m,f_n]=0\;.
  \end{split}
 \end{equation}
We see that the symmetry algebra includes not only the Virasoro algebra
$\hat{v}$, but also the Kac-Moody algebra
$\widehat{sl(2,\mathbb{R})}$,
which transforms under $\hat{v}$.
Note, that these
transformation properties are not the standard ones known from the
Sugawara construction (see appendix \ref{km} and \cite{Goddard:1986bp}).
However, this algebra reduces to the standard form upon the
change of basis given by $\hat{Q}^m=Q^m+mh^m$, such that it clearly defines
a consistent Lie algebra.

In summary, we can think of the scalar fields $\phi^n$ and $\varphi^n$
as parameterizing an
infinite-dimensional $\sigma$-model coset space
 \bea
  {\cal M}=\frac{\widehat{SL(2,\mathbb{R})}}{\widehat{SO(2)}}\;.
 \eea
Strictly speaking this is not full truth, since the
metric used to contract indices is actually algebra-valued.
Thus, here we have an algebra-valued generalization of a $\sigma$-model,
which in turn is the reason that it does not only have the
symmetries $\widehat{sl(2,\mathbb{R})}$, but instead
the whole algebra $\hat{v}\ltimes \widehat{sl(2,\mathbb{R})}$ defined by
(\ref{global}).

In total, the ungauged phase of the
effective Kaluza-Klein action without any truncation is therefore
given by
 \begin{equation}\label{ungauged}
  \begin{split}
   S=\int d^3 x \:&\big(-\varepsilon^{\mu\nu\rho}e_{\mu a}^{(n)}
   (\partial_{\nu}\omega_{\rho}^{a(-n)}-\partial_{\rho}
   \omega_{\nu}^{a(-n)}+\varepsilon^{abc}\omega_{\nu b}^{(m)}
   \omega_{\rho c}^{(-n-m)}) \\
   &+\frac{1}{2}eg^{\mu\nu}\phi^{-2}
   (\partial_{\mu}\phi\partial_{\nu}\phi+\partial_{\mu}\varphi
   \partial_{\nu}\varphi)\big)\;,
  \end{split}
 \end{equation}
where in the second term
the algebra multiplication defined in (\ref{scalar}) is implicit, or
in other words, where all fields are $\theta$-dependent and an integration
over $\theta$ is assumed.
The action is by construction invariant under spin-2 transformations.
Moreover, we have already seen that the scalar couplings are also invariant
under global Virasoro transformations. To see that this is also
the case for the generalized Einstein-Hilbert term,
we have to show that one can determine the
transformation rule for $\omega_{\mu}^{a(n)}$
such that the action stays invariant.
This is indeed possible, and one finds
 \bea\label{deltaomega}
  \delta_{\xi^5}\omega_{\mu}^{a(n)}=
  i\sum_k (n-k)\xi_k^5\omega_{\mu}^{a(n-k)}\;,
  \qquad \delta_{\xi^5}\omega_{\mu}^a=\xi^5\partial_{\theta}\omega_{\mu}^a\;\;.
 \eea
Equivalently, they can be computed by solving the $\omega_{\mu}^{a(n)}$
in terms of the vielbeins by use of (\ref{eom}) and then applying
a $\hat{v}$ transformation to this expression. Both results coincide.
Note that instead the full algebra-valued action (\ref{algaction})
transforms non-trivially under $\hat{v}$, namely as
 \bea
  \delta S^m = im\xi^5_n S^{(m-n)}.
 \eea
However, as we have already seen in sec.~3.2, it is sufficient to include
only the zero-component in (\ref{ungauged}), which is clearly
invariant.

\subsection{Dualities and gaugings}\label{dualgaugings}
So far we have determined the unbroken phase of the Kaluza-Klein theory in
a description where all propagating degrees of freedom
reside in scalar fields.
Before we turn to a gauging of a subgroup of the global symmetries
we have to ask whether it is still possible to assign all degrees of
freedom to scalars, since the introduction of gauge fields necessarily
seems to enforce the appearance of local degrees of freedom that
are instead carried by vectors.
However, in \ref{sugragauging} we have reviewed the peculiar
fact that in three-dimensional gauged supergravities all Yang-Mills-type
gaugings are on-shell equivalent to Chern-Simons gaugings with an
enlarged number of scalar fields \cite{Nicolai:2003bp,deWit:2003ja}.
Thus all bosonic degrees of freedom can still appear as scalar fields.
We are going to show that this duality also applies to
the present case.

To begin with, we note that in the gauged theory all partial derivatives
are replaced by covariant ones.
For a given field $\chi$
transforming in a representation $\Delta$ under $\hat{v}$
the covariant derivative reads
 \bea
  D_{\mu}\chi^{n}=\partial_{\mu}\chi^{n}-ig\sum_k (n-(1-\Delta)k)A_{\mu}^{k}
  \chi^{n-k}\;,
 \eea
where we have introduced the gauge coupling $g=M$.
Indeed, it transforms by construction covariantly under
local $\hat{v}$ transformations, $\delta_{\xi}(D_{\mu}\chi^{n})
=ig(n-(1-\Delta)k)\xi_kD_{\mu}\chi^{n-k}$, if we assume as usual that
$A_{\mu}^n$ transforms as a gauge field under the adjoint (i.e. as
the Kaluza-Klein vector in (\ref{variations}) with $\Delta=-1$).
Similarly, the non-abelian $\hat{v}$ field strength is given by
 \bea
  F_{\mu\nu}^n=\partial_{\mu}A_{\nu}^{n}-\partial_{\nu}A_{\mu}^{n}
  +ig \sum_m (n-2m)A_{\mu}^{n-m}A_{\nu}^{m}\;.
 \eea
These expressions are given for the Kaluza-Klein fields in
$\theta$-dependent notation by
 \begin{equation}\label{covariant}
  \begin{split}
   D_{\mu}\phi&=\partial_{\mu}\phi-gA_{\mu}\partial_{\theta}\phi
   -2g\phi\partial_{\theta}A_{\mu}\;, \\
   F_{\mu\nu}&=\partial_{\mu}A_{\nu}-\partial_{\nu}A_{\mu}-gA_{\mu}
   \partial_{\theta}A_{\nu}+gA_{\nu}\partial_{\theta}A_{\mu}\;.
  \end{split}
 \end{equation}

The part of the gauged action containing
scalar fields will be given by the covariantisation of the
action (\ref{matteraction}) according to (\ref{covariant}).
One easily checks that this transforms into a total $\theta$-derivative
under local
$\xi^5_k$-transformations, i.e.~defines an invariant action.
Furthermore, in appendix \ref{reduc} we show by
explicit reduction
that exactly these terms appear,
as well as an
explicit $\hat{v}$ gauge invariant mass term
for the spin-2 fields,
i.e. the action reads (see also \cite{Cho:1992rq,Cho:1991xk})
 \begin{equation}\label{scalaract}
  \mathcal{L}_{\text{scalar}}=\frac{1}{2}
  eg^{\mu\nu}\phi^{-2}D_{\mu}
  \phi D_{\nu}\phi-\frac{1}{4}e\phi^2g^{\mu\rho}g^{\nu\sigma}
  F_{\mu\nu}F_{\rho\sigma}+{\cal L}_{m}\;.
 \end{equation}

To show that this action is indeed on-shell equivalent to a Chern-Simons
gauged theory we introduce as in \ref{sugragauging}
new gauge fields for each of the former Yang-Mills fields,
or in other words, we enhance the gauge symmetry with nilpotent
shift symmetries (see also \cite{Hohm:2004rc}).
To explain this dualization procedure, let us consider
the Yang-Mills equation resulting from (\ref{scalaract})
 \bea
  D^{\mu}(\phi^2 F_{\mu\nu})=j_{\nu}\;,
 \eea
where $j_{\nu}$ denotes the current induced by the charged fields.
It implies integrability of the duality relation
 \bea\label{duality}
  \frac{1}{2} e^{-1}\varepsilon^{\mu\nu\rho}\phi^2F_{\nu\rho}
  =D^{\mu}\varphi+gB^{\mu}
  =:\mathcal{D}^{\mu}\varphi\;,
 \eea
where $\varphi$ will be the scalar field carrying the former degrees of
freedom of  $A_{\mu}$, and $B_{\mu}$ is the gauge field corresponding to
the enlargement of the gauge group.
>From the previous section we know already the transformation properties
of $\varphi$ under $\hat{v}$, and one may check explicitly
that it does not change in the gauged phase.
In particular, also the dual vector $B_{\mu}$ will transform
with $\Delta=2$. In other words, it transforms
under the dual of the adjoint representation of $\hat{v}$,
which will later on turn out to be important.\footnote{For a definition
see appendix \ref{km}.}
To define the dual action we add instead of the Yang-Mills term a
Chern-Simons-like term $B\wedge F$, where $F$ denotes the non-abelian
field strength, and get
 \bea\label{dualscalar}
  \mathcal{L}_{\text{scalar}}=\frac{1}{2}eg^{\mu\nu}\phi^{-2}
  \left(D_{\mu}\phi D_{\nu}\phi+\cal{D}_{\mu}\varphi\cal{D}_{\nu}
   \varphi\right)-\frac{1}{2}g\varepsilon^{\mu\nu\rho}B_{\mu}F_{\nu\rho}
   +{\cal L}_{m}\;.
 \eea
Indeed, varying with respect to $B_{\mu}$ one recovers
the duality relation (\ref{duality}),
and eliminating the dual scalar $\varphi$ by means
of this relation yields the Yang-Mills type theory (\ref{scalaract}).
Thus we have shown that the degrees of freedom of the $A_{\mu}^n$
can be assigned to new scalars $\varphi^n$, if at the same time
new topological gauge fields $B_{\mu}^n$ are introduced that promote
the former global shift transformations (i.e. the $e$-transformations
of $\widehat{sl(2,\mathbb{R})}$) to a local symmetry.

\section{Broken phase of the Kaluza-Klein theory}
Up to now we have determined the action (\ref{ungauged}) of the ungauged
theory, which
is invariant under global $\hat{v}\ltimes \widehat{sl(2,\mathbb{R})}$
transformations as well as local spin-2 transformations.
We argued that in order to get the full Kaluza-Klein action
one has to gauge the Virasoro algebra together with
the shift symmetries of (\ref{global}).
In the next section we discuss the
effect of this gauging on the topological fields. We will see that
they combine into a single Chern-Simons theory.
As we have seen in \ref{CSgravity} this is quite
analogous to gauged supergravities, where truncating to the
topological fields results in the Chern-Simons theories
of \cite{Achucarro:1987vz} for $AdS$-supergroups.
The scalars will be discussed thereafter.

\subsection{Local Virasoro invariance for
topological fields}\label{CSgauged}
As usual the gauging proceeds in several steps. First of all, one has to
replace all partial derivatives by covariant ones.
Let us start with the generalized Einstein-Hilbert term.
The covariant derivative for $\omega_{\mu}^{a(n)}$
in accordance with (\ref{deltaomega}) reads
 \bea
  D_{\mu}\omega_{\nu}^{a(n)}=\partial_{\mu}\omega_{\nu}^{a(n)}
  -ig\sum_m (n-m)A_{\mu}^m\omega_{\nu}^{a(n-m)}\;,
 \eea
or equivalently
 \bea
  D_{\mu}\omega_{\nu}^{a}=\partial_{\mu}\omega_{\nu}^a
  -A_{\mu}\partial_5 \omega_{\nu}^a\;.
 \eea
The covariantized Einstein-Hilbert action is then invariant under
local Virasoro transformations. In contrast it will no longer be invariant
under all spin-2 transformations, but only under three-dimensional
diffeomorphisms. This is due to the fact that the explicit
$\partial_5$ appearing
in the covariant derivatives will also act on the spin-2 transformation
parameter. Thus, the gauging will deform the spin-2 transformations.

Furthermore, we have already seen that
in order to guarantee that the resulting action will be
equivalent to the original Yang-Mills gauged theory,
one has to introduce a Chern-Simons term for the
Kaluza-Klein vectors $A_{\mu}^n$, whose propagating degrees of freedom are
now carried by the dual scalars $\varphi^n$,
as well as for the dual gauge fields $B_{\mu}^n$.
This implies that we do not have to gauge only the Virasoro algebra
$\hat{v}$, but instead the whole subalgebra of (\ref{global}), which
is spanned by $(Q^m,e_m)$, while the rigid symmetry given by
$h_m$ and $f_m$ will be broken explicitly.
Both gauge fields combine
into a gauge field for this larger algebra.
Moreover, in contrast to
$\hat{v}$ itself this algebra
carries a non-degenerate invariant
quadratic form, namely
 \bea\label{virform}
  \langle Q^m, e_n \rangle = \delta^{n,-m}\;,
 \eea
such that a Chern-Simons action can be defined. The existence
of this form is due to the fact that $e_m$ transforms actually
under the co-adjoint action of $\hat{v}$, as we have argued in 3.4.
We will see that the Chern-Simons action with respect to this quadratic
form indeed reproduces the correct $B\wedge F$-term in (\ref{dualscalar}).

It is tempting to ask, whether all topological fields, i.e.~the
gravitational fields together with the gauge fields for the Virasoro and
shift symmetry,
can be combined into a Chern-Simons theory for a larger algebra.
The latter would have to combine the affine Poincar\'e
algebra with the algebra spanned by $(Q^m,e_m)$.
Naively one would think that the semi-direct product
$\hat{v}\ltimes \widehat{iso(1,2)}$ defined in (\ref{kacmoody})
and extended by $e_m$ according to (\ref{global}) is the correct choice.
However, it does not reproduce the right Kaluza-Klein symmetry transformations,
and moreover, the algebra seems not to admit a non-degenerate and
invariant quadratic form.
To see that a Chern-Simons formulation nevertheless exists, we
observe that varying the total action consisting of the sum of
$\hat{v}$-covariantized Einstein-Hilbert action and $B\wedge F$
with respect to $A_{\mu}$, we get the non-abelian field strength for
$B_{\mu}$ plus terms of the form $e_{\mu}^a\partial_5\omega_{\nu a}$.
Thus, a Chern-Simons interpretation is only possible if the latter
terms are contained in the field strength of $B_{\mu}$, or in
other words, if the algebra also closes according to $[P,J]\sim e$.

Demanding consistency with the Jacobi identities and requiring that
$e_{\mu}^a$ and $\omega_{\mu}^a$ transform under the correct representation
of $\hat{v}$, the following Lie algebra is then uniquely fixed up to
a free parameter $\alpha$:
 \begin{equation}\label{bigalgebra}
  \begin{split}
   [P_a^m,J_b^n]&=\varepsilon_{abc}P^{c(m+n)}+i\alpha n\eta_{ab}e_{m+n}\;,
   \quad [J_a^m,J_b^n]=\varepsilon_{abc}J^{c(m+n)}\;, \\
   [P_a^m,P_b^n]&=0\;, \\
   [Q^m,Q^n]&=ig(m-n)Q^{m+n}\;, \qquad [Q^m,P_a^n]=ig(-m-n)P_a^{m+n}\;, \\
   [Q^m,J_a^n]&=-ignJ_a^{m+n}\;, \qquad [Q^m,e_n]=ig(-n-2m)e_{m+n}\;, \\
   [P_a^m,e_n]&=[J_a^m,e_n]=[e_m,e_n]=0\;.
  \end{split}
 \end{equation}
Here we have rescaled the $Q^m$ with the gauge coupling constant $g$
for later convenience.

We see that one gets an algebra which looks similar to the one
proposed in \cite{Dolan:1983aa} (see (\ref{kacmoody})),
except that it does not contain simply
the semi-direct product of $\hat{v}$ with the affine Poincar\'e algebra,
since the $P_a^n$ and $J_a^n$
transform in different representations of $\hat{v}$.
But in contrast to sec. \ref{nonlin}, where we observed a similar
phenomenon for the global symmetry algebra, there seems not to exist
an obvious change of basis which reduces the algebra to the standard form.
Namely, because of the different index structure the $Q^m$ can be
shifted neither by $P_a^m$ nor $J_a^m$.
In fact, that the algebra is consistent even in this non-standard form
is possible only because of the nilpotency of translations,
i.e. $[P_a^m,P_b^n]=0$.

Furthermore, we observe that the algebra admits a central
extension $e_m$ of the Poincar\'e algebra
even at the classical level. (Even though, strictly speaking, it is
only a central extension for the Poincar\'e subalgebra,
since the $e_m$ do not commute with the $Q^m$.)
Remarkably, it is exactly this modification of the algebra that
allows the existence of an invariant quadratic form. Namely, the
bilinear expression
 \bea
  W=P^{a(-m)}J_a^{(m)}+\frac{\alpha}{g} Q^m e_{-m}
 \eea
(in particular, $\langle Q^m,e_n\rangle
=\frac{g}{\alpha}\delta^{m,-n}$)
is invariant under (\ref{bigalgebra}).
The total Chern-Simons action constructed with respect to this quadratic
form, with the gauge field written as
 \bea\label{gaugefield}
  {\cal A}_{\mu}=e_{\mu}^{a(n)}P_a^{n}+\omega_{\mu}^{a(n)}J_a^{n}
  +A_{\mu}^n Q^n + B_{\mu}^ne_n\;,
 \eea
is then indeed given by
 \bea\label{gaugeCS}
  S_{CS}=\int d^3 x d\theta \big(\varepsilon^{\mu\nu\rho}e_{\mu a}
  (D_{\nu}\omega_{\rho}^a-D_{\rho}\omega_{\nu}^a
   +\varepsilon^{abc}\omega_{\nu b}\omega_{\rho c})
   +\frac{g}{\alpha}\varepsilon^{\mu\nu\rho}B_{\mu}F_{\nu\rho}\big)\;,
 \eea
i.e. consists of the $\hat{v}$-covariantized Einstein-Hilbert term
and the Chern-Simons action for $A_{\mu}$ and $B_{\mu}$.

Let us briefly comment on the reality constraints on (\ref{bigalgebra}).
Naively one would take (\ref{bigalgebra}) as real Lie algebra, and
correspondingly the gauge fields in (\ref{gaugefield}) would also be
real. However, the reality condition $(Q^*)^m=Q^m$ (and similarly for
all other generators) is not consistent,
since taking the complex conjugate of (\ref{bigalgebra}) changes
relative signs. Instead, only the
reality constraint $(Q^*)^m=Q^{-m}$ can be consistently imposed.
This is on the other hand also in accordance with the reality condition
for the original Kaluza-Klein fields in (\ref{fieldexp}),
and therefore the fields
in (\ref{gaugefield}) fulfill exactly the correct reality constraint.

The equations of motion for the Chern-Simons action in (\ref{gaugeCS})
again imply vanishing field strength,
 \bea
  {\cal F}_{\mu\nu}=R_{\mu\nu}^{a(n)}J_a^n+T_{\mu\nu}^{a(n)}P_a^n
  +F_{\mu\nu}^n Q^n + G_{\mu\nu}^n e_n=0\;,
 \eea
whose components can in turn be written as
 \begin{eqnarray}
   R_{\mu\nu}^{a(n)}&=&\partial_{\mu}\omega_{\nu}^{a(n)}-\partial_{\nu}
   \omega_{\mu}^{a(n)}+\varepsilon^{abc}\omega_{\mu b}^{(n-m)}\omega_{\nu c}
   ^{(m)} \nonumber \\
   &+&ig\sum_m (n-m)\omega_{\mu}^{a(n-m)}A_{\nu}^m
   -ig\sum_m mA_{\mu}^{n-m}\omega_{\nu}^{a(m)}\;, \nonumber \\
   T_{\mu\nu}^{a(n)}&=&D_{\mu}e_{\nu}^{a(n)}-D_{\nu}e_{\mu}^{a(n)}
   +\varepsilon^{abc}e_{\mu b}^{(n-m)}\omega_{\nu c}^{(m)}
   +\varepsilon^{abc}\omega_{\mu b}^{(n-m)}e_{\nu c}^{(m)} \;, \nonumber \\
   F_{\mu\nu}^n&=&\partial_{\mu}
   A_{\nu}^{n}-\partial_{\nu}A_{\mu}^{n}
   +ig\sum_m (n-2m)A_{\mu}^{n-m}A_{\nu}^{m}\;, \nonumber \\
   G_{\mu\nu}^n&=&\partial_{\mu}B_{\nu}^n-\partial_{\nu}B_{\mu}^n
   +ig\sum_m(m-2n)A_{\mu}^{n-m}B_{\nu}^m
   +ig\sum_m (n+m)B_{\mu}^{n-m}A_{\nu}^m \nonumber \\
   &+& i\alpha \sum_m me_{\mu}^{a(n-m)}\omega_{\nu a}^{(m)}
   - i\alpha\sum_m (n-m)\omega_{\mu}^{a(n-m)} e_{\nu a}^{(m)} \;.
 \end{eqnarray}
Here $D_{\mu}e_{\nu}^{a(n)}$ denotes the $\hat{v}$-covariant derivative
on $e_{\mu}^{a(n)}$, which is in $\theta$-notation given by
 \bea\label{ecov}
  D_{\mu}e_{\nu}^a=\partial_{\mu}e_{\nu}^a-A_{\mu}\partial_5e_{\nu}^a
  -e_{\nu}^a\partial_5 A_{\mu}\;.
 \eea
Moreover, all quantities can be rewritten in $\theta$-dependent
notation, e.g.~the non-abelian field strength for $B_{\mu}$ is given by
 \bea
  \begin{split}
   G_{\mu\nu}&=\partial_{\mu}B_{\nu}-\partial_{\nu}B_{\mu}
   +2(B_{\mu}\partial_5A_{\nu}-B_{\nu}\partial_5A_{\mu})
   -A_{\mu}\partial_5B_{\nu}+A_{\nu}\partial_5B_{\mu} \\
   &+\alpha \left(e_{\mu}^a\partial_{\theta}\omega_{\nu a}
   -e_{\nu}^a\partial_{\theta}\omega_{\mu a}\right) \;.
  \end{split}
 \eea

The gauge transformations for gauge parameter $u=\rho^{a(n)}P_a^n
+\tau^{a(n)}J_a^n+\xi^5_nQ^n+\Lambda^ne_n$ can be written as
 \begin{eqnarray}\label{bigtrans}
   \delta e_{\mu}^{a(n)}&=&\partial_{\mu}\rho^{a(n)}+\varepsilon^{abc}
   e_{\mu b}^{(n-m)}\tau_c^{(m)}+\varepsilon^{abc}\omega_{\mu b}^{(n-m)}
   \rho_c^{(m)}\nonumber \\
   &-&ignA_{\mu}^{n-m}\rho^{a(m)}+igne_{\mu}^{a(n-m)}\xi_m^5 \;,
   \nonumber  \\
   \delta \omega_{\mu}^{a(n)}&=&\partial_{\mu}\tau^{a(n)}+\varepsilon^{abc}
   \omega_{\mu b}^{(n-m)}\tau_c^{(m)} \nonumber \\
   &-&ig\sum_m mA_{\mu}^{n-m}\tau^{a(m)}
   +ig\sum_m (n-m)\omega_{\mu}^{a(n-m)}\xi_m^5 \nonumber \\
   \delta A_{\mu}^{n}&=&\partial_{\mu}\xi^{n}_5
   +ig\sum_m (n-2m)\xi_{m}^5 A_{\mu}^{n-m}\;, \nonumber \\
   \delta B_{\mu}^{n}&=&\partial_{\mu}\Lambda^{n}+ig\sum_m (m-2n)\Lambda^{m}
   A_{\mu}^{n-m}+ig\sum_m (n+m)\xi_{m}^5 B_{\mu}^{n-m} \nonumber \\
   &+&i\alpha\sum_m m e_{\mu}^{a(n-m)}\tau_a^{(m)}
   +i\alpha \sum_m (m-n)\omega_{\mu}^{a(n-m)}\rho_a^{(m)} \;.
 \end{eqnarray}

Let us now check, whether the Kaluza-Klein symmetries are included
in these gauge transformations.
First of all, it reproduces the correct transformation rule for $B_{\mu}$
under $\hat{v}$, as can be seen by rewriting the last equation
of (\ref{bigtrans}) in $\theta$-dependent notation
 \begin{eqnarray}
  \delta B_{\mu} &=& \partial_{\mu}\Lambda-2g\Lambda\partial_{\theta} A_{\mu}
  -gA_{\mu}\partial_{\theta}\Lambda
  +g\xi^5\partial_{\theta} B_{\mu}+2gB_{\mu}\partial_{\theta}\xi^5 \\ \nonumber
  &+&\alpha e_{\mu}^a\partial_{\theta}\tau_a-\alpha\rho_a
   \partial_{\theta}\omega_{\mu}^a\;.
 \end{eqnarray}
By comparing (\ref{bigtrans}) with (\ref{variations}) we
also see that the Virasoro gauge transformations
parameterized by $\xi^5$
are correctly reproduced for $e_{\mu}^a$ and $A_{\mu}$.
To compare with the spin-2 transformations
we define in analogy to (\ref{kkparam}) the transformation parameter
 \bea
  \rho^a=\xi^{\rho}e_{\rho}^a\;, \quad \tau^a=\xi^{\rho}\omega_{\rho}^a
  \;, \quad \xi^5=\xi^{\rho}A_{\rho}\;.
 \eea
Then one finds for the vielbein
 \bea
  \delta_{\xi}e_{\mu}^a=\xi^{\rho}\partial_{\rho}e_{\mu}^a
  +\partial_{\mu}\xi^{\rho}e_{\rho}^a + gA_{\rho}\partial_{\theta}
  \xi^{\rho}e_{\mu}^a
  -gA_{\mu}\partial_{\theta}\xi^{\rho}e_{\rho}^a-\xi^{\rho}T_{\rho\mu}^a\;,
 \eea
which implies that on-shell, i.e.~for $T_{\mu\nu}^a=0$, the gauge
transformations coincide with the Kaluza-Klein symmetries in (\ref{diff})
and (\ref{locomp}).
With the same transformation parameter and $\Lambda=\xi^{\rho}B_{\rho}$
we find for $A_{\mu}$ and $B_{\mu}$
the following transformation rules (again up to field strength terms)
 \begin{equation}\label{gaugetrans}
  \begin{split}
   \delta_{\xi}A_{\mu} &=\xi^{\rho}\partial_{\rho}A_{\mu}
   +\partial_{\mu}\xi^{\rho}A_{\rho}-gA_{\mu}\partial_{\theta}
   \xi^{\rho}A_{\rho}\;, \\
   \delta_{\xi}B_{\mu}&=\xi^{\rho}\partial_{\rho}B_{\mu}
   +\partial_{\mu}\xi^{\rho}B_{\rho}-gA_{\mu}\partial_{\theta}
   \xi^{\rho}B_{\rho}
   +2gB_{\mu}\partial_{\theta}\xi^{\rho}A_{\rho}
   +\alpha e_{\mu}^a\partial_{\theta}\xi^{\rho}\omega_{\rho a} \;,
  \end{split}
 \end{equation}
which reproduces for $A_{\mu}$ the same transformation as in
(\ref{variations}), up to the $\phi$-dependent term (which, of course,
cannot be contained in a Chern-Simons formulation).

As in the case of the pure gravity-spin-2 theory, the
topological phase of the Kaluza-Klein theory is given by a Chern-Simons
theory, and moreover the Kaluza-Klein symmetry transformations are on-shell
equivalent to the non-abelian gauge transformations determined by
(\ref{bigalgebra}). Even though this equivalence holds only on-shell,
the Kaluza-Klein transformations are separately an (off-shell) symmetry,
since $\delta_{\xi}{\cal A}_{\mu}=\xi^{\rho}{\cal F}_{\rho\mu}$
leaves the Chern-Simons action invariant, as can be easily checked with
(\ref{var}).

Finally, let us check that spin-2 transformations together with
the Virasoro transformations build a closed algebra,
as it should be at least on-shell, since they were constructed
as Yang-Mills gauge transformations.
For the vielbein, e.g., one finds\footnote{As before, we indicate Virasoro
transformations by a subscript $5$ on the transformation parameter.}
 \bea
  [\delta_{\xi},\delta_{\eta^5}]e_{\mu}^a=\delta_{(\eta\xi)}e_{\mu}^a
  -\delta_{(\xi\eta)^5}e_{\mu}^a\;,
 \eea
with the parameter given by
 \bea\label{param}
  (\eta\xi)^{\rho}=\eta^5\partial_5\xi^{\rho}\;, \qquad
  (\xi\eta)^5=\xi^{\rho}\partial_{\rho}\eta^5\;.
 \eea
The same formula holds for $A_{\mu}$ and $B_{\mu}$.
But, for $B_{\mu}$ one also has to check the closure of the shift
symmetries with spin-2 and here one finds
 \bea
  [\delta_{\xi},\delta_{\Lambda}]B_{\mu}=-\delta_{\tilde{\Lambda}}B_{\mu}
  -2\Lambda\partial_5\xi^{\rho}F_{\rho\mu}\;,
 \eea
where
 \bea\label{shiftclosure}
  \tilde{\Lambda}=\xi^{\rho}\partial_{\rho}\Lambda
  +2\Lambda\partial_5\xi^{\rho}A_{\rho}\;.
 \eea
Therefore the algebra closes only on-shell, i.e. if $F_{\mu\nu}=0$.

\subsection{Virasoro-covariantisation for scalars}\label{gaugediff}
To summarize the results of the last section,
we have seen that in the gauged phase the spin-2
transformations of sec. 3 are no longer a symmetry due to the substitution
of partial derivatives by covariant ones. Therefore
the spin-2 transformations have to be deformed by $g$-dependent
terms. For the topological fields we have seen that a Chern-Simons
formulation exists, which in turn yields modified spin-2
transformations, which are consistent by construction.

Next let us focus on the scalar fields. For them  we
have already noted the form of the covariant derivative
in (\ref{covariant}), and
the same formula holds for $\varphi$, but with the difference that
it also has to be covariant with respect to the local shift symmetries
gauged by $B_{\mu}$. The latter act as $\delta_{\Lambda}\varphi=-g\Lambda$,
i.e. the covariant derivative reads in $\theta$-notation
 \begin{equation}
  \mathcal{D}_{\mu}\varphi=\partial_{\mu}\varphi -A_{\mu}\partial_5\varphi
  -2\varphi\partial_5A_{\mu}+gB_{\mu}\;.
 \end{equation}

Altogether, replacing the partial derivatives in (\ref{ungauged})
by covariant ones and adding the Chern-Simons action constructed in the
last section as well as an explicit mass term which is known
to appear (see appendix \ref{reduc}), results in
 \bea\label{kkaction}
  \begin{split}
   S_{\text{KK}}=\int d^3 x d\theta \big[\varepsilon^{\mu\nu\rho}&
   \big(-e_{\mu}^a
   (D_{\nu}\omega_{\rho a}-D_{\rho}\omega_{\nu a}+\varepsilon_{abc}
   \omega_{\nu}^b\omega_{\rho}^c)
   -\frac{1}{2}gB_{\mu}F_{\nu\rho}\big) \\
   &+\frac{1}{2}eg^{\mu\nu}\phi^{-2}
   (D_{\mu}\phi D_{\nu}\phi+\mathcal{D}_{\mu}\varphi
   \mathcal{D}_{\nu}\varphi)+{\cal L}_{m}\big]\;.
  \end{split}
 \eea
Here we have determined the free parameter of the algebra
(\ref{bigalgebra}) to be $\alpha=2$
in order to get the correct Chern-Simons term for
$A_{\mu}$ and $B_{\mu}$ discussed in sec. 3.4.
There we have already observed
that varying this action with respect to $B_{\mu}$ one recovers
the duality relation (\ref{duality}).
In turn, the equations of motion for (\ref{kkaction}) and for the
Yang-Mills gauged action are equivalent.
This can be seen directly by imposing the gauge $\varphi = 0$
in (\ref{kkaction}) and then integrating out $B_{\mu}$, which
results exactly in the Kaluza-Klein action
containing the Yang-Mills term in (\ref{scalaract}).
Moreover, varying with respect to $\omega_{\mu}^a$ still implies
$T_{\mu\nu}^a=0$.
This shows that $\omega_{\mu}^a$
can be expressed in terms of $e_{\mu}^a$ as is standard, but with the
exception that all derivatives on $e_{\mu}^a$ are now
$\hat{v}$-covariant. In the second-order formulation this means that the
Einstein-Hilbert part looks formally the same as in sec.~\ref{dualgaugings},
but with all Christoffel symbols now containing $\hat{v}$-covariant
derivatives. This is on the other hand also what one gets by
direct Kaluza-Klein reduction in second-order form
\cite{Cho:1991xk,Aulakh:1985un}.
Thus we have shown,
that (\ref{kkaction}) is on-shell equivalent to the Kaluza-Klein
action which results from dimensional reduction.

In view of the fact that (\ref{kkaction}) is manifestly $\hat{v}$
and shift
invariant it remains the question how the spin-2 symmetries are realized.
As for the case of the topological fields,
also the $\sigma$-model action for the scalar fields will no longer
be invariant under the unmodified spin-2
transformations for the same reasons.
To find the deformed transformation rule for the scalars, one way
is to check the closure of the algebra. The unmodified spin-2
transformations do not build a closed algebra with the local $\hat{v}$
transformations. But, if we deform the
spin-2 transformation to
 \bea
  \delta_{\xi}\phi=\xi^{\rho}\partial_{\rho}\phi
  +2g\phi\partial_{\theta}\xi^{\rho}A_{\rho},
 \eea
the algebra closes according to
 \bea
  [\delta_{\xi},\delta_{\eta^5}]\phi=\delta_{(\eta\xi)}\phi
  -\delta_{(\xi\eta)^5}\phi\;,
 \eea
i.e.~exactly like in the case of the topological fields with
the parameters given in (\ref{param}).
The Kaluza-Klein transformations can therefore be entirely reconstructed by
requiring closure of the algebra.
The same transformation holds for the dual scalar~$\varphi$.

In the presence of matter fields we have to be careful about the
closure of the algebra also on the gauge fields.
Namely, for the pure Chern-Simons theory shift with spin-2 transformations
in (\ref{shiftclosure}) close on-shell  (as it
should), but for the theory constructed here the field strength does not
vanish.
Thus one way to get a closing algebra is to extend the
transformation rule according to
 \bea\label{Bmod}
  \delta^{\prime}B_{\mu}=-2\varphi\partial_{\theta}\xi^{\rho}F_{\mu\rho} \;,
 \eea
and all transformations close off-shell.

Therefore we see that in the full theory the transformation rules for the
vectors $A_{\mu}$ and $B_{\mu}$ get extended by scalar field dependent
terms. That is on the other hand also what we already know from the symmetry
variations in
(\ref{variations}) for $A_{\mu}^n$,
and these terms will be needed in order for
the full action to be spin-2 invariant. This is in complete analogy
to the construction of gauged supergravities, where the procedure
of gauging is only consistent with supersymmetry, if additional
couplings like mass terms are added, while the supersymmetry
variations are supplemented by scalar-dependent terms.
However, in the present case the invariance of the
Yang-Mills gauged Kaluza-Klein theory
is guaranteed by construction, which in turn implies that the
on-shell equivalent dual theory (\ref{kkaction}) is also invariant
(if one assumes transformation rules for $B_{\mu}$, which
are on-shell given by the variation of the left-hand side of
(\ref{duality})).
In view of our aim to construct the Kaluza-Klein theories for
more general backgrounds, it would however
be important to find a systematic
procedure to determine the scalar-dependent corrections for gaugings of
arbitrary diffeomorphism Lie algebras.
This we will leave for future work, but here let us
just show how the scalar-dependent correction in (\ref{Bmod})
ensures the invariance under spin-2 for a subsector.

For this it will be convenient to separate from the spin-2
transformations those
parts which represent already a symmetry for each term separately.
To do so we remember that to realize the
spin-2 transformations on the topological fields as gauge
transformations we had to switch on also the Virasoro transformations
with parameter $\xi^5=\xi^{\rho}A_{\rho}$. Now we will turn the logic
around and apply a spin-2 transformation followed by a Virasoro
transformation with parameter $\xi^5=-\xi^{\rho}A_{\rho}$.
Since Virasoro invariance is manifest, this is a symmetry if and only
if spin-2 is a symmetry. One may easily check that on $e_{\mu}^a$
and $\phi$ (as well as $\varphi$) this transformation is given by
 \bea\label{gaugeddiff}
  \begin{split}
   \delta_{\xi}\phi&=\xi^{\rho}D_{\rho}\phi\;, \\
   \delta_{\xi}e_{\mu}^a&=\xi^{\rho}D_{\rho}e_{\mu}^a
   +D_{\mu}\xi^{\rho}e_{\rho}^a\;.
  \end{split}
 \eea
Here we have used (\ref{ecov})
and also introduced a Virasoro covariant derivative
for the spin-2 transformation parameter (of which we may
think as transforming as
${\delta_{\eta^5}\xi^{\mu}=\eta^5 \partial_5\xi^{\mu}}$),
 \bea
  D_{\mu}\xi^{\rho}=\partial_{\mu}\xi^{\rho}-A_{\mu}\partial_5\xi^{\rho} \;.
 \eea
We see that we get transformation rules which look formally like
a diffeomorphism symmetry, except that all appearing derivatives
are $\hat{v}$-covariant. In the following we will refer to these
transformations as `gauged diffeomorphisms'.
In contrast, the gauge fields $A_{\mu}$ and $B_{\mu}$ transform as
 \bea\label{Adiff}
  \begin{split}
   \delta_{\xi}A_{\mu}&=\xi^{\rho}F_{\rho\mu}\, \\
   \delta_{\xi}B_{\mu}&=
   \xi^{\rho}D_{\rho}B_{\mu}+D_{\mu}\xi^{\rho}B_{\rho}
   +2\hspace{0.1em}e_{\mu}^a\partial_{\theta} \xi^{\rho}\omega_{\rho a}\;.
  \end{split}
 \eea

It remains the question whether actions can be constructed that are
manifestly invariant under these transformations.
To analyze this let us start with an action constructed from a scalar
Lagrangian given by
 \bea
  S=\int d^3 x d\theta \hspace{0.2em}e\hspace{0.2em} \mathcal{L}\;,
 \eea
and moreover being invariant under local Virasoro transformations.
Put differently, this means that the Lagrangian varies as
$\delta_{\xi^5}{\cal L}=\xi^5\partial_5{\cal L}-2{\cal L}\partial_5\xi^5$
under $\hat{v}$ (because then it transforms together with the
vielbein determinant, whose symmetry variation reads
$\delta_{\xi^5}e=\xi^5\partial_5 e +3e\partial_5\xi^5$,
into a total $\theta$-derivative).
By use of the $\hat{v}$-covariant derivative
given by
 \bea
  D_{\mu}\mathcal{L}=\partial_{\mu}{\cal L}-A_{\mu}\partial_5{\cal L}
  +2{\cal L}\partial_5 A_{\mu}\;,
 \eea
we can then evaluate the variation of the action under gauged
diffeomorphisms and find
 \bea
  \begin{split}
   \delta_{\xi} S &=\int d^3 x d\theta \left[(\xi^{\rho}D_{\rho}e
   +eD_{\rho}\xi^{\rho})\mathcal{L}+e\xi^{\rho}D_{\rho}\mathcal{L}\right]
   = \int d^3 x d\theta D_{\rho}(e\xi^{\rho}\mathcal{L})\\
   &=\int d^3 x d\theta\left[\partial_{\rho}(e\xi^{\rho}\mathcal{L})
   -\partial_5(e\xi^{\rho}A_{\rho}\mathcal{L})\right]=0\;.
  \end{split}
 \eea
Thus, if one constructs an action from a Lagrangian that
transforms as a scalar under gauged diffeomorphisms, then the action
is invariant under these gauged diffeomorphisms if and only if it
is also invariant under local Virasoro transformations.
The latter requirement is satisfied in our theory by construction.
Thus it remains to be checked whether the Lagrangian transforms as a
scalar. However, using (\ref{gaugeddiff}), (\ref{Adiff})
and $[D_{\mu},D_{\nu}]\phi=
-2\phi\partial_5 F_{\mu\nu}-\partial_5\phi F_{\mu\nu}$, one proves
that the covariant derivative $D_{\mu}\phi$
transforms under gauged diffeomorphisms as
 \bea\label{covdiff}
  \delta_{\xi}(D_{\mu}\phi)=D_{\mu}\xi^{\rho}D_{\rho}\phi
  +\xi^{\rho}D_{\rho}D_{\mu}\phi-2\phi\partial_5\xi^{\rho}F_{\rho\mu}\;,
 \eea
i.e. it does not transform like a one-form, but requires
an additional piece proportional to $F_{\mu\nu}$,
which again shows that corrections have
to be added to the transformation rules.

Let us consider the subsector of the theory where we rescale
 \bea
  e_{\mu}^a\rightarrow \kappa e_{\mu}^a\;, \qquad
  \varphi \rightarrow \kappa ^{-1/2}\varphi\;,
 \eea
and then take the limit $\kappa \rightarrow 0$.
The action then reads
 \bea\label{subsec}
  S_{\kappa\rightarrow 0}=\frac{1}{2}\int d^3 xd\theta\left(-g
  \varepsilon^{\mu\nu\rho}B_{\mu}F_{\nu\rho}+eg^{\mu\nu}\phi^{-2}
  {\cal D}_{\mu}\varphi{\cal D}_{\nu}\varphi\right)\;.
 \eea
However, in view of the fact that the term $\sim e_{\mu}^a
\partial_{\theta} \xi^{\rho}\omega_{\rho a}$ in (\ref{Adiff})
disappears in this limit
and with the additional contribution (\ref{Bmod}) in
the $B_{\mu}$ variation, the extra term in (\ref{covdiff}) is
canceled, and the kinetic term for $\varphi$ is therefore separately
invariant. The Chern-Simons term on the other hand transforms according
to (\ref{var}) as
  \bea
  \delta_{\xi}S_{\kappa\rightarrow 0}
  =-g\int\varepsilon^{\mu\nu\rho}F_{\sigma\mu}F_{\nu\rho}\varphi
  \partial_5\xi^{\sigma}=0\;,
 \eea
where we have used that a totally antisymmetric object in four indices
vanishes in $D=3$.
Thus we have shown that the scalar field modification in (\ref{Bmod})
is sufficient in order to restore the spin-2 invariance of this
subsector of the theory.

\section{Spin-2 symmetry for general matter fields}\label{spin2matter}
As we have discussed in the last section the gauging of global
symmetries requires a deformation of the spin-2 transformations,
which in turn induces a spin-2 mass term. This is in analogy to the
gauging of supergravity, where the spin-3/2 transformations
(i.e.~the supersymmetry variations) have to be modified by $g$-dependent
terms, which similarly induces gravitino mass terms. However, as we have seen
in sec.~\ref{sugragauging}, the gauging
of supergravity generically induces also a potential for the
scalar fields. Therefore one may wonder whether a similar phenomenon
can happen for spin-2 theories. In fact, so far we discussed only the theory
corresponding to a Kaluza-Klein reduction of the Einstein-Hilbert term.
For the gauging of more general theories -- exhibiting the
Kaluza-Klein action for general matter couplings already
in the higher-dimensional
theory -- we will however see that
a scalar potential naturally appears.

To show this let us discuss the simplest example of a scalar field
$\eta$ in $D=4$ coupled to gravity,
 \bea\label{simpmatter}
  {\cal L}=-ER + \frac{1}{2}Eg^{MN}\partial_M\eta\partial_N\eta\;.
 \eea
The ungauged phase of this theory simply consists of the
algebra-valued Einstein-Hilbert term in sec.~\ref{infspin-2} coupled
to a $\theta$-dependent (or algebra-valued) kinetic scalar term
of canonical form $\ft12\partial_{\mu}\eta\partial^{\mu}\eta$
(compare eq.~(\ref{scalar})).
This theory is invariant under the local spin-2 symmetries in
\ref{infspin-2}, and rigid Virasoro transformations, which act
on $\eta$ as
 \bea
  \delta_{\xi}\eta=\xi^{\rho}\partial_{\rho}\eta\;, \qquad
  \delta_{\xi^5}\eta=\xi^5\partial_{\theta}\eta\;.
 \eea
Thus $\eta$ transforms in the $\Delta =0$ representation of $\hat{v}$.
The gauging of $\hat{v}$ again requires a minimal substitution,
which reads in the given case
 \bea
  \partial_{\mu}\eta \longrightarrow D_{\mu}\eta =
  \partial_{\mu}\eta -gA_{\mu}\partial_{\theta}\eta\;.
 \eea
Under the gauged diffeomorphism of sec.~\ref{gaugediff} this
covariant derivative transforms as
 \bea
  \delta_{\xi}(D_{\mu}\eta)=\xi^{\rho}D_{\rho}D_{\mu}\eta
  +D_{\mu}\xi^{\rho}D_{\rho}\eta\;,
 \eea
which can be shown along the lines of (\ref{covdiff}). We see
that it transforms like a 1-form. As we have argued in the last
section the corresponding action just for $\eta$ will
therefore be invariant under the
gauged diffeomorphisms $\delta_{\xi}\eta=\xi^{\rho}
D_{\rho}\eta$. However, we know that the invariance of the
Einstein-Hilbert term requires an additional $\phi$-dependent
variation for $A_{\mu}$ (see (\ref{locomp})). This in turn will spoil
the invariance of the scalar couplings. Namely, under
$\delta^{\prime}_{\xi}A_{\mu}=-\phi^{-2}\partial_5\xi^{\rho}g_{\rho\mu}$
one has
 \bea\label{variation}
  \delta_{\xi}^{\prime}\left(eg^{\mu\nu}D_{\mu}\eta
  D_{\nu}\eta\right)=2e\phi^{-2}\partial_5\xi^{\rho}D_{\rho}\eta
  \partial_5\eta\;.
 \eea
To compensate for this one can add a scalar potential of the
form
 \bea\label{potential}
  eV(\phi,\eta)=e\phi^{-2}(\partial_5\eta)^2\;,
 \eea
which transforms as
 \bea
  \delta_{\xi}(eV)=\partial_{\rho}\left(e\xi^{\rho}\phi^{-2}
  (\partial_5\eta)^2\right)+2e\phi^{-2}\partial_5\xi^{\rho}D_{\rho}
  \eta\partial_5\eta\;,
 \eea
and therefore cancels (\ref{variation}) up to a total derivative.
Altogether we have shown
that the deformed spin-2 transformations require a scalar potential,
and the matter couplings in the gauged phase are given by
 \bea\label{gaugedsimpmatter}
  S[\phi,\eta]=\frac{1}{2}\int d^3xd\theta e\left(g^{\mu\nu}
  D_{\mu}\eta D_{\nu}\eta - V(\phi,\eta)\right)\;,
 \eea
with the potential in (\ref{potential}). One may also check
explicitly that the Kaluza-Klein reduction of (\ref{simpmatter}) leads exactly
to (\ref{gaugedsimpmatter}).

\section[Consistent truncations and extended supersymmetry]
{Consistent truncations and extended \\supersymmetry}\label{truncation}

After constructing the consistent gravity--spin-2 couplings appearing
in Kaluza-Klein theories in the broken and unbroken phase, one may ask
the following question:
Is a truncation to a finite subset of spin-2 fields possible?
To answer this let us first review in which sense the
truncation to the massless modes is justified.

Since in standard Kaluza-Klein compactifications the higher
Kaluza-Klein modes are much heavier than the zero-modes,
they can be integrated out in an effective description.
For this one usually assumes that this is equivalent to just
setting the massive modes equal to zero. Apart from the
question whether this is really the correct approach of
`integrating out' degrees of freedom, it is not guaranteed
that this is a consistent truncation in the Kaluza-Klein sense.
The latter requires that the truncated theory is still compatible
with the higher-dimensional equations of motion.
More precisely, this demands that
each solution of the truncated theory
can be lifted to a solution of the full theory.
To illustrate the latter, let us consider the massless theory
resulting from an $S^1$ compactification (see (\ref{zero-mode}) above).
Setting the dilaton $\phi$ to a constant (as was done originally
by Kaluza) implies by its equations of motion $F^{\mu\nu}F_{\mu\nu}=0$.
This is of course not consistent with a generic solution of
the Yang-Mills equations for $\phi=\text{const}$.
Thus the truncation of the dilaton is not consistent in the Kaluza-Klein
sense, even though it is perfectly consistent by itself (just describing
the usual Einstein-Maxwell system).
In contrast it is usually assumed that the truncation
of the massive modes is consistent. This can be seen directly
for compactifications on tori. For them the zero-modes are
simply characterized by the requirement of being independent of
the internal coordinates. Therefore the field equations and
symmetry variations in (\ref{variations}) do not mix zero-modes
with massive modes, and the truncation is consistent.
However, for compactifications on generic manifolds
this is a highly non-trivial statement, and in general
actually not true \cite{Duff:1986hr,Duff:1984hn}.
In fact, an explicit proof requires several elaborate field
redefinitions and a large number of miraculous identities have
to be satisfied
\cite{deWit:1983vq,deWit:1984nz}.\footnote{Such a proof
has been done explicitly
for the case of Kaluza-Klein reduction of 11-dimensional supergravity
on $AdS_4\times S^7$ in \cite{deWit:1986iy}, based on results in
\cite{deWit:1986mz}.
See also \cite{Nastase:1999kf,Nastase:1999cb,Cvetic:2000dm,Cvetic:2000eb}.}

Let us now comment on the question whether truncations
to a finite number of spin-2 fields might be possible.
First of all, such a
truncation is known not to be consistent in the strict
Kaluza-Klein sense \cite{Duff:1986hr}.
Nevertheless it has been
proposed in the early literature on Kaluza-Klein theories that
this truncation might still be consistent by itself, thus
providing a circumvention of the no-go theorem also for the
case of a finite number of spin-2 fields \cite{Nappi:1989ny}.
The situation has been
clarified in \cite{Duff:1989ea} and can be rephrased by use
of the analysis in \ref{infspin-2} as follows.
The truncation to a finite number
of spin-2 fields containing, say, all fields with level $|n| \leq N$
for fixed $N$, is not consistent in the Kaluza-Klein sense since
the corresponding subset of the Kac-Moody algebra is simply not a
subalgebra. This is actually just a different manifestation of the fact
that the naive truncation of the corresponding Chern-Simons theory
would allow for solutions that cannot be lifted to solutions of
the full theory. However, one might still hope to get a theory
which is consistent by itself. This would be the case if and only if
the subset resulting from the Kac-Moody algebra by setting all
generators with $|n| > N$ to zero would result in a consistent Lie
algebra (albeit not being a subalgebra). This turns out not to be
the case. Rather one finds that the Jacobi identity is violated, e.g.
 \bea\label{Jacobi}
  [[J_a^{(1)},J_b^{(1)} ],J_c^{(-1)} ]+[[J_c^{(-1)},J_a^{(1)} ],J_b^{(1)} ]
  +[[J_b^{(1)},J_c^{(-1)} ],J_a^{(1)} ] =2\eta_{c[a}J_{b]}^{(1)}\;.
 \eea
Correspondingly, the resulting Chern-Simons theory would be
inconsistent.\footnote{Note that the truncation to, e.g., $n=1$
does result in a consistent Lie algebra, on which, however, the
quadratic form (\ref{quadform}) degenerates. Instead there exists
an alternative invariant form, whose Chern-Simons theory leads to
the ghost-like gravity -- spin-2 coupled system discussed in
\cite{Hindawi:1995an,Wald:1986dw}.}
Moreover, (\ref{Jacobi}) shows that these inconsistencies appear
in the ungauged but also in the gauged phase, as can be seen from
(\ref{kacmoody}) and (\ref{bigalgebra}).
In analogy to the given reasoning it has been argued in
\cite{Duff:1989ea} that the
consistency problems related to Kaluza-Klein truncations are
basically the same as those related to `higher-spin' couplings.
Put differently, a `higher-spin' theory resulting from
Kaluza-Klein reduction represents a consistent Kaluza-Klein truncation
if and only if it is consistent by itself.

Let us now discuss the similar problem for massive spin-3/2 couplings.
In general the infinite tower of spin-3/2 fields appearing in Kaluza-Klein
theories would require also a Kac-Moody-like extension of a superalgebra
as symmetry group. The resulting field theory would than have an
infinite number of supercharges (${\cal N}=\infty$), among
which all but finitely many are spontaneously broken.\footnote{For
instance, the topological subsector of
this Kaluza-Klein supergravity could then presumably be constructed as
a Chern-Simons theory for the corresponding super-Kac-Moody algebra.}
Even though we will not analyze the structure of
these superextensions in this thesis, we can already draw some
conclusions from the theory constructed in sec.~\ref{spin32con}.
There we have seen that Kaluza-Klein supergravity on
$AdS_3\times S^3\times S^3$ consists of a tower of ${\cal N}=8$
supermultiplets (in accordance with the supersymmetry that is
preserved by the background), which contains two spin-3/2 multiplets.
While the multiplets containing fields up to spin-1
are described by gauged ${\cal N}=8$ supergravities,
each of the two spin-3/2 multiplets required already ${\cal N}=16$
supersymmetry. Thus the appropriate symmetry algebra will be some
infinite-dimensional extension of the ${\cal N}=8$ AdS superalgebra,
which contains two ${\cal N}=16$ superalgebras as consistent
truncations. Since it is natural to assume that the above reasoning
for spin-2 couplings also applies to spin-3/2 systems, we are lead
to expect that a truncation is consistent in the Kaluza-Klein
sense if it is consistent by itself.
Since the ${\cal N}=16$ theory defined in \ref{spin32con} is by
construction consistent, this can already be interpreted as evidence
that the corresponding truncation is consistent in the strict
Kaluza-Klein sense. This in turn would mean that the infinite-dimensional
extension of the ${\cal N}=8$ algebra contains at least two ${\cal N}=16$
superalgebras as (consistent) subalgebras.

\chapter{Applications for the AdS/CFT correspondence}\label{adschapter}

\section{The AdS/CFT dictionary}\label{adscft}

In the introduction we mentioned the AdS/CFT correspondence
as a realization of the holographic principle and emphasized
the importance of massive Kaluza-Klein modes. After discussing
the construction of effective actions for massive Kaluza-Klein states,
we are going to discuss potential applications for the AdS/CFT
correspondence.

First of all we have to explain the AdS/CFT duality more
precisely. Maldacena originally conjectured the correspondence by
considering a number $N$ of $D3$ branes, which are on the one hand
described by a $U(N)$ Born-Infeld gauge theory, and on the other hand
as a (solitonic) solution of supergravity. In the limit
$\alpha^{\prime}\rightarrow 0$ both descriptions leave free gravity
in the bulk together with ${\cal N}=4$ super-Yang-Mills theory in
$3+1$ dimensions on the one side and type IIB supergravity on
$AdS_5\times S^5$ (as the near-horizon limit) on the
other side \cite{Maldacena:1997re,Aharony:1999ti}.
Then, Maldacena concluded, both theories have to be equivalent.
However, what does it mean exactly that
these theories are `dual'? This has been clarified by
Witten in \cite{Witten:1998qj} and by Gubser, Klebanov and Polyakov in
\cite{Gubser:1998bc}, which we will briefly explain in the
following.

Let us first try to understand in which sense a Minkowski space
can be interpreted as the boundary of AdS.
As in \cite{Witten:1998qj} we are going to discuss for simplicity
reasons the duality in the case of euclidean AdS and Minkowski
spaces.
The $d+1$-dimensional AdS space can be defined as an open unit
ball $\sum_{i=0}^d y_i^2 <1$
in $\mathbb{R}^{d+1}$ (with coordinates $y_0,...,y_d$) with metric
 \bea\label{adsmetric}
  ds^2=\frac{4}{(1-|y|^2)^2} \sum_{i=0}^d dy_i^2\;.
 \eea
The boundary of this space, namely the sphere $S^d$ defined by
$\sum_{i=0}^dy_i^2=1$, can in turn be viewed as the conformal
compactification of euclidean $d$-dimensional Minkowski space
(where a point at infinity has been added).
However, the metric defined in (\ref{adsmetric}) does not induce a
metric on the boundary, since it becomes singular for $|y|=1$.
In order to get a metric on the boundary which is related to
(\ref{adsmetric}) one may pick a function $f$ on the closed
unit ball (i.e.~on the open ball together with its boundary),
which has a first order zero on the boundary (as, e.g.,
$f(y)=1-|y|^2$), and then consider the metric $d\tilde{s}^2=f^2ds^2$.
The latter extends to a well defined metric on the boundary.
But, there exists no natural choice for the function $f$ required
for defining this metric, and therefore the latter is only well-defined
up to conformal transformations. In fact, any rescaling of $f$ will induce
a conformal rescaling of the metric on the boundary. Thus, the
given metric on AdS defines a conformal structure on the boundary.

We have seen in which way the boundary of euclidean AdS
may be viewed as (the conformal compactification of) euclidean
Minkowski space. And moreover we have argued that a given metric
on AdS induces a conformal structure on the Minkowski space.
This is in agreement with the claim that a gravity
theory on AdS is dual to a conformal field theory on the
boundary. Let us now examine the question how exactly two such
theories might be related. As a first simple example we consider a
free massless scalar field on AdS, i.e. a field $\phi$
obeying the Laplace equation
${\nabla^{\mu}\nabla_{\mu}\phi =0}$.\footnote{The notion
of mass is actually more subtle on AdS than recognized by
this equation. For a more precise definition see the discussion
of $AdS$ representation in sec. \ref{kkspec}.}
It is a well-known fact that for a given function $\phi_0$
on the sphere there exists a unique solution of the Laplace
equation on the ball which reduces to $\phi_0$ on the boundary.
Thus in the case of a scalar field there exists a one-to-one
correspondence with functions $\phi_0$ on $S^d$ --
which will later be interpreted as sources in the CFT -- and
solutions of the Laplace equation on AdS. Similar results can be
derived for gauge fields and also for the metric itself.
For the latter the result is known as the Graham-Lee
theorem. In analogy to the discussion above it
states that any conformal structure on $S^d$
is induced by a unique metric on AdS solving the Einstein equations
with negative cosmological constant.
In total we can conclude that for all massless fields appearing
in a supergravity theory there exists a unique solution of their
equations of motion obeying a given set of boundary conditions,
which are in turn interpreted as the data of a conformal field
theory.

Let us note that in accordance with this picture the
symmetries on both sides of the duality match, since the
conformal group in $D$ dimensions, $SO(2,D)$, coincides with
the AdS isometry group in $D+1$ dimensions.
However, the $D=3$ case is exceptional as the $AdS_3$ group
$SL(2,\mathbb{R})\times SL(2,\mathbb{R})$ at the boundary
gets enhanced to two copies of the Virasoro algebra \cite{Brown:1986nw},
corresponding to the infinite-dimensional conformal
symmetry in two dimensions.

As a next step we have to understand how the dynamics on both
sides of the correspondence might be related.
As the boundary data $\phi_0$ should be
interpreted as sources in the CFT, there will be couplings to a
conformal field/operator $\mathcal{O}$ of the
form $\int_{S^d}\phi_0\mathcal{O}$.
The precise form of the correspondence, as developed in
\cite{Witten:1998qj,Gubser:1998bc}, now claims that
the correlation functions in the CFT are encoded in the
supergravity action $S$ via the relation\footnote{Of course,
this equation has to be suitably regularized, since generically
both sides are simply infinite \cite{Witten:1998qj}.
See, e.g., \cite{Bianchi:2001kw,Skenderis:2002wp} for the notion of
`holographic renormalization'.}
 \bea
  \vev{\exp \int_{S^d}\phi_0\mathcal{O}}_{\text{CFT}}
  =\exp (-S(\phi))\;.
 \eea
Here the supergravity action is evaluated on those solutions of
its equations of motion that obey the required boundary conditions.

So far we have discussed the AdS/CFT correspondence in case of
massless fields in the supergravity theory.
On the CFT side these fields correspond to sources $\phi_0$
that have conformal dimension zero. This is simply due to
the fact that the function $f$ defining the conformal class mentioned
above does not show up at all in the definition of $\phi_0$.
Since $\phi_0$ has conformal dimension zero, it follows that
the conformal dimension of the CFT field $\mathcal{O}$ is $d$.

Let us now turn to the case of a scalar field satisfying the
massive Klein-Gordon equation $(\nabla^{\mu}\nabla_{\mu}+m^2)\phi=0$.
For the analysis of this equation it is convenient to choose
a coordinate system in which we introduce $z$ according to
$|y|= \text{tanh}(\frac{z}{2})$ as a new coordinate.
The boundary at $|y|=1$ then corresponds to $z\rightarrow \infty$.
For large $z$ the Klein-Gordon equation can be written as
 \bea
  \left(-e^{-dz}\frac{d}{dz}e^{dz}\frac{d}{dz}+m^2\right)\phi=0\;.
 \eea
Making the ansatz $\phi \sim e^{(\Delta-d)z}$ this equation reduces to
 \bea\label{root}
  \Delta = \frac{d}{2}\pm \sqrt{\frac{d^2}{4} +m^2}\;,
 \eea
where the mass is given in units of the AdS length scale $L_0$.
Therefore, in the massless case there are the two independent
solutions $\phi\sim 1$ and $\phi \sim e^{-dz}$.
The existence of the constant solution is in turn the reason
for the existence of a unique solution satisfying the given
boundary conditions defined by $\phi_0$. (Roughly speaking,
in an expansion of $\phi$ into harmonics each partial wave yields
a given constant at infinity and adding these up one gets the unique
solution of the Laplace equation \cite{Witten:1998qj}.)
In contrast, the massive case yields two independent solutions of
the form $\phi \sim e^{\Delta_{\pm}z}$. Thus, one cannot find a solution
of the massive equations of motion that approaches a given constant
at infinity and so there exists no unique solution on AdS.
As for the definition of the conformal structure
on the boundary we may again take
a function $f$ that has a simple zero at the boundary.
Then one can look for solutions of the equations of motion
that behave instead like
 \bea\label{phi0}
  \phi \sim f^{d-\Delta}\phi_0\;.
 \eea
(An obvious example would be $f \sim e^{-z}$ which has a zero for
$z\rightarrow\infty$.)
The definition of $\phi_0$ in (\ref{phi0}) depends on the
choice of $f$. As we have seen above, a rescaling $f\rightarrow
e^w f$ induces a conformal transformation on the boundary.
Since $\phi$ cannot be affected by such a transformation,
$\phi_0$ has to transform at the same time according to
$\phi_0\rightarrow e^{w(\Delta -d)}\phi_0$. Thus we can conclude
that the associated conformal operator $\mathcal{O}$
has conformal dimension $\Delta$.
In summary, massive fields in the supergravity theory correspond
to fields in the CFT whose conformal dimension is given in terms
of the mass as a real root of (\ref{root}).\footnote{Stability of
supergravity on an Anti-de Sitter space does not require
$m^2$ to be strictly positive, but just bounded from below,
in accordance with the condition $m^2\geq -d^2/4$ implied
by the reality of (\ref{root}).
See \cite{Breitenlohner:1982jf,Breitenlohner:1982bm} and also
the discussion in \cite{Klebanov:1999tb}.}

So far we discussed the AdS/CFT correspondence in case of
an unbroken conformal symmetry on the CFT side. However, if
both theories are equivalent, the duality should also persist
in case that certain scalars get a vev and break some part of
the symmetry spontaneously \cite{Klebanov:1999tb}.
Even more, a duality is expected to hold also in case
of an explicit breaking of the conformal symmetry by the
addition of mass terms \cite{Freedman:1999gp}. Those deformations
take the form
 \bea
  {\cal L}^{\prime}={\cal L}_{\rm CFT}+ m {\cal O}\;,
 \eea
where ${\cal O}$ generically denotes some operator in the CFT.
Correspondingly, the conformal symmetry will be broken as well as
some supercharges and possibly some part of the gauge group.
Since the conformal transformations correspond to isometries
in the AdS bulk, a breaking of the conformal symmetry
will also lead to geometrical deformations of the AdS space.
Those deformations are given by so-called domain wall solutions,
which have a reduced isometry group \cite{Bianchi:2001de}.
As the dual field theory is no longer scale-invariant,
RG flows are possible. The domain-wall solutions provide
in turn an effective tool for the analysis of RG flows and thus of
more realistic aspects of field theories.

Apart from that the focus has more recently turned to
another type of generalized AdS/CFT duality.
These are the so-called marginal deformations, which have been
considered in an interesting work by Lunin and
Maldacena \cite{Lunin:2005jy}.
These deformations preserve the conformal symmetry completely,
but break some amount of supersymmetry and the gauge symmetry.
Therefore the gravity duals
are still AdS theories, but with a reduced internal symmetry.

In the following we are going to discuss those marginal
deformations for the gravity dual on $AdS_3\times S^3\times S^3$.
This is motivated by the fact that the dual conformal field theory
is much less understood than, e.g., in the $AdS_3\times S^3$ case.
To begin with, let us briefly summarize where the different versions of
the $AdS_3/CFT_2$ duality come from.
The $AdS_3\times S^3$ geometry is -- instead of the system of D3
branes mentioned at the beginning of this section -- realized as
the near-horizon limit of a system of parallel D1 and D5 branes.
The dual field theory is relatively well understood and
given by a non-linear $\sigma$-model,
whose target space is a symmetric product
orbifold {\sf Sym}$^{N}(M_4)$.
In contrast, the $AdS_3\times S^3\times S^3$ background arises
as the near-horizon geometry of the so-called double D1-D5 system.
Its dual field theory is required to  realize a \textit{large}
${\cal N}=4$ superconformal algebra. The latter
contains two instead of one affine $\widehat{SU(2)}$ subalgebras,
corresponding to the isometries on the second $S^3$.
Surprisingly, this complicates the direct determination of the dual
CFT. See \cite{Gukov:2004ym,Gukov:2004fh} for possible approaches to
this problem.
A better understanding of the gravity dual might be helpful
and so we turn now to the discussion of marginal deformations in
terms of the theories constructed in chapter~\ref{spin32chapter}.

In sec.~\ref{4,0} we discuss the effect of turning on some
scalars in the spin-3/2 multiplet, which will further break
half of the supercharges and some part of the gauge group.
Similarly we discuss in sec.~\ref{3,3} a marginal deformation in
the Yang-Mills multiplet, which will break the gauge group to
its diagonal.

\section{Marginal ${\cal N}=(4,0)$ deformations}\label{4,0}
In this section we will focus on marginal deformations in the
spin-3/2 multiplet on $AdS_3\times S^3\times S^3$, whose
effective Kaluza-Klein supergravity was constructed in
chapter~\ref{spin32chapter}.

There we have computed the scalar potential for the gauge
group singlets for the spin-3/2 multiplet $(\ft12,\ft12,\ft12,\ft12)_S$
on $AdS_3\times S^3\times S^3$ and have moreover shown that the
ground state with all scalars
having vanishing vacuum expectation value
breaks already half of the supersymmetry.
In order to study deformations of the dual CFT, it would be interesting
to analyze different ground states, which break further symmetries and
where in addition some of the scalar fields get a vev.
For this one has to compute and to minimize the scalar potential on
these fields.
In order to be sure that such a minimum of a truncated sector is also a
ground state of the full theory, the truncation has to include all scalars
that are singlets under a subgroup of the
symmetry group \cite{Warner:1983vz}.

Here this is the case, e.g., for the six scalars transforming as
$(0,0,1,0)\oplus (0,0,0,1)$ (in addition to the two singlets we have
already discussed), because these are singlets under $SO(4)_L$.
Let us give a vev to all of these six scalar fields.
Then in turn
we can use a $SO(3)^+\times SO(3)^-$ rotation to bring them into the
form\footnote{We will show explicitly that even in the case of a
non-linearly realized symmetry such a transformation always exists.}
 \begin{equation}\label{vev}
  \vev{\vec{\phi}_+}=\begin{pmatrix} \phi_3 \\ 0 \\ 0 \end{pmatrix}, \qquad
  \vev{\vec{\phi}_-}=\begin{pmatrix} \phi_6 \\ 0 \\ 0 \end{pmatrix}.
 \end{equation}
This ground state breaks spontaneously $SO(3)\times SO(3)\rightarrow
SO(2)\times SO(2)$, which implies that four of the gauge fields
will become massive in a Higgs effect. This in turn requires the
existence of four scalar fields that can act as Goldstone bosons.
For the choice (\ref{vev}) of the ground state these Goldstone bosons
are given by the fluctuations around (\ref{vev}) in the 2/3 component,
whereas the fluctuations around the 1-component describe the Higgs fields,
i.e. we get in total two Higgs fields and four Goldstone bosons,
exactly as required. Moreover, we expect the scalar potential not to depend
on the Goldstone bosons, i.e. on the 2/3 component.
But we could have chosen each component as the Higgs field initially,
while the other two would then act as Goldstone bosons,
not entering the scalar potential. This in turn implies that
the potential will also not depend on these scalars, in particular not
on the Higgs field, which therefore stays massless.
This one can also verify directly, by computing the $T$-tensor for
$\vec{\phi}_+$ and $\vec{\phi}_-$ and proceeding as in sec. \ref{sec:pot}.

\subsection{Non-linear realization of $SO(3)^+\times SO(3)^-$}

In the following we are going to analyze how the spontaneously
broken $SO(3)^+\times SO(3)^-$
symmetry is realized. For this we compute first the metric for
the $\sigma$-model scalar manifold, which is spanned by the eight
scalar fields that are singlets under $SO(4)_L$ --
denoted by $\phi_1$,...,$\phi_8$.
This can be done by computing the
corresponding $E_{8(8)}$-valued group element
${\cal V}$ and extracting from the
non-compact part $P_{\mu}^A$ of the current ${\cal V}^{-1}D_{\mu}
{\cal V}$ the metric via
 \begin{equation}
  \frac{1}{2}g_{ij}(\phi)D^{\mu}\phi^iD_{\mu}\phi^j
  = \frac{1}{4}P^{\mu A}P_{\mu}^A\;,
  \qquad i,j=1,...,8\;.
 \end{equation}
One finds a metric of the form
 \begin{equation}
  ds^2=15\left(d\phi_1^2 +d\phi_2^2 +2\cosh^2(\frac{\phi_1-\phi_2}{2})
  ds_6^2\right),
 \end{equation}
which is a warped product of a flat two-dimensional space with
the six-dimensional space given by
 \begin{eqnarray}\label{bigmetric}
   ds_6^2 &=& \cosh^2\phi_4\cosh^2\phi_5\cosh^2\phi_6\cosh^2\phi_7
   \cosh^2\phi_8 d\phi_3^2 \\ \nonumber
   &+&\cosh^2\phi_5\cosh^2\phi_6\cosh^2\phi_7\cosh^2\phi_8
   d\phi_4^2\\ \nonumber
   &+&\cosh^2\phi_6\cosh^2\phi_7\cosh^2\phi_8 d\phi_5^2 \\ \nonumber
   &+&\cosh^2\phi_7\cosh^2\phi_8 d\phi_6^2 \\ \nonumber
   &+&\cosh^2\phi_8 d\phi_7^2 \\ \nonumber
   &+&d\phi_8^2\;.
 \end{eqnarray}
We are going to show that this space is nothing else than the coset space
$SO(1,6)/SO(6)$ or in other words the euclidean six-dimensional
Anti-de Sitter space.
The latter can be defined as a hypersurface in a
seven-dimensional Minkowski space spanned by $X_0,...,X_6$ and carrying
the $SO(1,6)$ invariant metric
 \begin{equation}\label{mink}
  ds^2=-dX_0^2+dX_1^2+dX_2^2+dX_3^2+dX_4^2+dX_5^2+dX_6^2\;.
 \end{equation}
This six-dimensional hypersurface is given by
 \begin{equation}\label{hyper}
  -X_0^2+X_1^2+X_2^2+X_3^2+X_4^2+X_5^2+X_6^2=-1\;.
 \end{equation}
If we parametrize the surface (\ref{hyper}) by the following coordinates
 \begin{equation}\label{coordinates}
  \begin{split}
   X_0&=\cosh\psi\cosh\omega\cosh t \cosh\phi\cosh\theta\cosh\chi\; , \\
   X_1&=\cosh\psi\sinh\omega\cosh\phi\cosh\theta\cosh\chi\; , \\
   X_2&=\sinh\psi\cosh\phi\cosh\theta\cosh\chi\; ,  \\
   X_3&=\cosh\psi\cosh\omega\sinh t\cosh\phi\cosh\theta\cosh\chi\; ,\\
   X_4&=\sinh\phi\cosh\theta\cosh\chi\; , \quad
   X_5=\sinh\theta\cosh\chi\;, \quad X_6=\sinh\chi\; .
  \end{split}
 \end{equation}
and identify $\phi_3=t$, $\phi_4=\omega$, $\phi_5=\psi$, $\phi_6=\phi$,
 $\phi_7=\theta$ and $\phi_8=\chi$,
the metric which is induced by (\ref{mink}) on (\ref{hyper})
is given exactly by (\ref{bigmetric}).

This manifold contains two three-dimensional submanifolds,
which correspond to the two triplets in $(0,0,1,0)\oplus (0,0,0,1)$.
Namely, if we set $\phi=\theta=\chi=0$,
this defines a submanifold with metric
 \begin{equation}\label{smallmetric}
  ds_3^2=\cosh^2\phi_5\cosh^2\phi_4d\phi_3^2
  +\cosh^2\phi_5 d\phi_4^2+d\phi_5^2\; .
 \end{equation}
The latter is also a coset space, namely $SO(1,3)/SO(3)$, which can be
seen in complete analogy to the manifold considered above: If we define
a three-dimensional hypersurface in a four-dimensional Minkowski space,
which is parametrized by (\ref{coordinates}) with $\phi=\theta=\chi=0$,
the induced metric is given by (\ref{smallmetric}).
The same is true for the second triplet, i.e. the submanifold given
by $t=\omega=\psi=0$ is also $SO(1,3)/SO(3)$.

Now we can also examine how the $SO(3)^{\pm}$ acts as an isometry on
the scalar fields.\footnote{Note, that these scalar fields do not build
a linear representation of $SO(3)$.}
It is sufficient to discuss the case of one triplet of scalar fields,
i.e. the case where the coset space is reduced to $SO(1,3)/SO(3)$.
Then the isometry group is clearly given by $SO(1,3)$ and its
subgroup $SO(3)$ is characterized
by the requirement that $X_0$ remains invariant. This implies in
particular that the combination $\cosh\psi\cosh\omega\cosh t$
is invarariant under the non-linear action of $SO(3)$, as can be seen
from (\ref{coordinates}).\footnote{One
may also check explicitly that the action of the non-linear
$SO(3)$ coset space symmetries explained in sec.
\ref{gaugedsugra} leaves this combination invariant.}
Furthermore it is the only invariant, since it is equal to
$X_1^2+X_2^2+X_3^2+1$ and apart from the latter there are no other
independent $SO(3)$ invariants.
Similarly we can derive the following:
Making an $SO(3)$ transformation on $\vec{X}=(X_1,X_2,X_3)$ to bring it
into the form $\vec{X}=(0,0,|\vec{X}|)$ implies for the coordinates
of $SO(1,3)/SO(3)$
 \begin{equation}
  t=\text{Arsinh}\left(\frac{1}{\sqrt{60}}|\vec{X}|\right),
  \qquad \omega=0\;, \qquad \psi=0\;.
 \end{equation}
This in turn implies that it is sufficient to evaluate all expressions
only on one of the scalar fields, say $\phi_3=t$, because by use of the
$SO(3)$ symmetry the others can be set to zero (as one would expect for
Goldstone bosons). Afterwards all expressions containing only $t$
can be `covariantized' by use of the rule
 \begin{equation}
  t \longrightarrow \text{Arsinh}
  \sqrt{\cosh^2\psi\cosh^2\omega\cosh^2t-1}\;.
 \end{equation}
In the following we will therefore evaluate all expressions only for
the case ${\phi_4=\phi_5=0}$ and similarly for the second triplet.

\subsection{Resulting ${\cal N}=(4,0)$ spectrum}

That the $SO(4)_R$ symmetry is broken implies that also
some amount of the
supersymmetry of the right factor of the supergroup will be broken,
because ${\cal N}=4$ supersymmetry in the AdS background
is not consistent without the
required internal symmetries. To determine the amount of unbroken
supersymmetry, we use that the number of solutions of
$\vev{\delta\psi_{\mu}}=0$ in the Anti-de Sitter background is
given by the number of eigenvalues of $A_1$ that are equal to
$\pm\frac{1}{2}$. One finds
 \bea
  \label{A1values}
   \ft32+2h_{+} &&(\#2)\nonumber\\
   \ft32+2h_{-}&&(\#2)\nonumber\\
   -\ft12-2h_{+}&&(\#2)\nonumber\\
   -\ft12-2h_{-}&&(\#2)\nonumber\\
   -\ft32 && (\#4) \nonumber\\
   \ft12 && (\#4)
   \;,
 \eea
with
 \bea
  h_{\pm}&=& -\frac12+\frac{1}{2\sqrt{1+\alpha^{2}}}\,
  \Big[\cosh^{2}\hat{\phi}_3
  +\alpha^{2}\cosh^{2}{\phi}_6
   \pm2\alpha\sinh\hat{\phi}_3\sinh\phi_6\Big]^{\ft12}\;.
  \label{hpm}
 \eea
We have also made a field redefinition given by
 \begin{equation}
  \sinh\hat{\phi}_3=\sinh\phi_3\cosh\phi_6\;.
 \end{equation}

This implies that all supercharges of the right factor are spontaneously
broken, i.e. the whole supergroup is broken to
 \begin{equation}
  D^1(2,1|\alpha)_L\times D^1(2,1|\alpha)_R\rightarrow
  D^1(2,1|\alpha)_L\times
  SL(2,\mathbb{R})_R\times SO(2)_R^+\times SO(2)_R^-\;,
 \end{equation}
and the theory has a residual ${\cal N}=(4,0)$ supersymmetry.

We will now turn to the question, how the supermultiplets rearrange under
the reduced symmetry. So far everything has been expressed in terms of
short multiplets of $D^1(2,1|\alpha)$. But due to the fact that some
additional fields get massive we also have to expect the appearance
of long multiplets. The general structure of these multiplets in terms
of their highest-weight states $h_0$ and $h_0=\ft12$
is given in tab.~\ref{longtable}
 \begin{table}[b]
 \centering
  \begin{tabular}{c|c}
   $h$ & $(\ell^+,\ell^-)$ \\ \hline
   $h_0$ & $(0,0)$ \\
   $h_0+\ft12$ & $(\ft12,\ft12)$\\
   $h_0+1$ & $(1,0)\oplus (0,1)$ \\
   $h_0+\ft32$ & $(\ft12,\ft12)$  \\
   $h_0+2$ & $(0,0)$
  \end{tabular} 
  \qquad \qquad \quad
  \begin{tabular}{c|c}
   $h$ & $(\ell^+,\ell^-)$ \\ \hline
   $\ft12$ & $(\ft12,\ft12)$ \\
   $1$ & $(1,0)\oplus(0,1)\oplus (0,0)$\\
   $\ft32$ & $(\ft12,\ft12)$ \\
   $2$ & $(0,0)$
  \end{tabular}\caption{Long and short
   ${\cal N}=(4,0)$ multiplets}\label{longtable}
 \end{table}
Let us denote them by $[h_0]_l$ and $[0]_s$, respectively.
Due to the fact that no supersymmetry of the right factor survives,
the $16^2=256$ degrees of freedom will assemble into $16$ supermultiplets,
each of them containing $16$ degrees of freedom.
The precise form of the resulting
${\cal N}=(4,0)$ multiplets, in particular the shifted values for the
conformal dimensions, can be extracted from the mass spectrum of the spin-3/2
and vector fields. The former we have already computed in (\ref{A1values}),
while the latter can be extracted from the eigenvalues of the non-compact
part of the T-tensor. Altogether one finds that the spin-$\ft32$
multiplet $(\ft12,\ft12,\ft12,\ft12)_S$ decomposes under ${\cal N}=(4,4)
\rightarrow {\cal N}=(4,0)$ into
 \begin{equation}
  \begin{split}
   (\ft12,\ft12,\ft12,\ft12)_S\rightarrow 2[h^+]_l\otimes (h^+ +\ft12)
   &\Oplus 2[h^-]_l\otimes (h^- +\ft12) \\
   \Oplus 2[h^+]_l\otimes (h^+ +\ft32)
   &\Oplus 2[h^-]_l\otimes (h^- +\ft32) \\
   \Oplus 2[h^3]_l\otimes (h^3+1)
   &\Oplus 2[h^4]_l\otimes (h^4+1) \\
   \Oplus [0]_s\otimes (2)
   &\Oplus 3[0]_s\otimes (1)\;,
  \end{split}
 \end{equation}
where each supermultiplet of $D^1(2,1|\alpha)_L$ is tensored with a
representation of $SL(2,\mathbb{R})$, characterized by its conformal
dimension $h$.
Here the multiplets in the first two lines are massive spin-3/2 multiplets,
containing one gravitino, 4 vectors, 7 spin-1/2 fermions and 4
scalars. The multiplets in the third and fourth line are massive spin-1
multiplets, containing 2 vectors, 8 spin-1/2 fermions and 6 scalars.
Finally, the multiplets in the last line are a massive spin-3/2
multiplet, containing four gravitinos, 7 vectors, 4 spin-1/2 fermions
and one scalar, and three massive spin-1 multiplets, each of them
containing one vector, 8 spin-1/2 fermions and 7 scalars.
Altogether the field content is given apart from the supergravity multiplet
by 12 massive spin-3/2 fields, 50 massive vectors, 116 spin-1/2 fermions
and 78 scalars. This is exactly the field content expected from the
original multiplet $(\ft12,\ft12,\ft12,\ft12)_S$, if one takes into
account that 4 scalars are eaten by some vectors, whereas 4 fermions
get eaten by the gravitinos.
Explicitly, the conformal dimensions are given by
(\ref{hpm}) and moreover by
\begin{equation}
 \begin{split}
    h^3&=-\frac{1}{2}+\frac{1}{2}\sqrt{1+16\sinh^2(\hat{\phi}_3)
  \frac{\alpha^2}{1+\alpha^2}}\;, \\
  h^4&=-\frac{1}{2}+\frac{1}{2}\sqrt{1+16\sinh^2(\phi_6)
  \frac{1}{1+\alpha^2}}\;,
 \end{split}
\end{equation}

With the explicit form of the ${\cal N}=(4,0)$ supermultiplets at hand
we can also check another symmetry, namely the discrete symmetry which
interchanges the two spheres $S^{\pm}$, or in other words which exchanges
the two scalar field triplets. For the ratio $\alpha$ of the two spheres radii
this symmetry acts as $\alpha\rightarrow 1/\alpha$, and we have to
check whether the spectrum reflects this symmetry -- as it should, since
$D^1(2,1|\alpha)\cong D^1(2,1|\frac{1}{\alpha})$.
Under this symmetry one has
 \begin{equation}
  \frac{1}{1+\alpha^2} \rightarrow \frac{\alpha^2}{1+\alpha^2}\;, \qquad
  \frac{2\alpha}{1+\alpha^2}\rightarrow \frac{2\alpha}{1+\alpha^2}\;.
 \end{equation}
We can therefore conclude that this discrete symmetry acts on the
scalar fields as
 \begin{equation}\label{discaction}
  \phi_6\rightarrow -\hat{\phi}_3\;, \quad \hat{\phi}_3\rightarrow \phi_6\;,
  \qquad\text{ or } \qquad
  \sinh\phi_6\rightarrow -\sinh\phi_3\cosh\phi_6\;,
 \end{equation}
because then the conformal dimensions transform into each other,
 \begin{equation}
  h^+ \rightarrow h^-, \qquad h^3\rightarrow h^4,
 \end{equation}
such that the spectrum is invariant.
Moreover, it is also possible to describe this symmetry in terms of
the $SO(6)\subset SO(1,6)$ isometries, which are realized on the full coset
space $SO(1,6)/SO(6)$. Namely, for the `embedding coordinates' $X_1$, ...,
$X_6$ the transformation (\ref{discaction}) implies
 \begin{equation}
  X_1^2+X_2^3+X_3^2=X_4^{\prime 2}+X_5^{\prime 2}+X_6^{\prime 3}\;,
 \end{equation}
where the primed coordinates correspond to the transformed scalar fields.
After fixing the $SO(3)^+\times SO(3)^-$ symmetry -- which we have done
by rotating everything into the $\phi_3$, $\phi_6$ - direction --
this specifies the transformation which rotates both vectors $X^+$ and $X^-$
into each other,
or to be more precise, $X_3\rightarrow X_4$ and $X_4\rightarrow -X_3$.

\subsection{Lifting the deformation to $D=10$}\label{lifting}

Finally we will discuss the question, whether the considered deformation
corresponds to a higher-dimensional geometry, or in other words whether
the deformed theory can also be obtained from a Kaluza-Klein reduction
on a deformation of $AdS_3\times S^3\times S^3$.
Since, as we argued in sec.~\ref{truncation},
there is some evidence that the spin-3/2
multiplet describes a consistent truncation, such a 10-dimensional
solution has to exist. Constructing this solution would
actually yield further evidence for the consistency of the
truncation.

The deformed background geometry can be identified through its isometries.
Namely, due to the fact that the gauge group is broken from
$SO(4)_L\times SO(3)_R^+\times SO(3)_R^-$ to $SO(4)_L\times
SO(2)_R^+\times SO(2)_R^-$,
the isometry group will be reduced similarly.
Put differently, both 3-spheres will be replaced by a three-dimensional
manifold whose isometry group is $SO(3)\times SO(2)$
and which is a smooth deformation
of a $S^3$. Such a geometry indeed exists and is given by a
so-called `squashed' sphere \cite{Duff:1986hr,Awada:1982pk,Duff:1983nu}.
If the latter has radius $R$ its line element is given by
 \begin{equation}\label{squashed}
  \begin{split}
   ds^2&=\frac{R^2}{4}\left[\sigma_1^2+\sigma_2^2+
   \frac{1}{1+q}\sigma_3^2\right] \\
   &=\frac{R^2}{4}\left[d\theta^2+\sin^2\theta d\phi^2+\frac{1}{1+q}
   \left(d\psi^2+2\cos\theta d\psi d\phi+\cos^2\theta d\phi^2\right)\right],
  \end{split}
 \end{equation}
where $\sigma_i$ are the left-invariant one forms on $S^3$
and $q\in (-1,\infty)$ is the `squashing' parameter.
This geometry is by definition invariant under the left $SO(3)_L$, but
breaks the right $SO(3)_R$ to $SO(2)_R$ for $q\neq 0$.
Thus the isometry group is $SO(3)_L\times SO(2)_R$ for both
spheres and the case $q=0$ corresponds to a undeformed 3-sphere.
This manifold is topologically
still a 3-sphere, but with a squashed $S^1$ fibration over $S^2$.
In order to show that this geometry arises as a solution of
type IIB supergravity, one would have to define a similar deformation
of the 3-form flux, which gives rise to (\ref{squashed}) via the
Einstein equations. We will leave this for future work.

\section{Marginal ${\cal N}=(3,3)$ deformations}\label{3,3}
In this section we are going to discuss marginal deformations, which
result in a partial supersymmetry breaking
${\cal N}=(4,4)\rightarrow {\cal N}=(3,3)$.\footnote{This
section is based on work done with
Marcus Berg and Henning Samtleben \cite{Berg}.}
These explorations are motivated by a system of intersecting
D-branes, which has been considered in \cite{Gukov:2004ym}.
More specifically, the original $AdS_3\times S^3\times S^3$
background is the near horizon limit of a double D1/D5 system,
containing in particular a D5/D5' system of branes.
In case that the number of D5 and D5' branes are equal, $Q_5^+ = Q_5^-$,
such a configuration can be deformed via joining the D5-
and D5'-branes into a single set of $Q_5$ D5-branes along a
4-manifold. This manifold is in turn characterized by a parameter $\rho$,
of which we therefore may think as a deformation parameter,
describing the deformation away from the original background.
The deformed system breaks the $SO(4)\times SO(4)$ to a diagonal
subgroup \cite{Gukov:2004ym}.

In the effective supergravities such a deformation corresponds to
giving a vev to certain scalars in such a way that the gauge
symmetry gets spontaneously broken (i.e.~here to the diagonal
subgroup). If the required scalars are contained in one of the
lowest multiplets of chapter \ref{spin32chapter},
this spontaneous symmetry breaking should be visible within one
of the effective supergravities discussed there.
In the following we are going to argue that the deformation
considered in \cite{Gukov:2004ym} can indeed by seen in one of
the YM multiplets.

\subsection{Deformations in the Yang-Mills multiplet}
We have to identify a deformation in the YM multiplets that
breaks the gauge group $SO(4)\times SO(4)$ to a diagonal subgroup,
breaking the supersymmetry at the same time as
${\cal N}=(4,4)\rightarrow {\cal N}=(3,3)$.
More specifically, we consider a breaking of the gauge group
$SO(4)_L\times SO(4)_R$
to $SO(3)^{(D)}_L\times SO(3)^{(D)}_R$, where $SO(3)^{(D)}_L$ and
$SO(3)^{(D)}_R$ represent the diagonal of the two factors in $SO(4)_L$
and $SO(4)_R$, respectively.  Moreover, we set $\alpha=1$ in the
following.

One sees from tab. \ref{spinspin}
that under this subgroup each YM multiplet contains two
scalar singlets, i.e. we have a four-dimensional manifold of scalars
invariant under $SO(3)^{(D)}_L\times SO(3)^{(D)}_R$.
At the origin, these scalars come
in two pairs with square masses $0$ and $3$,
i.e.~they correspond to
operators of conformal dimensions $2$ and $3$.  In particular, there
are two marginal operators.

\begin{figure}[tb]
  \begin{center}
  \epsfxsize=65mm
  \epsfysize=45mm
  \epsfbox{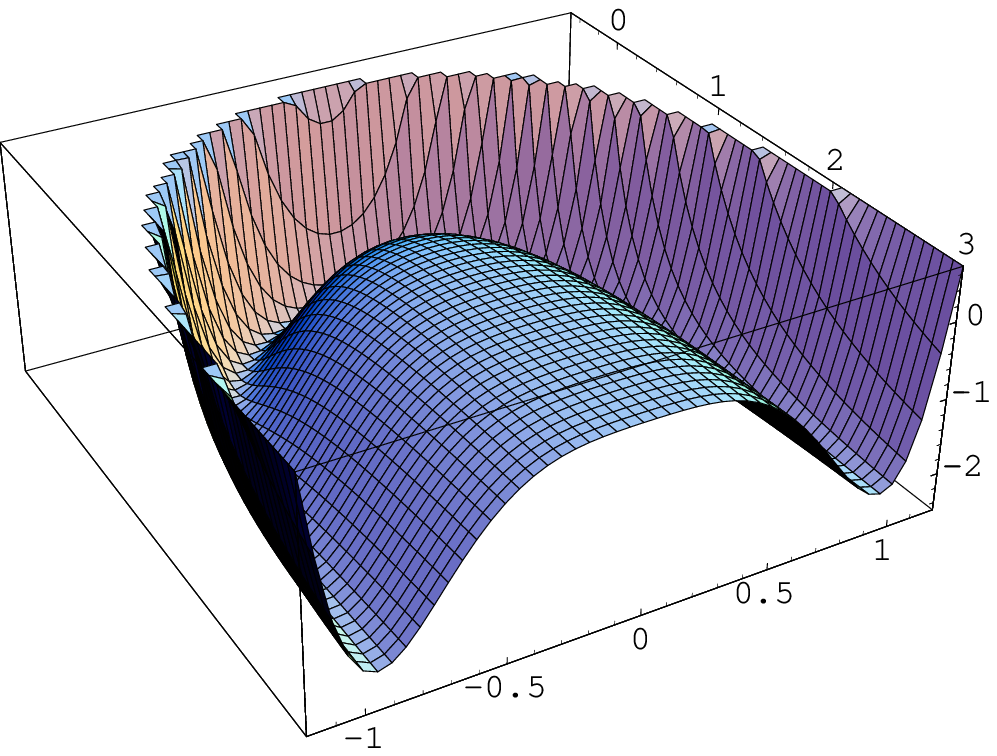}
  \qquad
  \epsfxsize=65mm
  \epsfysize=45mm
  \epsfbox{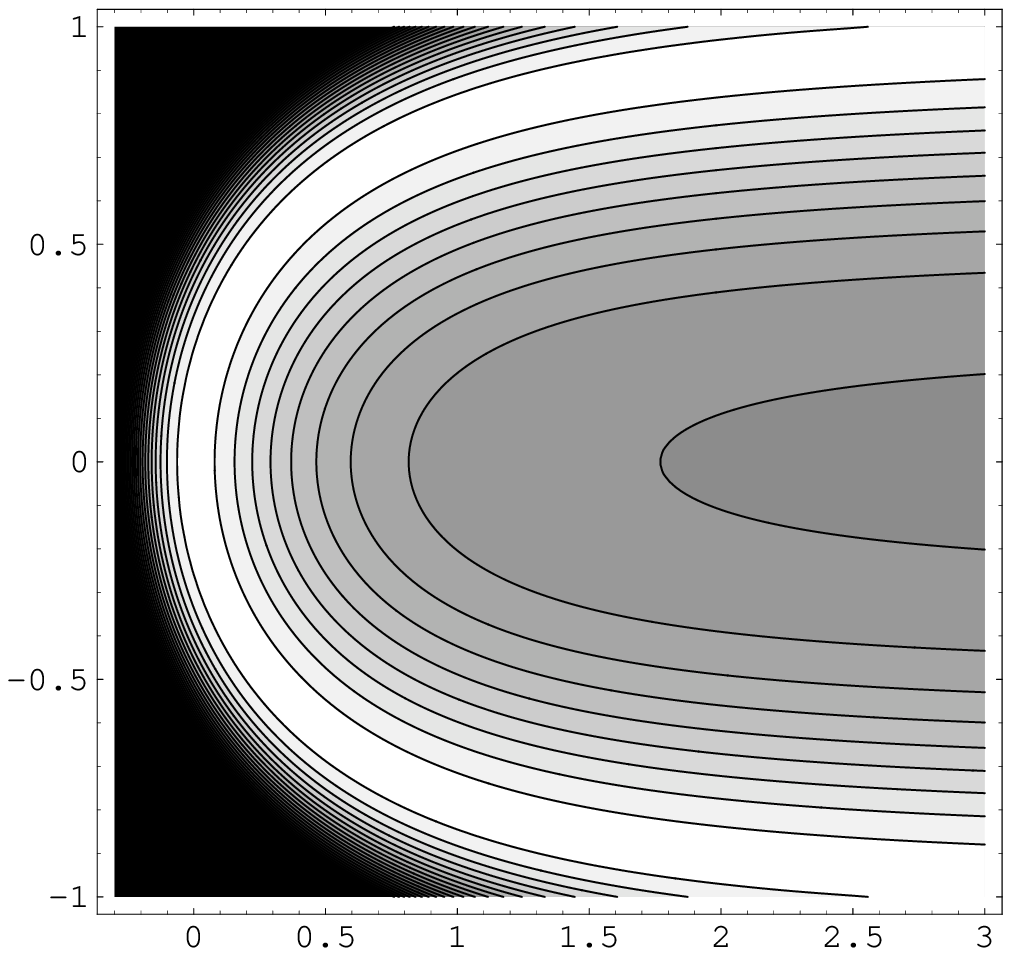}
  \end{center}
  \caption{{\small Flat valley in the potential, the ${\cal N}=(4,4)$
  origin is located at $(0,0)$. Coordinates here are
  $x=\sin\phi\sinh z\,,
  y=\cos\phi\sinh z$.}}
  \label{potentials}
 \end{figure}

Let us now consider the truncation of the Lagrangian
on this four-dimensional target space manifold.
The effective action was given by a ${\cal N}=8$ supergravity,
and the scalar fields take values in the coset space
$SO(8,8)/SO(8)\times SO(8)$.
Accordingly we can parametrize them by a $SO(8,8)$ matrix ${\cal S}$ as
 \bea
  {\cal S}&=& \exp \left(
  \begin{array}{cccc}
  0&0&v_{1}&w_{2}\\
  0&0&w_{1}&v_{2}\\
  v_{1}&w_{1}&0&0\\
  w_{2}&v_{2}&0&0
  \end{array}\right) \;,
  \label{4dsubspace}
 \eea
where each entry represents a multiple of the
$4\times4$ unit matrix.
In particular, the two gauge group singlets are parametrized
by $v_{1}$, $v_{2}$, while the truncation to a single YM multiplet
is given by $w_{1}=w_{2}=0$.
In the following, we further truncate to the two-dimensional
subspace defined by $v_{1}=v_{2}$, $w_{1}=-w_{2}$.
Similar to the analysis above, Lagrangian and scalar potential
can now be computed.
In terms of the new variables
 \bea
  z^{2}=v_{1}^{2}+w_{1}^{2}\;, \quad
  \phi=\arctan (w_{1}/v_{1})\;,
 \eea
the Lagrangian reads
 \bea
  e^{-1}{\cal L}=\partial_{\mu}z\,\partial^{\mu}z+
  \sinh^{2}\!z\,\partial_{\mu}\phi\,\partial^{\mu}\phi -V \;,
 \eea
with the scalar potential
 \bea
  V=-2+8\sinh^2\!z \, (\sinh\!z-\cos\phi \cosh\!z)^2\,
  (1 + 2\cosh2z- 2\cos\phi\sinh2z)\;.
 \eea
This scalar potential is bounded from below by $V\ge-2$
and takes this value along the curve
 \bea
  z={\rm arctanh}(\cos\phi) \;,
 \eea
which thus constitutes a flat direction in the potential,
see Figure~\ref{potentials}.
Explicit computation shows that this extends to a flat direction
in the four-dimensional target space~(\ref{4dsubspace})
and thus of the full scalar potential.
This deformation breaks the gauge
group down to $SO(3)^{(D)}_L\times SO(3)^{(D)}_R$.

\subsection{Resulting ${\cal N}=(3,3)$ spectrum}
Let us now turn to the analysis of the residual supersymmetries.
As above they can be extracted from the gravitino mass spectrum.
This is determined by the eigenvalues of the tensor $A_{1}$,
for which one finds
 \bea
  m_{i}=\pm\ft12\quad (\#3)\;,
  \qquad m_{i}=\pm\ft12\sqrt{8\cosh2z -7}\quad (\#1)\;.
 \eea
Supersymmetry is thus broken from ${\cal N}=(4,4)$ down
to ${\cal N}=(3,3)$.
For the corresponding conformal dimensions $\Delta=\ft12+|m|$
this yields
 \bea
  \Delta_{\rm gravitino} &=&
  \{ \ft32, \ft32, \ft32, \ft32, \ft32, \ft32, 1+\ft12\sqrt{8\cosh2z -7},
  1+\ft12\sqrt{8\cosh2z -7}\} \;.
 \eea

In order to determine the reorganization into ${\cal N}=(3,3)$
supermultiplets, we also have to compute some bosonic masses.
In the present case it is actually possible to diagonalize the full
scalar potential around the deformation to get the scalar masses.
One finds
 \bea
  m_{i}^{2}&=&\left\{
  \begin{array}{ll}
  -1 & (\#  9)\\
  0 & (\# 34)\\
  3 & (\# 1)\\
  \ft12\,(-1+4\cosh2z - 3\sqrt{8\cosh2z -7}) & (\# 1)\\
  \ft12\,(-5+4\cosh2z -\sqrt{8\cosh2z -7}) & (\# 9)\\
  \ft12\,(-5+4\cosh2z +\sqrt{8\cosh2z -7}) & (\# 9)\\
  \ft12\,(-1+4\cosh2z - 3\sqrt{8\cosh2z -7}) & (\# 1)
  \end{array}\right.
 \eea
For the associated conformal dimensions
$\Delta=1+\sqrt{1+m^{2}}$ this implies
 \bea
  \Delta_{i}&=&\left\{
  \begin{array}{ll}
  1 & (\#  9)\\
  2 & (\# 34)\\
  3 & (\# 1)\\
  \ft12\,(-1+\sqrt{8\cosh2z -7}) & (\# 1)\\
  \ft12\,(1+\sqrt{8\cosh2z -7}) & (\# 9)\\
  \ft12\,(3+\sqrt{8\cosh2z -7}) & (\# 9)\\
  \ft12\,(5+\sqrt{8\cosh2z -7}) & (\# 1)
  \end{array}\right.
  \label{scalardeltas}
 \eea
>From these values we can infer the entire spectrum
in terms of ${\cal N}=(3,3)$ supermultiplets.
The spectrum is organized under
the supergroup $O\!Sp(3|2,\mathbb{R})_{\rm L}\otimes
O\!Sp(3|2,\mathbb{R})_{\rm R}$, whose supermultiplets we have
to shortly review in the following.

A short $O\!Sp(3|2,\mathbb{R})$ supermultiplet is defined by its highest
weight state $(\ell)^{h_{0}}$, where $\ell$ labels the $SO(3)$ spin
and $h_{0}=\ell/2$ is the charge under the Cartan subgroup
~$SO(1,1)\subset SL(2,\mathbb{R})$.
The corresponding supermultiplet, which we will denote by
$(\ell)_{\rm S}$, is
generated from the highest weight state by the action of
two out of the three supercharges and carries $8\ell$ degrees of freedom.
Its $SO(3)^{\pm}$ representation content is summarized in Table~\ref{short3}.

\begin{table}[b]
 \centering
  \begin{tabular}{c | c  }
  $h$ & \\
  \hline
   $h_{0}$ &  $(\ell)$  \\[.5ex]
   $h_{0}+\frac{1}{2}$ & $(\ell)+(\ell\!-1\!)$  \\[.5ex]
   $h_{0}+1$  & $(\ell\!-1\!)$
  \end{tabular}
  \caption{\small The generic
  short supermultiplet $(\ell)_{\rm S}$ of $O\!Sp(3|2,\mathbb{R})$,
  with  $h_{0}=\ell/2$.}
  \label{short3}
\end{table}

The generic long multiplet $(\ell)_{\rm long}$
instead is built from the action
of all three supercharges
on the highest weight state and correspondingly carries
$8(2\ell+1)$ degrees of freedom.
Its highest weight state satisfies the unitarity bound
$h\ge\ell/2$.
In case this bound is saturated, the long multiplet decomposes
into two short multiplets (\ref{short3}) according to
\bea
 (\ell)_{\rm long} &=&
 (\ell)_{\rm S} \oplus (\ell\!+\!1)_{\rm S} \;.
 \label{long3}
\eea
 A semishort ${\cal N}=4$ multiplet $(\ell^+,\ell^-)_{{\rm S}}$
breaks according to
 \bea
  (\ell^+,\ell^-)_{{\rm S}} &=&
  (\ell^+\!+\ell^-)_{{\rm S}} +
  (\ell^+\!+\ell^-\!\!-\!1)_{{\rm long}} + ~\dots~ +
  (|\ell^+\!-\ell^-|)_{{\rm long}}\;,
 \eea
into semishort and long ${\cal N}=3$ multiplets.

The ${\cal N}=(3,3)$ spectrum can now be summarized as follows.
Let us first note, that the two YM-multiplets
$(0,1;0,1)_{\rm S}$ and $(1,0;1,0)_{\rm S}$ reduce to
the same short ${\cal N}=(3,3)$ supermultiplet~$(1;1)_{\rm S}$,
given in table~\ref{N3}.
Comparing the conformal dimensions~(\ref{scalardeltas}) to table~\ref{N3}
we observe that along the deformation a linear combination of the two
YM multiplets remains in the short multiplet~$(1;1)_{\rm S}$, whereas the
other fields combine into the long massive ${\cal N}=(3,3)$
supermultiplet~$(0;0)_{\rm long}$ with $h=\ft14\,(-1+\sqrt{8\cosh2z -7})$
summarized in table~\ref{N3long}. In total, the two
${\cal N}=(4,4)$ YM multiplets decompose into
${\cal N}=(3,3)$ multiplets according to
 \bea
  (0,1;0,1)_{\rm S}\Oplus (1,0;1,0)_{\rm S}\longrightarrow
  (1,1)_{\rm S}\Oplus (0,0)_{\rm long}\;.
 \eea
\begin{table}[bt]
\centering
  \begin{tabular}{c||c|c|c|}
   \raisebox{-1.25ex}{$h_L$} \raisebox{1.25ex}{$h_R$} &
   $\frac{1}{2}$ & $1$
   & $\frac{3}{2}$
   \rule[-2ex]{0pt}{5.5ex}\\
    \hline\hline
   $\frac{1}{2}$ & $(1;1)$  & $(1;0)+(1;1)$ & $(1;0)$
   \rule[-1.5ex]{0pt}{4ex}\\
    \hline
   $1$ &  $(0;1)+(1;1)$
   & $(0;0)+(0;1)+(1;0)+(1;1)$ &
   $(0;0)+(1;0)$  \rule[-1.5ex]{0pt}{4ex} \\
     \hline
   $\frac{3}{2}$ &  $(0;1)$ & $(0;0)+(0;1)$
   & $(0;0)$ \rule[-1.5ex]{0pt}{4ex} \\
  \hline
  \end{tabular}
      \caption{\small The short ${\cal N}=(3,3)$ multiplet $(1;1)_{\rm S}$.}
\label{N3}
\end{table}
\begin{table}[bt]
\centering
  \begin{tabular}{c||c|c|c|c|}
   \raisebox{-1.25ex}{$h_L$} \raisebox{1.25ex}{$h_R$} &
   $h$ & $h+\frac12$ & $h+1$ & $h+\frac{3}{2}$
   \rule[-2ex]{0pt}{5.5ex}\\
    \hline\hline
   $h$ & $(0;0)$  & $(0;1)$ & $(0;1)$ & $(0;0)$
   \rule[-1.5ex]{0pt}{4ex}\\
    \hline
   $h+\frac{1}{2}$ & $(1;0)$  & $(1;1)$ & $(1;1)$ & $(1;0)$
   \rule[-1.5ex]{0pt}{4ex}\\
    \hline
   $h+1$ & $(1;0)$  & $(1;1)$ & $(1;1)$ & $(1;0)$
   \rule[-1.5ex]{0pt}{4ex}\\
    \hline
   $h+\frac{3}{2}$ & $(0;0)$  & $(0;1)$ & $(0;1)$ & $(0;0)$
   \rule[-1.5ex]{0pt}{4ex}\\
  \hline
  \end{tabular}
      \caption{\small The long ${\cal N}=(3,3)$ multiplet $(0;0)_{\rm long}$.} \label{N3long}
\end{table}
As the deformation is switched off, we find the ${\cal N}=(3,3)$
multiplet shortening:
 \bea
  (0;0)_{\rm long} &\rightarrow&
  (0;0)_{\rm S} +(1;0)_{\rm S} +(0;1)_{\rm S} + (1;1)_{\rm S} \;,
 \eea
where $(0;0)_{\rm S}$, $(1;0)_{\rm S}$, and $(0;1)_{\rm S}$ denote
unphysical multiplets without
propagating degrees of freedom, e.g.:
\bea
  \begin{tabular}{c||c|c|c|}
   \raisebox{-1.25ex}{$h_L$} \raisebox{1.25ex}{$h_R$} &
   $0$ & $\frac12$ & $1$
   \rule[-2ex]{0pt}{5.5ex}\\
    \hline\hline
   $0$ & $(0;0)$  &  & $-(0;0)$
   \rule[-1.5ex]{0pt}{4ex}\\
    \hline
   $\frac{1}{2}$ &  & &
   \rule[-1.5ex]{0pt}{4ex}\\
    \hline
   $1$ & $-(0;0)$  &  &  $(0;0)$
   \rule[-1.5ex]{0pt}{4ex}\\
    \hline
  \end{tabular}
  \qquad\quad
    \begin{tabular}{c||c|c|c|}
   \raisebox{-1.25ex}{$h_L$} \raisebox{1.25ex}{$h_R$} &
   $0$ & $\frac12$ & $1$
   \rule[-2ex]{0pt}{5.5ex}\\
    \hline\hline
   $\frac{1}{2}$ & $(1;0)$  & & $-(1;0)$
   \rule[-1.5ex]{0pt}{4ex}\\
    \hline
   $1$ & $(0;0)+(1;0)$  & &  $-(0;0)-(1;0)$
   \rule[-1.5ex]{0pt}{4ex}\\
    \hline
   $\frac{3}{2}$ & $(0;0)$  &  & $-(0;0)$
   \rule[-1.5ex]{0pt}{4ex}\\
  \hline
  \end{tabular}
  \nonumber
\eea
Negative states have to be interpreted here as in sec.~\ref{kkspec}.
For further details see \cite{Berg}.

\chapter{Outlook and Discussion}
In this thesis we analyzed massive states in Kaluza-Klein theories
through their spontaneously broken symmetries. We focused in
particular on the local `higher-spin' symmetries which are
required for the consistency of the spin-3/2 and spin-2 couplings.
For Kaluza-Klein supergravity on
$AdS_3\times S^3\times S^3\times S^1$ we discussed the effective
theories for the lowest (spin-1/2 and spin-1) supermultiplets
and for a spin-3/2 multiplet. While the former can be described
as gauged ${\cal N}=8$ supergravities -- in accordance with the
amount of supersymmetry that is preserved by the background --
the latter requires an enhancement of supersymmetry to ${\cal N}=16$.
We constructed this theory as a new example of a gauged
maximally-supersymmetric $AdS_3$ supergravity.
It was shown that this theory does not possess a phase,
where all supercharges are unbroken. Rather we found that the
vacuum at the origin of the scalar potential in fig. \ref{graph}
breaks already half of the supersymmetry, giving rise to
eight massive spin-3/2 fields via a super-Higgs mechanism.
Thus we confirmed the general expectation of
sec. \ref{higherspin} that the massive spin-3/2 states
appearing in Kaluza-Klein supergravity have to be accompanied
by spontaneously broken supersymmetries.
However, the puzzle remains how an infinite tower of
spin-3/2 fields could be coupled consistently to gravity.
According to the given reasoning this needs an infinite
number of supercharges, which naively would then lead to states of arbitrary
high spin, in conflict with the fact that Kaluza-Klein
supergravities contain only fields up to spin 2.
The constructed theory suggests, however, the following resolution.
The aforementioned no-go theorem actually relies on the assumption that
the theory admits a phase, where the entire supersymmetry is
unbroken, such that the field content can be organized into
representations of the required superalgebra. If this theory, instead,
does not possess such a phase, the no-go theorem simply does not apply.

The analogous problems for spin-2 states were discussed in chapter 4,
where we focused on Kaluza-Klein compactification of four-dimensional
gravity on an $S^1$.
There we showed that the unbroken phase, in which the spin-2 fields
are massless,  corresponds to the `decompactification limit',
i.e. the phase where the radius of the internal manifold
goes to infinity. Correspondingly, there is an infinite-dimensional
gauge symmetry, which ensured consistency of the gravity -- spin-2
couplings. The resulting three-dimensional theory has been constructed
as a Chern-Simons theory based on the affine Poincar\'e algebra.
We have moreover shown the existence of a geometrical interpretation
for these spin-2 theories, which can be applied to arbitrary
dimensions. This is an extension of the Riemannian
geometry underlying ordinary general relativity to a notion
of algebra-valued differential geometry.
While the Chern-Simons formulation is special to $D=3$, the latter
geometrical formulation exists in any dimension. Correspondingly,
we show in appendix~\ref{B} that the Chern-Simons theory related
to an arbitrary internal manifold is equivalent to the generalized
Einstein-Hilbert action based on algebra-valued differential geometry.
Finally we discussed the `broken phase', in which the spin-2
fields are supposed to become massive via a novel
spin-2-Higgs mechanism. It has been shown that this phase results
from a gauging of a subgroup of the rigid symmetry group, which
in turn deforms the spin-2 symmetries and induces a spin-2 mass term.
The entire construction had a deep analogy to the gauging of
supergravity:
First of all, the spin-2 symmetries in (\ref{variations}) are
`supersymmetries' in the sense that they transform fields of different
spin into each other, even though they do this in a purely bosonic
theory. Like in supergravity (see the discussion in \ref{higherspin})
here the consistency relies on the fact that also the metric transforms
under the spin-2 symmetries.
Concerning the gauging we observed that it is possible to
start in the ungauged phase from a formulation,
where all degrees of freedom are carried by scalars.
These scalars span in turn a (generalized) non-linear $\sigma$-model
manifold carrying an enhanced global symmetry, in this case the affine
extension of the Ehlers group. The gauge vectors in turn appeared
as purely topological gauge fields and combined with the
spin-2 fields and the metric into a Chern-Simons theory for an
extended algebra. This parallels the case of gauged supergravities,
where the compact gauge vectors combine with the metric and the
gravitinos into one of the Chern-Simons theories based on
AdS supergroups discussed in
sec.~\ref{gaugedsugra}.\footnote{Similarly, in \cite{Blencowe:1988gj}
consistent couplings of an infinite tower of higher-spin fields to
gravity have been constructed as a Chern-Simons theory of a
higher-spin algebra. The Chern-Simons theory of sec.~\ref{CSgauged}
provides the analogue for an infinite tower of spin-2 fields.}
Moreover, we argued in sec.~\ref{spin2matter} that generically
the gauging is only consistent with the spin-2 symmetries if
a scalar potential is introduced, which also parallels the
gauging of supergravity.

In chapter 5 we considered marginal deformations of the theories
constructed in chapter 3. We analyzed the effects of switching on
certain scalar fields in the spin-3/2 multiplet
$(\ft12,\ft12;\ft12,\ft12)_S$, which results in a spontaneous
partial supersymmetry breaking ${\cal N}=(4,4)\rightarrow {\cal N}=(4,0)$
as well as a breaking of part of the gauge group.
We have moreover determined the geometry of
the moduli space spanned by the scalars participating
in this symmetry breaking.
Similarly we examined a marginal deformation in the spin-1 multiplets,
which implied supersymmetry breaking to ${\cal N}=(3,3)$.
The required Higgs and
super-Higgs effects were analyzed as well as the resulting spectrum
and the reorganization into representations of the reduced
AdS supergroups. These explorations provide the first step of a
more detailed analysis of the dual field theories.

The presented work can be extended into various directions.
First of all it would be interesting to find the 10-dimensional
ancestors for the marginal deformations discussed in chapter 5.
We introduced already a possible 10-dimensional geometry in
sec.~\ref{lifting}, which would give rise to the required breaking
of the gauge group. It remains to be checked that
this metric can be embedded into an exact solution of
type IIB supergravity. The analogous question for the ${\cal N}=(3,3)$
deformation will be discussed in a forthcoming paper \cite{Berg}.

Furthermore, there are a variety of directions, in which the
investigation of the spin-2 symmetries of chapter 4 can be
extended.
First, the analysis of the spin-2 symmetries in the gauged phase
has only been performed for the pure Chern-Simons theory
(in which the Kaluza-Klein symmetries are realized as Yang-Mills
gauge transformations) and a particular subsector of the matter-coupled
theory. In order to apply this program to more complicated internal
manifolds, a systematic analysis of possible gaugings and the
resulting deformations of the spin-2 symmetries is necessary.
In addition, it would be interesting to study these questions
also for higher-dimensional geometries. A first step for this can be
found in appendix~\ref{B}.

Moreover, since we are ultimately interested in compactifications on AdS
spaces, the Kac-Moody algebra based on the Poincar\'e
group should be generalized such that it contains the
affine extension of the AdS group.
However, as $AdS_3\times S^1$ is not a solution of four-dimensional
AdS-gravity, this enforces already the introduction of a non-flat
compact manifold. For instance, one could analyze the $AdS_3\times S^3$
case, in which the diffeomorphism group of $S^3$ should appear as
a gauge group. One may expect the appearance of an extended algebra
of the form (\ref{bigalgebra}) and an associated Chern-Simons
description.

Finally, it would be crucial to study the supersymmetric case.
We argued already in sec.~\ref{truncation} that this could
presumably be achieved via the introduction of a super-Kac-Moody
algebra, whose Chern-Simons theory would possess an infinite
number of supersymmetries (${\cal N}=\infty$). Applied to the
case of, e.g.,  $AdS_3\times S^3$, this algebra should carry an AdS
supergroup as subgroup.
Apart from the finite number of supersymmetries preserved by the
background, all supercharges will be broken spontaneously.

In analogy to the spin-2 case one could think that any truncation
to a finite number of spin-3/2 fields larger than expected from
the background supergroup (corresponding to the `zero-modes')
is inconsistent. However, the theory constructed in chapter~3 shows
already that this is not true in general. As discussed in
\ref{truncation}, the infinite-dimensional symmetry underlying the
entire Kaluza-Klein tower on $AdS_3\times S^3\times S^3$ has to
contain not only the ${\cal N}=8$ supergroup determined by the
background, but also ${\cal N}=16$ algebras associated to the
additional massive spin-3/2 fields.
In contrast to compactifications on tori, the existence of consistent
subalgebras larger than the symmetry group of the zero-modes
can therefore not be excluded a priori.
It is amusing to speculate about the possibility that consistent
subalgebras may exist which have even more than $32$ real
supercharges. For instance, the next massive spin-3/2
multiplet on $AdS_3\times S^3\times S^3$, namely $(1,1;1,1)_S$
in (\ref{tower}), contains $72$ massive spin-3/2 fields, thus
requiring ${\cal N}=80$ supersymmetry!
Even though this sounds unlikely at first sight, there seems to be no
convincing argument excluding the existence of such a truncation.

\chapter{Appendices}
\appendix

\renewcommand{\thesection}{\Alph{section}}
\renewcommand{\theequation}{\Alph{section}.\arabic{equation}}

\section{The different faces of $E_{8(8)}$}\label{A}

The maximal supergravity theories in three dimensions are organized
under the exceptional group $E_{8(8)}$. In particular, their scalar
sector is given by a coset space $\sigma$ model with target space
$E_{8(8)}/SO(16)$. In this appendix, we describe the Lie algebra
$\mathfrak{e}_{8(8)}$ in different decompositions relevant
for the embedding of the gauge group and for the construction
of the embedding tensor in the main text.

\subsection{$E_{8(8)}$ in the $SO(16)$ basis}\label{A1}

The $248$-dimensional Lie algebra of $E_{8(8)}$ may be characterized
starting from its $120$-dimensional maximal compact subalgebra $\mathfrak{so}(16)$,
spanned by generators $X^{IJ}=X^{[IJ]}$ with commutators
 \bea
  [X^{IJ},X^{KL}]&=&\delta^{JK}X^{IL}-\delta^{IK}X^{JL}-\delta^{JL}X^{IK}
  +\delta^{IL}X^{JK}\;,
 \eea
where $I, J=1, \dots, 16$ denote $SO(16)$ vector indices.
The $128$-dimensional non-compact part of $\mathfrak{e}_{8(8)}$ is spanned by
generators $Y^A$ which transform in the fundamental spinorial representation of $SO(16)$,
i.e.~which satisfy commutators
 \bea
   [X^{IJ},Y^A]&=&-\ft12\,\Gamma_{AB}^{IJ}\,Y^B\;,
   \qquad
   [Y^A,Y^B]~=~\ft{1}{4}\,\Gamma_{AB}^{IJ}\,X^{IJ}\;.
 \eea
Here $A, B=1, \dots, 128$ label the spinor representation of $SO(16)$
and $\Gamma^{IJ}=\Gamma^{[I}\Gamma^{J]}$
denotes the antisymmetrized product of $SO(16)$ $\Gamma$-matrices.
Moreover, we use indices
$\dot{A}, \dot{B}=1, \dots, 128$ to label the conjugate spinor
representation of $SO(16)$.
In the main text, indices $\mathcal{M}, \mathcal{N} =1, \dots, 248$
collectively label the full Lie algebra of $E_{8(8)}$, i.e.~
$\{t^{\cM}\} = \{X^{IJ},Y^{A}\}$ with
\bea
[\,t^{\cM},t^{\cN}\,] &=& f^{\cM\cN}_{\hspace{1.5em}\cK}\,t^{\cK}\;.
\eea
The Cartan-Killing form finally is given by
 \bea\label{cartan}
  \eta^{\cM\cN}&=&\frac{1}{60}\,\text{tr}\,(t^{\cM}\,t^{\cN})~=~
  \frac{1}{60}\,
  f^{\cM\cK}_{\hspace{1.5em}\cL}\,f^{\cN\cL}_{\hspace{1.5em}\cK}\;.
 \eea

\subsection{$E_{8(8)}$ in the $SO(8,8)$ basis}\label{A2}

Alternatively, $\mathfrak{e}_{8(8)}$ may be built starting
from its maximal
subalgebra $\mathfrak{so}(8,8)$ spanned by 120 generators $X^{{\tt IJ}}$
with commutators
 \bea
  [X^{{\tt IJ}},X^{\tt KL}]&=&\eta^{\tt JK}X^{\tt IL}
  -\eta^{\tt IK}X^{\tt JL}-\eta^{\tt JL}X^{\tt IK}
  +\eta^{\tt IL}X^{\tt JK}\;,
 \eea
where $\tt{I}, \tt{J}, \dots$ now denote vector indices of $SO(8,8)$
and $\eta_{\tt IJ}=\text{diag}(-1,\dots,-1,1,\dots,1)$
is the $SO(8,8)$ invariant tensor.
Similarly to the above, the full $\mathfrak{e}_{8(8)}$
is obtained by adding 128 generators $\hat Q_{\tt A}$, ${\tt A} =1, \dots, 128$,
transforming in the spinor representation of $SO(8,8)$
\bea\label{rel1}
  [X^{\tt IJ},\hat Q_{\tt A}]&=&
  -\ft12 \Gamma^{{\tt IJ}}{}_{{\tt A}}{}^{{\tt B}}\,\hat Q_{\tt B}\;,
  \qquad
    [\hat Q_{{\tt A}},\hat Q_{{\tt B}}]~=~\ft14\,\eta_{{\tt IK}}\eta_{{\tt JL}}
  \eta_{{\tt BC}}\,
  \Gamma^{{\tt KL}}{}_{{\tt A}}{}^{{\tt C}}\,X^{{\tt IJ}}\;.
\eea
Here $\ga{{\tt IJ}}{{\tt B}}{{\tt A}}$
denote the (rescaled) $SO(8,8)$-generators in the spinor representation, i.e.
 \bea
  \Gamma^{{\tt IJ}}{}_{{\tt A}}{}^{{\tt B}}
   =\ft12(\ga{{\tt I}}{\dot{{\tt C}}}{{\tt A}}\,\gaquer{{\tt J}}{{\tt B}}
   {\dot{{\tt C}}}-\ga{{\tt J}}{\dot{{\tt C}}}{{\tt A}}\,
  \gaquer{{\tt I}}{{\tt B}}{\dot{{\tt C}}})\;,
 \eea
where the gamma matrices satisfy
 \bea
 \label{gammaso88}
  \ga{{\tt I}}{\dot{{\tt C}}}{{\tt A}}\,\gaquer{{\tt J}}{{\tt B}}{\dot{{\tt C}}}+
  \ga{{\tt J}}{\dot{{\tt C}}}{{\tt A}}\,\gaquer{{\tt I}}{{\tt B}}{\dot{{\tt C}}}
  =2\eta^{{\tt IJ}}\,\delta_{{\tt A}}^{\hspace{0.3em}{\tt B}}\;,
 \eea
with the transpose $\overline{\Gamma}$, and where $\tt{A}, \tt{B}, \dots$,
denote spinor indices and $\dot{\tt{A}}, \dot{\tt{B}}, \dots$,
conjugate spinor indices of $SO(8,8)$.
It is important to note that in contrast to the $SO(16)$ decomposition
described above, spinor indices in these equations are raised
and lowered not with the simple $\delta$-symbol but with the
corresponding $SO(8,8)$ invariant tensors $\eta_{\tt AB}$, $\eta_{\dot{\tt A}\dot{\tt B}}$
of indefinite signature (cf.~(\ref{eta88}) below).

\subsection{$E_{8(8)}$ in the $SO(8)\times SO(8)$ basis}\label{A22}

According to (\ref{embbb}), (\ref{16vs88})
in the main text, the two decompositions of sections~\ref{A1} and \ref{A2}
may be translated into each other upon further breaking down to
$SO(8)_{L}\times SO(8)_{R}$.
To this end, we use the decomposition $SO(8,8)\rightarrow SO(8)_{L}\times SO(8)_{R}$
with
\bea\label{so88}
   \f{16}_V&\rightarrow& (\f{8}_V,\f{1})\oplus (\f{1},\f{8}_V)\;, \nonumber\\
   \f{128}_S&\rightarrow& (\f{8}_S,\f{8}_C)\oplus (\f{8}_C,\f{8}_S)\;,\qquad
   \f{128}_C~\rightarrow~ (\f{8}_S,\f{8}_S)\oplus (\f{8}_C,\f{8}_C)\;,
\eea
corresponding to the split of $SO(8,8)$ indices:
 \bea
  {\tt I} = (\li{a},b), \qquad {\tt A} = (\alpha\dot{\beta}\;,
  \dot{\gamma}\delta)\;, \qquad
  \dot{{\tt A}}=(\alpha \beta, \dot{\gamma}\dot{\delta})\;.
 \eea
Here, $\li{a}, \li{b}, \dots$ and $a, b, \dots$ denote vector indices for
the left and the right $SO(8)$ factor, respectively, while
$\alpha, \beta, \dots$ and $\dot{\alpha}, \dot{\beta}, \dots$ denote
spinor and conjugate spinor indices, respectively, for both $SO(8)$ factors.
The invariant tensors
$\eta^{{\tt IJ}}$, $\eta_{{\tt AB}}$ and $\eta_{\dot{{\tt A}}\dot{{\tt B}}}$
in this $SO(8)$ notation take the form
 \begin{equation}\label{eta88}
  \begin{split}
   \eta^{{\tt IJ}}=\left(\begin{array}{cc} -\delta^{\li{a}\li{b}} & 0 \\
   0 & \delta^{ab} \end{array}\right)\;, \qquad
   \eta_{{\tt AB}}&=\left(\begin{array}{cc} \eta_{\alpha\dot{\alpha},
   \beta\dot{\beta}} & 0 \\ 0 &
   \eta_{\dot{\alpha}\beta,\dot{\gamma}\delta}\end{array}\right)
   =\left(\begin{array}{cc} \delta_{\alpha\beta}\delta_{\dot{\alpha}
   \dot{\beta}} & 0 \\ 0 & -\delta_{\dot{\alpha}\dot{\gamma}}
   \delta_{\beta\delta}\end{array}\right)\;, \\[1ex]
   \eta_{\dot{{\tt A}}\dot{{\tt B}}}
   &=\left(\begin{array}{cc} \eta_{\alpha\beta,\gamma\delta} & 0 \\
   0 & \eta_{\dot{\alpha}\dot{\beta},\dot{\gamma}\dot{\delta}}
   \end{array}\right)
   =\left(\begin{array}{cc} \delta_{\alpha\gamma}\delta_{\beta\delta} & 0 \\
   0 & -\delta_{\dot{\alpha}\dot{\gamma}}
   \delta_{\dot{\beta}\dot{\delta}}\end{array}\right)\;.
  \end{split}
 \end{equation}
It is straightforward to verify, that the $SO(8,8)$ gamma matrices~(\ref{gammaso88}) can be
expressed in terms of the $SO(8)$ gamma matrices $\Gamma^a_{\alpha\dot{\gamma}}$
as (see also~\cite{Nicolai:1986jk})
 \bea
 \ga{a}{\delta\epsilon}{\beta\dot{\gamma}} &=&
 \delta_{\beta\delta}
   \Gamma^a_{\epsilon\dot{\gamma}}
\;,
\qquad
 \ga{a}{\dot{\gamma}\dot{\delta}}{\dot{\alpha}\beta}
   ~=~
   -\delta_{\dot{\alpha}\dot{\gamma}}\Gamma_{\beta\dot{\delta}}^a\;,
   \nonumber\\
   \ga{\li{a}}{\gamma\delta}{\dot{\alpha}\beta}
   &=&
   \delta_{\beta\delta}
   \Gamma_{\gamma\dot{\alpha}}^a
   \;,
   \qquad
   \ga{\li{a}}{\dot{\gamma}\dot{\delta}}{\alpha\dot{\beta}}~=~
   -\delta_{\dot{\beta}\dot{\delta}}
   \Gamma^a_{\alpha\dot{\gamma}}\;.
 \eea

With the results from the previous section,
$\mathfrak{e}_{8(8)}$ can now explicitly
be given in the $\mathfrak{so}(8)_{L} \oplus \mathfrak{so}(8)_{R}$ basis.
Generators split according to $\{X^{ab}, X^{\li{a}\li{b}},X^{a\li{b}},
\hat Q_{\alpha\dot{\beta}},\hat Q_{\dot{\gamma}\delta}\}$
with the commutation relations
\bea
 \label{e81}{}
   [X^{ab},X^{cd}]&=&\delta^{bc}X^{ad}
   -\delta^{ac}X^{bd}-\delta^{bd}X^{ac}
   +\delta^{ad}X^{bc}\;, \nonumber\\[.5ex] {}
   [X^{\li{a}\li{b}},X^{\li{c}\li{d}}]&=&
   -\delta^{\li{b}\li{c}}X^{\li{a}\li{d}}
   +\delta^{\li{a}\li{c}}X^{\li{b}\li{d}}
   +\delta^{\li{b}\li{d}}X^{\li{a}\li{c}}
   -\delta^{\li{a}\li{d}}X^{\li{b}\li{c}}\;, \nonumber\\[.5ex] {}
   [X^{ab},X^{c\li{d}}]&=&\delta^{bc}X^{a\li{d}}-\delta^{ac}X^{b\li{d}}\;,
   \nonumber\\[.5ex] {}
   [X^{\li{a}\li{b}},X^{c\li{d}}]&=&\delta^{\li{a}\li{d}}X^{c\li{b}}
   -\delta^{\li{b}\li{d}}X^{c\li{a}}\;, \nonumber\\[.5ex] {}
   [X^{a\li{b}},X^{c\li{d}}]&=&\delta^{\li{b}
   \li{d}}X^{ac}-\delta^{ac}
   X^{\li{b}\li{d}}\;,\nonumber\\[1.5ex] {}
{}      [X^{ab},\hat Q_{\alpha\dot{\beta}}]&=&-\ft12\overline{\Gamma}{}^{[a}
   _{\dot{\beta}\epsilon}\Gamma_{\epsilon\dot{\delta}}^{b]}
   \hat Q_{\alpha\dot{\delta}}\;,
   \qquad [X^{ab},\hat Q_{\dot{\alpha}\beta}]~=~-\ft12\Gamma_{\beta\dot{\zeta}}
   ^{[a}\overline{\Gamma}{}_{\dot{\zeta}\delta}^{b]}\hat Q_{\dot{\alpha}\delta}\;,
   \nonumber\\[.5ex] {}
   [X^{\li{a}\li{b}},\hat Q_{\alpha\dot{\beta}}]&=&\ft12\Gamma_{\alpha
   \dot{\epsilon}}^{[a}\overline{\Gamma}{}_{\dot{\epsilon}\gamma}^{b]}
   \hat Q_{\gamma\dot{\beta}}\;, \qquad
   [X^{\li{a}\li{b}},\hat Q_{\dot{\alpha}\beta}]~=~
   \ft12\overline{\Gamma}{}_{\dot{\alpha}\epsilon}^{[a}
   \Gamma_{\epsilon\dot{\gamma}}^{b]} \hat Q_{\dot{\gamma}\beta}\;, \nonumber\\[.5ex] {}
   [X^{a\li{b}},\hat Q_{\alpha\dot{\beta}}]&=&
   \ft12\Gamma_{\delta\dot{\beta}}^{a} \Gamma_{\alpha\dot{\gamma}}^{b}
   \hat Q_{\dot{\gamma}\delta}\;,
   \qquad
   [X^{a\li{b}}\;,\hat Q_{\dot{\alpha}\beta}]~=~
   \ft12\Gamma_{\beta\dot{\delta}}^{a} \Gamma_{\gamma\dot{\alpha}}^{b}
   \hat Q_{\gamma\dot{\delta}}\;,
   \nonumber\\[1.5ex] {}
      [\hat Q_{\alpha\dot{\alpha}},\hat Q_{\beta\dot{\beta}}]&=&\ft14\delta_{\alpha\beta}
   \overline{\Gamma}{}^{[a}_{\dot{\alpha}\gamma}\Gamma^{b]}_{\gamma\dot{\beta}}
   X^{ab}-\ft14\delta_{\dot{\alpha}\dot{\beta}}\Gamma^{[a}_
   {\alpha\dot{\gamma}}\overline{\Gamma}{}_{\dot{\gamma}\beta}^{b]}
   X^{\li{a}\li{b}}\;, \nonumber\\[.5ex] {}
   [\hat Q_{\alpha\dot{\alpha}},\hat Q_{\dot{\beta}\beta}]&=&
   -\ft12\Gamma_{\beta\dot{\alpha}}^{a}
   \Gamma_{\alpha\dot{\beta}}^{b} X^{a\li{b}}\;, \nonumber\\[.5ex] {}
   [\hat Q_{\dot{\alpha}\alpha},\hat Q_{\dot{\beta}\beta}]
   &=& \ft14\delta_{\alpha\beta}\overline{\Gamma}{}_{\dot{\alpha}\gamma}^{[a}
   \Gamma_{\gamma\dot{\beta}}
   ^{b]}X^{\li{a}\li{b}}
   -\ft14\delta_{\dot{\alpha}\dot{\beta}}\Gamma_{\alpha\dot{\gamma}}^{[a}
   \overline{\Gamma}{}_{\dot{\gamma}\beta}^{b]}X^{ab}\;.
 \eea

Moreover, the Cartan-Killing form (\ref{cartan}) can be computed
in the $SO(8)\times SO(8)$ basis by
use of this explicit form of the structure constants.
The result is
 \begin{equation}
\label{CK2}
  \begin{split}
   \eta^{ab,cd}&=-\delta^{[ab],[cd]}\;, \qquad \eta^{\hat{a}\hat{b},
   \hat{c}\hat{d}}=-\delta^{[\hat{a}\hat{b}],[\hat{c}\hat{d}]}\;, \qquad
   \eta^{a\hat{b},c\hat{d}}=\delta^{ac}\delta^{\hat{b}\hat{d}}\;, \\
   \eta_{\alpha\dot{\beta},\gamma\dot{\delta}}&=\delta_{\alpha\gamma}
   \delta_{\dot{\beta}\dot{\delta}}\;, \qquad
   \eta_{\dot{\alpha}\beta,\dot{\gamma}\delta}=-\delta_
   {\dot{\alpha}\dot{\gamma}}\delta_{\beta\delta}\;,
  \end{split}
 \end{equation}
while all other components vanish.

Finally let us identify explicitly the $SO(16)$ subalgebra in this
$SO(8,8)$ basis. With respect to the $SO(8)\times SO(8)$ decomposition
of $SO(16)$ in (\ref{so16deco}) the indices split according to
 \bea
  I=(\dot{\alpha},\beta)\;, \qquad A=(\alpha\dot{\beta},\hat{a}b)\;,
  \qquad \dot{A}=(\alpha a,\hat{b} \dot{\beta})\;.
 \eea
Correspondingly, the $SO(16)$ generators $X^{IJ}$ decompose into
$X^{\alpha\beta}$, $X^{\dot{\alpha}\dot{\beta}}$
and $X^{\alpha\dot{\beta}}$, and can be written in terms of the
compact $E_{8(8)}$ generators by
 \bea\label{basechange}
  X^{\alpha\beta}=\ft{1}{2}\Gamma_{\alpha\beta}^{ab}X^{ab}\;, \qquad
  X^{\dot{\alpha}\dot{\beta}}=-\ft{1}{2}
  \Gamma_{\dot{\alpha}\dot{\beta}}^{\hat{a}\hat{b}}X^{\hat{a}\hat{b}}\;,
  \qquad
  X^{\alpha\dot{\beta}}= \hat Q_{\dot{\beta}\alpha}\;.
 \eea
That these satisfy the $SO(16)$ algebra can be verified explicitly
by use of standard gamma matrix identities. The noncompact generators $Y^{A}$
are identified as
\bea
Y^{\alpha\dot{\beta}} =  \hat{Q}_{\alpha\dot{\beta}}
\;,\qquad
Y^{\hat{a}b} = X^{b\hat{a}} \;.
\eea
One immediately verifies that this split into compact and noncompact generators
is in agreement with the eigenvalues of the Cartan-Killing form~(\ref{CK2}).

\subsection{$E_{8(8)}$ in the $SO(4)\times SO(4)$ basis}\label{A3}

To explicitly describe the embedding of the gauge group
$G_{0}={G}_{\rm c} \ltimes (\hat{T}_{34}, {T}_{12})$
described in section~\ref{embed}, we finally need the decomposition
of $E_{8(8)}$ under the $SO(4)_{L}\times SO(4)_{R}$ from~(\ref{so34}).
This is obtained from the previous section
upon further decomposition according
to~(\ref{embedding33}), (\ref{so4embed}).
In $SO(8)_{R}$ indices $a$, $\alpha$, $\dot{\alpha}$, this corresponds to the splits
 \bea\label{indexsplit}
  a=([ij],0,\bar0)\;, \qquad \alpha =(i,j)\;, \qquad
  \dot{\alpha}=(i,j)\;,
 \eea
and similarly for $SO(8)_{L}$. Here, $i, j, \dots$ denote $SO(4)$ vector indices.
The $SO(8)$ gamma matrices can then be expressed
in terms of the invariant $SO(4)$ tensors $\delta^{ij}$ and $\varepsilon^{ijkl}$ as
 \bea\label{gamma}
  \Gamma^{ij}=\left(\begin{array}{cc} \varepsilon^{ij} & 2\delta^{ij} \\
  -2\delta^{ij} & \varepsilon^{ij}\end{array}\right)\;, \qquad
  \Gamma^{0}=\left(\begin{array}{cc} \textbf{1} & 0 \\
  0 & -\textbf{1}\end{array}\right)\;, \qquad
  \Gamma^{\bar0}=\left(\begin{array}{cc} 0 & \textbf{1} \\
  \textbf{1} & 0 \end{array}\right)\;,
 \eea
 with the $4 \times 4$ matrices
  \bea
 {\bf 1}_{kl} &=& \delta_{kl}\;,\qquad
  (\varepsilon^{ij})_{kl}~=~\varepsilon^{ijkl}\;, \qquad
  (\delta^{ij})_{kl}~=~\delta^{ij}_{kl}~=~
  \delta^{i[k}\delta^{l]j}\;.
 \eea
It is straightforward to check that the matrices~(\ref{gamma}) satisfy the standard
Clifford algebra, making use of the relations
 \begin{equation}
  \begin{split}
   \delta^{ij}(\delta^{mn})^t
+   \delta^{mn}(\delta^{ij})^t
+ \varepsilon^{ij}(\varepsilon^{mn})^t
+ \varepsilon^{mn}(\varepsilon^{ij})^t
   &~=~2\delta^{ij,mn}\,\textbf{1}, \\
   \varepsilon^{ij}(\delta^{mn})^t+\varepsilon^{mn}(\delta^{ij})^t
    - \delta^{ij}(\varepsilon^{mn})^t- \delta^{mn}(\varepsilon^{ij})^t&~=~0,
  \end{split}
 \end{equation}
which can be proved using the identity $\varepsilon^{[ijkl}\delta^{m]}_n=0\,$.
Next we have to decompose these $\Gamma$-matrices into selfdual and
anti-selfdual parts, corresponding to (\ref{indexsplit}),
 \bea
   \Gamma^{ij}_{\pm}=\frac{1}{\sqrt{2}}
   \big(\Gamma^{ij}\pm\frac{1}{2}\varepsilon^{ijkl}\Gamma^{kl}\big),
 \eea
such that $\tilde{\Gamma}^{ij}_{\pm}
:=\frac{1}{2}\varepsilon^{ijkl}\Gamma^{kl}_{\pm}=\pm\Gamma^{ij}_{\pm}$.
Inserting the representation (\ref{gamma}) of $\Gamma$-matrices
into the structure constants in~(\ref{e81}) yields the
decomposition of
$\mathfrak{e}_{8(8)}$
in the $\mathfrak{so}(4)_{L} \oplus \mathfrak{so}(4)_{R}$ basis.

\section{Kac-Moody and Virasoro algebras}\label{km}
Here we give a short introduction into the subject of
Kac-Moody algebras and the Virasoro algebra.
For the special case that their central charges vanish
they have a simple geometrical meaning, and therefore
we discuss this case first.

A Kac-Moody algebra $\hat{\frak{g}}$ is associated to an
ordinary finite-dimensional Lie group\footnote{This one is usually,
but not in this thesis, taken to be semi-simple and compact.} $G$
as the Lie algebra of the so-called loop group.
The latter is defined as the set of smooth maps from the unit
circle $S^1$ into the group $G$. Parametrizing the $S^1$ by
an angle $\theta$ as in the main text, the loop group ${\cal G}$
consists of periodic maps
 \bea
  \theta \longrightarrow \gamma(\theta) \in G \;,
 \eea
whose group structure is given by point-wise multiplication.
It defines an infinite-dimensional Lie group.
To determine the Lie algebra $\hat{\frak{g}}$ of this group,
we introduce a basis  $t^a$ for the Lie algebra $\frak{g}$ of
$G$. The loop group elements can then be written as
 \bea
  \gamma(\theta)=\exp [ \alpha_a(\theta) t^a ] \;,
 \eea
or, near the identity, as
 \bea\label{lieexp}
  \gamma(\theta)\approx 1 + t^a\sum_{n=-\infty}^{\infty} \alpha_a^n
  e^{in\theta}\;,
 \eea
where we have performed a Fourier expansion of $\alpha$.
This relation implies that the generators of the loop group
can be identified with
 \bea\label{gendef}
  t_n^a=t^a e^{in\theta}\;,
 \eea
since then the expansion (\ref{lieexp}) reads
 \bea
  \gamma \approx 1 + \sum_{n,a}\alpha_a^n t_n^a\;.
 \eea
With (\ref{gendef}) the Lie algebra $\hat{\frak{g}}$ can be computed
to be
 \bea\label{kmalgebra}
  [t_m^a,t_n^b] = f^{ab}_{\hspace{0.8em}c}\hspace{0.3em} t_{m+n}^c\;,
  \qquad m,n = -\infty,...,\infty \;,
 \eea
where $f^{ab}_{\hspace{0.8em}c}$ are the structure constants of $G$.
(\ref{kmalgebra}) defines the so-called Kac-Moody algebra or the
affine extension of $G$. The Lie algebra $\frak{g}$ of $G$ is
embedded as the subalgebra $\hat{\frak{g}}_0$ spanned by the
generators $t_0^a$. Moreover, the generators satisfy
$(t^a_n)^{\dagger}=t^a_{-n}$.

Next let us turn to the Virasoro algebra. While the loop group
can be defined as the group of maps from $S^1$ into a Lie group
$G$, the Virasoro algebra $\hat{v}$ can be introduced as the Lie
algebra of the Diffeomorphism group of $S^1$. This
diffeomorphism group ${\cal V}=\text{Diff}(S^1)$ consists
of smooth and invertible maps $S^1\rightarrow S^1$, whose
group multiplication is defined by composition
 \bea
  (\xi_1 \cdot \xi_2)(\theta) = \xi_1(\xi_2(\theta))\;.
 \eea
Infinitesimally they are given by
 \bea
  \theta\longrightarrow \theta - \alpha (\theta)\;,
 \eea
where $\alpha$ is periodic in $\theta$. On functions $f$ on $S^1$
these diffeomorphisms act as
 \bea
  f(\theta) \longrightarrow f(\theta) - \alpha(\theta)\frac{d}{d\theta}
  f(\theta)\;.
 \eea
Upon expanding $\alpha$ in Fourier components, one can
read of the generators of $\hat{v}$:
 \bea\label{visgen}
  Q_n = -e^{in\theta}\frac{d}{d\theta}\;.
 \eea
The resulting Lie algebra is then given by
 \bea\label{vhat}
  [ Q_m,Q_n ] = i(m-n)Q_{m+n}\;.
 \eea
This is the so-called Virasoro algebra $\hat{v}$.
In analogy to $\hat{\frak{g}}$ the generators fulfill the
reality constraint $Q_n^{\dagger}=Q_{-n}$.

Since both type of algebras, the Kac-Moody algebras and the
Virasoro algebra, are defined geometrically with regard to
$S^1$, there exists an obvious way to interrelate both algebras.
In fact, any diffeomorphism $\xi\in {\cal V}$ can act on a loop group
element $\gamma\in {\cal G}$ as
 \bea
  (\xi \cdot \gamma)(\theta) = \gamma(\xi(\theta))\;.
 \eea
This defines a semi-direct product $\hat{v}\ltimes \hat{\frak{g}}$
with elements $(\xi,\gamma)$, whose Lie algebra can be computed
by use of (\ref{gendef}) and (\ref{visgen}) to be given by
(\ref{kmalgebra}), (\ref{vhat}) and
 \bea\label{semi-direct}
  [ Q_m,t_n^a ] = -n \hspace{0.3em} t_{m+n}^a\;.
 \eea

Let us now turn to the more general case of non-vanishing
central charges. One may check explicitly that consistent
Lie algebras (in the sense of fulfilling the Jacobi identities)
can be defined with the following central extensions. The
latter are defined as generators $c$ that commute with
all generator, e.g.~$[Q_m,c]=0$.
The centrally extended Virasoro and Kac-Moody algebras then read
 \bea\label{centralex}
  \begin{split}
   [ Q_m,Q_n ] &= i(m-n)Q_{m+n} + \frac{c}{12}m(m^2-1)\delta_{m,-n}\;, \\
   [t_m^a,t_n^b] &= f^{ab}_{\hspace{0.8em}c}\hspace{0.3em} t_{m+n}^c
   +km\delta^{ab}\delta_{m,-n}\;.
  \end{split}
 \eea
In the case of vanishing central charge $c=0$ the Virasoro algebra
is also called Witt algebra.
Moreover, the semi-direct product structure (\ref{semi-direct})
is consistent also with the central extensions in (\ref{centralex}).
The structure of this semi-direct product can also be
obtained via the so-called Sugawara construction, in which
the Virasoro generators are realized as bilinears of Kac-Moody
generators \cite{Goddard:1986bp}.
It should be noted that for generic Lie groups $G$
this is a unique product between the associated Kac-Moody algebra
and $\hat{v}$, in contrast to the affine Poincar\'e algebra
$\widehat{iso(1,2)}$ discussed in the main text.

\subsubsection{Representations of $\hat{v}$}
Let us briefly summarize the representations of the Witt algebra,
which are relevant in the main text \cite{Cho:1991xk}.
First, the adjoint representation
can clearly be defined also for infinite-dimensional
Lie algebras and is here given by
 \bea
  \delta \chi^n = -\sum_{k,m}\xi^k (t_k)^n_{\hspace{0.2em}m}\chi^m=
  -\sum_{k,m}\xi^k f^n_{\hspace{0.2em}km}\chi^m
  = i\sum_{k}(n-2k)\xi^k\chi^{n-k}\;,
 \eea
where $f^n_{\hspace{0.2em}km}$ are the structure constants of $\hat{v}$
defined by (\ref{vhat}).
In analogy to this a much broader class of representations
can be defined according to
 \bea\label{virrep}
  \delta \chi^{n}=i \sum_k (n-(1-\Delta)k)\xi^k \chi^{n-k}\;.
 \eea
One can check explicitly, that these are representations of $\hat{v}$:
 \bea
  [\delta_{\xi^m},\delta_{\xi^n}]\chi^k=i(m-n)\delta_{\xi^{m+n}}
  \chi^k\;, \qquad \xi^{m+n} = \xi^m\xi^n\;.
 \eea
The representations of $\hat{v}$ can therefore be labeled
by a number $\Delta$, which we call conformal dimension
in analogy to conformal field theories. The adjoint representation
is included in (\ref{virrep}) as $\Delta=-1$.

Among these representations is the dual of the adjoint representation,
which can be characterized as follows.
For each
representation $\rho$ on a vector space $V$ one has the dual representation
$\rho^{*}$ on the dual space $V^{*}$, which is defined by the requirement
$\left<\rho^*(g)(v^*),\rho(g)(v)\right>=\left<v^*,v\right>$, where
$\left<\hspace{0.1em},\right>$
denotes the natural pairing between vectors in $V$ and
$V^*$ and $g$ is a group element. This implies
$\rho^*(g)=\trans{\rho}(g^{-1})$ or at the level of the Lie algebra
$\rho^*(X)=-\trans{\rho}(X)$, where $X\in \frak{g}$.
Since the adjoint representation
is given by $(t_m)^n_{\hspace{0.3em}k}=f_{mk}^n$,
the co-adjoint representation matrices read
$(t_k^*)^{\hspace{0.4em}m}_n=-(t_k)^m_{\hspace{0.4em}n}=-f^{m}_{kn}$.
Applied to the Witt algebra, the co-adjoint action is
 \bea
  \delta\chi^*_n = -\sum_{k,m}\xi^k(t^*_k)_n^{\hspace{0.4em}m}\chi^*_m
  =i\sum_k (k-n)\xi_k \chi^*_{n+k}\;,
 \eea
or, defining $\chi^n:=\chi_{-n}^*$,
 \bea
  \delta\chi^n = i\sum_k (n+k)\xi^k\chi^{n-k}\;,
 \eea
which coincides with the representation (\ref{virrep}) for $\Delta =2$.

\newpage

\section{Kaluza-Klein action on $\mathbb{R}^3\times S^1$ with Yang-Mills
type gauging}\label{reduc}

As we have already emphasized, the inclusion of all Kaluza-Klein
modes in the effective action for reductions on $S^1$ (or in general
on arbitrary tori) can also be done explicitly,  and in fact have been
done for reductions to $D=4$
(see \cite{Cho:1992rq,Cho:1991xk,Cho:1992xv,Aulakh:1985un,Aulakh:1984qx}).
Here we will show this for the $D=4\rightarrow D=3$ reduction of
four-dimensional gravity, which yields the Kaluza-Klein theory with
Yang-Mills gauge fields.

In practice the computation is significantly simplified
by use of the vielbein formalism. More specifically, the
Einstein-Hilbert action
 \bea\label{Einst}
  S_{\text{EH}}=-\int d^4 x E R = -\int d^4 x E E^M_A E_B^N
  R_{MN}^{\hspace{0.5em}AB}
 \eea
can be computed from the components of the spin-connection
$\omega_M^{\hspace{0.3em}AB}$ by use of
 \bea
   R_{MN}^{\hspace{0.5em}AB} = 2\partial_{[M}\omega_{N]}^{\hspace{0.3em}AB}
   +2 \omega_{[M}^{\hspace{0.3em}AC}\omega_{N]C}^{\hspace{0.5em}B}\;,
 \eea
where the spin connection in flat indices is given by
 \bea\label{connection}
  \begin{split}
   \omega_{ABC}&=\frac{1}{2}(\Omega_{ABC}-\Omega_{BCA}+\Omega_{CAB})\;, \\
   \Omega_{ABC}&=2E_{[A}^M E_{B]}^N\partial_M E_{NC}\;.
  \end{split}
 \eea
It is convenient to express the Einstein-Hilbert action
entirely in terms of $\Omega_{ABC}$. Inserting (\ref{connection})
into (\ref{Einst}) and performing several partial integrations one
gets
 \bea\label{EHomega}
  S_{\text{EH}}=-\int d^4 x E\left[ -\frac{1}{4}\Omega^{ABC}\Omega_{ABC}
  +\frac{1}{2}\Omega^{ABC}\Omega_{BCA}
  +\Omega_{C\hspace{0.3em}B}^{\hspace{0.4em}B}
  \Omega_{\hspace{0.4em}A}^{C\hspace{0.3em}A}\right]\;.
 \eea
Computing the $\Omega_{ABC}$ by use of the vielbein (\ref{metric})
and its inverse
 \bea\label{invmetric}
  E_A^M=\left(\begin{array}{cc} \phi^{1/2}e^{\mu}_a &
  -\phi^{1/2} e_a^{\rho}A_{\rho} \\ 0 & \phi^{-1/2} \end{array}\right)\;,
 \eea
one gets the following (still $\theta$-dependent) coefficients
 \bea
  \begin{split}\label{coeff}
   \Omega_{abc}&=\phi^{1/2}\left[ \hat{\Omega}_{abc}^{(3)}-
   e_{[a}^{\mu}\eta_{b]c}\phi^{-1}D_{\mu}\phi\right]\;, \\
   \Omega_{ab}^{\hspace{0.6em}5}&=\phi^{3/2}e^{\mu}_ae^{\nu}_bF_{\mu\nu}\;, \\
   \Omega_{5b}^{\hspace{0.6em}5}&=-\frac{1}{2}
   \phi^{-1/2}e_b^{\mu}D_{\mu}\phi \;, \\
   \Omega_{5b}^{\hspace{0.6em}c}&=\phi^{-1/2}e^{\mu}_bD_5e_{\mu}^c\;.
  \end{split}
 \eea
All expressions appear already in a $\hat{v}$-covariant
fashion.\footnote{Note, that this is not the case for the
$\omega_M^{\hspace{0.4em}AB}$, where one index has been transformed into
a space-time index by use of the vielbein (\ref{metric}).}
Specifically,
 \bea
  \hat{\Omega}_{abc}^{(3)}=2e_{[a}^{\mu}e_{b]}^{\nu}D_{\mu}e_{\nu c}\;,
 \eea
with the $\hat{v}$-covariant derivative $D_{\mu}e_{\nu c}$ defined
in (\ref{ecov}).
Furthermore, we defined following \cite{Cho:1991xk} a covariantized
$x^5$-derivative, which is given by
 \bea
  D_5 e_{\mu}^a=\partial_5 e_{\mu}^a-\frac{1}{2}\phi^{-1}\partial_5\phi
  \hspace{0.1em}e_{\mu}^a\;.
 \eea
The denotation `covariant derivative' is justified by the fact
that $D_5e_{\mu}^a$
transforms in contrast to $\partial_5 e_{\mu}^a$ covariantly under
local $\hat{v}$ transformations:
\bea
  \delta_{\xi^5}(D_5 e_{\mu}^a)=\xi^5\partial_5(D_5 e_{\mu}^a)
  +2\partial_5\xi^5 D_5 e_{\mu}^a\;.
 \eea
This means, that $D_5e_{\mu}^a$ transforms covariantly,
even though it transforms under a different representation
as $e_{\mu}^a$ itself.

Inserting (\ref{coeff}) into the Einstein-Hilbert action in the
form (\ref{EHomega}) one finds after some computations
 \bea
   S_{\text{EH}}=\int d^3 xd\theta e\left[-R^{(3),\text{cov}}
   -\frac{1}{4}\phi^2F^{\mu\nu}F_{\mu\nu}+\frac{1}{2}\phi^{-2}g^{\mu\nu}
   D_{\mu}\phi D_{\nu}\phi + {\cal L}_m\right]\;,
 \eea
where $R^{(3),\text{cov}}$ denotes the generalized Ricci scalar
with respect to the covariantized connection in (\ref{coeff}).
Moreover, ${\cal L}_m$ contains the spin-2 mass term, which is induced by
the gauging, and reads
 \bea\label{spin2mass}
   {\cal L}_m= \frac{1}{4}\phi^{-2}g^{\mu\nu}
   g^{\rho\sigma}(D_5g_{\mu\rho}D_5g_{\nu\sigma}-D_5g_{\mu\nu}
   D_5g_{\rho\sigma})
   -e^{a\mu}e^{b\nu}F_{\mu\nu}e_b^{\rho}D_5e_{\rho a}\;.
 \eea
One may check explicitly that ${\cal L}_m$ is invariant under
local $\hat{v}$ transformations. In particular, the power of $\phi$
in front of the spin-2 mass term can be entirely determined
from the requirement that the action stays invariant.

That (\ref{spin2mass}) gives mass to the spin-2 fields in the
Kaluza-Klein vacuum (\ref{kkvacuum}) can be seen as follows.
The mass term in the free spin-2 theory (\ref{free})
already given by Pauli and Fierz reads
 \bea\label{PFmass}
  \mathcal{L}_{\text{mass}} = \frac{M^2}{2}\left(h^{\mu\nu}h_{\mu\nu}-
  (\eta^{\mu\nu}h_{\mu\nu})^2\right).
 \eea
In fact, expanding (\ref{spin2mass}) around the Kaluza-Klein vacuum
(\ref{kkvacuum})
 \bea
  g_{\mu\nu}^n(x)=\eta_{\mu\nu}+\kappa h_{\mu\nu}^n(x)\;,
 \eea
and integrating out $\theta$ one gets the leading contribution
 \bea
  \frac{1}{2\kappa^2}\int d\theta{\cal L}_m=\frac{1}{2}
  \sum_{n=-\infty}^{\infty}
  M_n^2\eta^{\mu\nu}\eta^{\rho\sigma}
  \left(h_{\mu\rho}^nh_{\nu\sigma}^{-n}
  -h_{\mu\nu}^nh_{\rho\sigma}^{-n}\right)+O(\kappa^2)\;,
 \eea
where
 \bea
  M_n^2=\left(\frac{n}{R}\right)^2\;.
 \eea
This in turn shows that in lowest order in $\kappa$ (\ref{spin2mass})
reduces to an infinite sum of Pauli-Fierz spin-2 mass
terms (\ref{PFmass}). Furthermore it can be shown that the vector and
scalar modes $A_{\mu}^n$ and $\phi^n$ can be absorbed via a
field redefinition into the spin-2 fields, such that in total
the latter become massive in a Higgs effect.
(At the linearized level this analysis has been performed, e.g.,
in \cite{Cho:1992rq}.)

\section{Spin-2 theory for arbitrary
internal manifold}\label{B}
In the main text we have shown that the Chern-Simons gauge theory
of the affine Poincar\'e algebra describes a consistent
gravity-spin-2 coupling. This is on the other hand also equivalent to
Wald's algebra-valued generalization of the Einstein Hilbert action,
where the algebra is given by the algebra of smooth functions
on $S^1$.
We are going to show that this picture generalizes to the case of
an arbitrary internal manifold.

Let $M$ be an arbitrary compact Riemannian manifold and
$\{e_m\}$ a complete set of spherical harmonics (where generically
$m$ now denotes a multi-index), which we also take as a basis for
the algebra of smooth functions on $M$.
The infinite-dimensional extension
of the Poincar\'e algebra is no longer given by a Kac-Moody algebra, but
instead is spanned by generators $P_a^n=P_a\otimes e_n$ and
$J_a^n=J_a \otimes e_n$, which satisfy the Lie algebra (compare the
algebra in \cite{Boulanger:2002bt})
 \bea\label{genalg}
  \begin{split}
   [P_a^m,P_b^n]&=0\;, \qquad \quad
   [J_a^m,J_b^n]=\varepsilon_{abc}J^{c}\otimes (e_m \cdot e_n)\;,  \\
   [J_a^m,P_b^n]&=\varepsilon_{abc}P^{c}\otimes (e_m \cdot e_n)\;,
  \end{split}
 \eea
Here $\cdot$ denotes ordinary multiplication of functions.
Note, that this algebra reduces for the case $M=S^1$ to the
Kac-Moody algebra $\widehat{iso(1,2)}$ in (\ref{kacmoody}).
There exists also an inner product on the space of functions, which
is given by
 \bea\label{L2}
  (e_m,e_n)=\int_M \text{dvol}_M e_m e_n\;,
 \eea
such that a non-degenerate quadratic form on (\ref{genalg})
exists:
 \bea
  \langle P_a^m,J_b^n \rangle =\eta_{ab} (e_m,e_n)\;.
 \eea
In complete analogy to sec. \ref{infspin-2} a Chern-Simons theory can then
be defined, whose equations of motion read
 \begin{equation}
  \begin{split}
   \partial_{\mu}e_{\nu}^{a(n)}-\partial_{\nu}e_{\mu}^{a(n)}
   +a^n_{mk}\varepsilon^{abc}\left(e_{\mu b}^{(m)}\omega_{\nu c}^{(k)}
   +\omega_{\mu b}^{(m)}e_{\nu c}^{(k)}\right)&=0\;, \\
   \partial_{\mu}\omega_{\nu}^{a(n)}-\partial_{\nu}\omega_{\mu}^{a(n)}
   +a^n_{mk}\varepsilon^{abc}\omega_{\mu b}^{(m)}\omega_{\nu c}^{(k)}&=0\;,
  \end{split}
 \end{equation}
while they are invariant under
 \bea
   \delta e_{\mu}^{a(n)}=\partial_{\mu}\rho^{a(n)}+
    a^n_{mk}\left(\varepsilon^{abc}
    e_{\mu b}^{(m)}\tau_c^{(k)}+\varepsilon^{abc}\omega_{\mu b}^{(m)}
    \rho_c^{(k)}\right),
 \eea
where $a^n_{mk}$ defines the algebra structure with respect to
the basis $\{e_m\}$.
If one defines the transformation parameter to be
 \bea\label{kkparam2}
  \rho^{a(n)}=a^n_{mk}\xi^{\mu(m)}e_{\mu}^{a(k)}, \qquad
  \tau^{a(n)}=a^n_{mk}\xi^{\mu(m)}\omega_{\mu}^{a(k)},
 \eea
one can show using the equations of motion and the associativity
(\ref{asso}) of the algebra, that
 \bea
  \delta e_{\mu}^{a(n)}=a_{mk}^n\left(\xi^{\rho(m)}\partial_{\rho}
  e_{\mu}^{a(k)}
  +\partial_{\mu}\xi^{\rho (m)}e_{\rho}^{a(k)}\right).
 \eea
For the algebra-valued metric defined by
$g_{\mu\nu}^n=a_{mk}^ne_{\mu}^{a(m)}e_{\nu a}^{(k)}$
this implies
 \begin{equation}
  \begin{split}
   \delta_{\xi}g_{\mu\nu}^n&=\partial_{\mu}\xi^{\rho (l)}
   g_{\rho\nu\hspace{0.3em}l}^{\hspace{0.6em}n}+\partial_{\nu}
   \xi^{\rho (l)} g_{\rho\mu\hspace{0.3em}l}^{\hspace{0.6em}n}
   +\xi^{\rho (l)}\partial_{\rho}
   g_{\mu\nu\hspace{0.3em}l}^{\hspace{0.6em}n} \\
   &=\nabla_{\mu}\xi_{\nu}^n+\nabla_{\nu}\xi_{\mu}^n\;,
  \end{split}
 \end{equation}
where again (\ref{asso}) has been used.
Altogether, the gauge transformations of the Chern-Simons theory
for (\ref{genalg}) coincide with the algebra-diffeomorphisms.
Thus we have shown that also for arbitrary internal manifolds
the Chern-Simons description based on the algebra (\ref{genalg})
is equivalent to Wald's algebra-valued multi-graviton theory.
This is turn confirms, that the spin-2 theories appearing in
Kaluza-Klein actions resulting from compactifications to
arbitrary dimensions can also be treated within Wald's framework.

\providecommand{\href}[2]{#2}\begingroup
\raggedright\endgroup


\end{document}